\documentclass[12pt]{article}
\usepackage[utf8]{inputenc}
\usepackage{setspace}

\newcommand{\blind}{0} 

\usepackage{latexsym,amsmath,amssymb,amsfonts,amsthm}
\usepackage{bm} 
\usepackage{bbm} 
\usepackage{mathrsfs}
\usepackage{graphicx}
\graphicspath{{./images/}}
\usepackage{booktabs}
\usepackage{multirow}
\usepackage{makecell}
\usepackage{adjustbox}
\usepackage{longtable}
\usepackage{rotating}

\usepackage{natbib}
\setcitestyle{authoryear,round,citesep={;},aysep={}}
\usepackage{xr-hyper}
\usepackage[colorlinks,citecolor=blue,urlcolor=blue]{hyperref}
\makeatletter
\newcommand*{\addFileDependency}[1]{\typeout{(#1)}\@addtofilelist{#1}\IfFileExists{#1}{}{\typeout{No file #1.}}}
\makeatother

\usepackage{xurl}

\usepackage{authblk}
\usepackage{float}
\usepackage{algorithm}
\usepackage{algpseudocode}
\usepackage{xcolor}
\usepackage{textcomp}

\newcommand{\samethanks}[1][\value{footnote}]{\footnotemark[#1]}

\title{Deep Classifier Kriging for Probabilistic Spatial Prediction of Air Quality Index}

\if0\blind
\author[1]{Junyu Chen\thanks{These authors contributed equally to this work.}}
\author[2]{Pratik Nag\samethanks}
\author[3]{Huixia Judy Wang}
\author[4]{Ying Sun}

\affil[1]{Department of Statistics, The George Washington University, Washington, DC, USA}
\affil[2]{Department of Statistics, University of Wisconsin-Madison, USA}
\affil[3]{Department of Statistics, Rice University, Houston, Texas, USA}
\affil[4]{CEMSE Division, Statistics Program, King Abdullah University of Science and Technology, Thuwal, Saudi Arabia}
\else
\author{}
\fi

\begin{document}
\date{}
\maketitle
\vspace{-2.5em} 

\begin{abstract}
Accurate spatial interpolation of the air quality index (AQI), computed from concentrations of multiple air pollutants, is essential for regulatory decision-making. AQI fields are inherently non-Gaussian and often exhibit complex nonlinear spatial structures that are not adequately captured by classical statistical modeling frameworks. Additionally, AQI measurements obtained from weather stations (`gold standard') are sparsely available, making it difficult for data-intensive neural network-based models to optimize effectively. To address these issues, we first propose a nonparametric \textit{data fusion} mechanism that enables the modeling of non-collocated bivariate processes and the integration of heterogeneous air pollution data sources. Second, we introduce \textit{deep classifier kriging} (DCK), a flexible, distribution-free deep learning framework for estimating full predictive distribution functions for univariate and bivariate spatial processes. Through extensive simulation studies, we validate the proposed \textit{data fusion} approach and DCK against several benchmark models. We then apply DCK to probabilistic spatial prediction of AQI by fusing sparse but high-quality station observations with spatially continuous yet biased auxiliary model outputs, yielding spatially resolved predictive distributions that support downstream tasks such as exceedance and extreme-event probability estimation for regulatory risk assessment and policy formulation.
\end{abstract}


\section{Introduction}\label{sec:intro}

Spatial prediction and uncertainty quantification are central objectives in environmental statistics, particularly when multiple related variables are observed over space. In many scientific and regulatory applications, interest lies not only in predicting individual spatial processes but also in jointly modeling multivariate responses to enable coherent probabilistic inference and conditional prediction. In the spatial setting, this joint dependence is learned at each unobserved location by borrowing information about dependence from nearby locations in the spatial fields, thereby enabling coherent conditional prediction and principled uncertainty quantification. From a kriging perspective, this corresponds to multivariate or co-kriging frameworks \citep{cressie1990origins}, where cross-dependence between processes is explicitly modeled to support joint and conditional probabilistic prediction. Such joint modeling is especially critical when variables are observed on different spatial supports or exhibit nonlinear and non-Gaussian relationships, circumstances that frequently arise in environmental and atmospheric sciences.

This need for joint probabilistic modeling is directly motivated by contemporary air-quality assessment. The air quality index (AQI) is the primary metric used in regulatory frameworks and public communication to summarize ambient air pollution levels and associated health risks. Defined on a standardized 0-500 scale, the AQI aggregates information from key pollutants regulated under the U.S. Clean Air Act, most notably PM$_{2.5}$, O$_3$, NO$_2$, SO$_2$, CO, and PM$_{10}$. Among these, fine particulate matter (PM$_{2.5}$) often dominates elevated AQI levels due to its strong and well-established links to adverse respiratory and cardiovascular outcomes \citep{lung2025peaks, board2024inhalable}. As a result, AQI and PM$_{2.5}$ are strongly related through EPA-defined, nonlinear breakpoint functions, making PM$_{2.5}$ a natural auxiliary variable for enhancing AQI prediction.

Therefore, fusing multiple complementary data sources becomes advantageous for achieving more reliable estimation of spatially resolved AQI fields \citep{fuentes2005model}. In practice, these sources consist of regulatory monitoring networks based on the Federal Reference Method (FRM), operated and maintained by the U.S. Environmental Protection Agency \citep{epaAQS}, and numerical atmospheric models. FRM monitors provide high-fidelity point-level measurements that are transformed into AQI values; however, these observations are sparse. In contrast, numerical atmospheric models such as the Community Multiscale Air Quality (CMAQ) \citep{epaCMAQdata} model from the U.S. Environmental Protection Agency, generate spatially complete PM$_{2.5}$ concentration fields by simulating emissions, transport, and atmospheric chemistry \citep{appel2020community}. Although CMAQ outputs offer rich spatial coverage, they are subject to structural and parametric biases. From a statistical perspective, this setting naturally motivates \textit{data fusion} and conditional prediction: combining accurate but sparse AQI observations with biased but spatially complete CMAQ model outputs to infer the conditional distribution of AQI given the CMAQ PM$_{2.5}$ fields. Joint modeling of these two processes provides a principled mechanism for leveraging their complementary strengths and directly addresses the scientific objective of spatially resolved, probabilistically calibrated AQI prediction by explicitly modeling their dependence structure.

Although conditional prediction naturally arises from joint modeling, fully specified joint spatial models can be challenging to construct when data are collected from heterogeneous sources. Over the past decades, several approaches have been developed to address this challenge. Classical approaches, such as co-kriging and Bayesian hierarchical spatial models, rely on Gaussian process assumptions to characterize multivariate spatial dependence and to provide closed-form predictors and uncertainty quantification \citep{cressie2015statistics}. In the context of air quality index prediction, \citet{fuentes2005model} proposed a hierarchical Bayesian framework that combines ground-based station measurements with numerical model outputs for spatial prediction. Building on this line of work, \citet{berrocal2010space} developed two modeling frameworks that integrate monitoring data with outputs from multiple numerical models, as well as historical station observations. While these approaches are theoretically elegant, their practical effectiveness depends critically on the specification of parametric cross-covariance structures and the assumption of Gaussianity. In practice, both PM$_{2.5}$ concentrations and AQI exhibit skewness, heavy tails, threshold effects, and complex nonlinear relationships, which can undermine these assumptions and lead to miscalibrated uncertainty, particularly for extremes that are most relevant for public health decision-making. Moreover, extending classical co-kriging approaches to large data sets may become infeasible due to their substantial computational complexity.

To address the rigidity of classical formulations and distributional assumptions, particularly in univariate and bivariate settings, deep learning approaches have recently been proposed as flexible alternatives for spatial and spatio-temporal modeling of air quality. Convolutional and recurrent neural networks have demonstrated strong empirical performance in capturing complex nonlinear patterns in atmospheric data \citep{cheng2021deep, zhang2018deep}. Methods such as \textit{DeepKriging} and its extensions embed spatial processes through basis representations that are passed through deep neural networks, enabling spatial dependence to be learned in a data-driven manner \citep{NAG2023100773, chen2020deepkriging, nag2025bivariate}. Although their approach is computationally efficient and provides valid predictions with uncertainty quantification, it is limited to collocated data and therefore cannot be readily applied in settings such as ours, where heterogeneous data sources must be fused through a fully data-driven mechanism without reliance on distributional assumptions.

Building on the limitations of existing approaches, the contributions of this paper are as follows. First, we propose a novel nonparametric \textit{data fusion} mechanism that enables joint modeling of bivariate spatial processes with non-collocated observations. Second, we introduce a novel deep learning-based framework, termed \textit{deep classifier kriging} (DCK), which bridges the gap between classical co-kriging methods that rely on restrictive distributional and model assumptions and existing deep learning approaches that lack a principled mechanism for joint probabilistic prediction from heterogeneous data sources.

As one real data application, we implement the proposed DCK model along with the \textit{data fusion} mechanism to jointly model FRM-based AQI observations and CMAQ-simulated PM$_{2.5}$ concentrations. The approach leverages the accuracy of regulatory observations together with the spatial richness of numerical models, resulting in improved predictive performance and calibrated uncertainty. We demonstrate the practical utility of the method through comprehensive analyses over three representative regions of the United States-California, the Northeast, and the Southeast-each exhibiting distinct and well-documented spatial pollution regimes.

The remainder of the paper is organized as follows. Section \ref{sec:exploratory_analysis} provides some exploratory analysis on the FRM-based AQI observations and CMAQ-simulated PM$_{2.5}$ concentrations. Section~\ref{sec:method} presents the \textit{data fusion} mechanism and the DCK model in detail. Section~\ref{sec:simulation} evaluates the proposed frameworks through controlled simulation studies. Section~\ref{sec:application} applies DCK to national-scale AQI prediction across the United States. Section~\ref{sec:discussion} concludes with a discussion of the main findings and directions for future research. Supplementary Material provides additional results and implementation details.

\section{Exploratory data analysis}\label{sec:exploratory_analysis}

\begin{figure}[t!]
\centering

\begin{tabular}{c}
\includegraphics[width=1\textwidth]{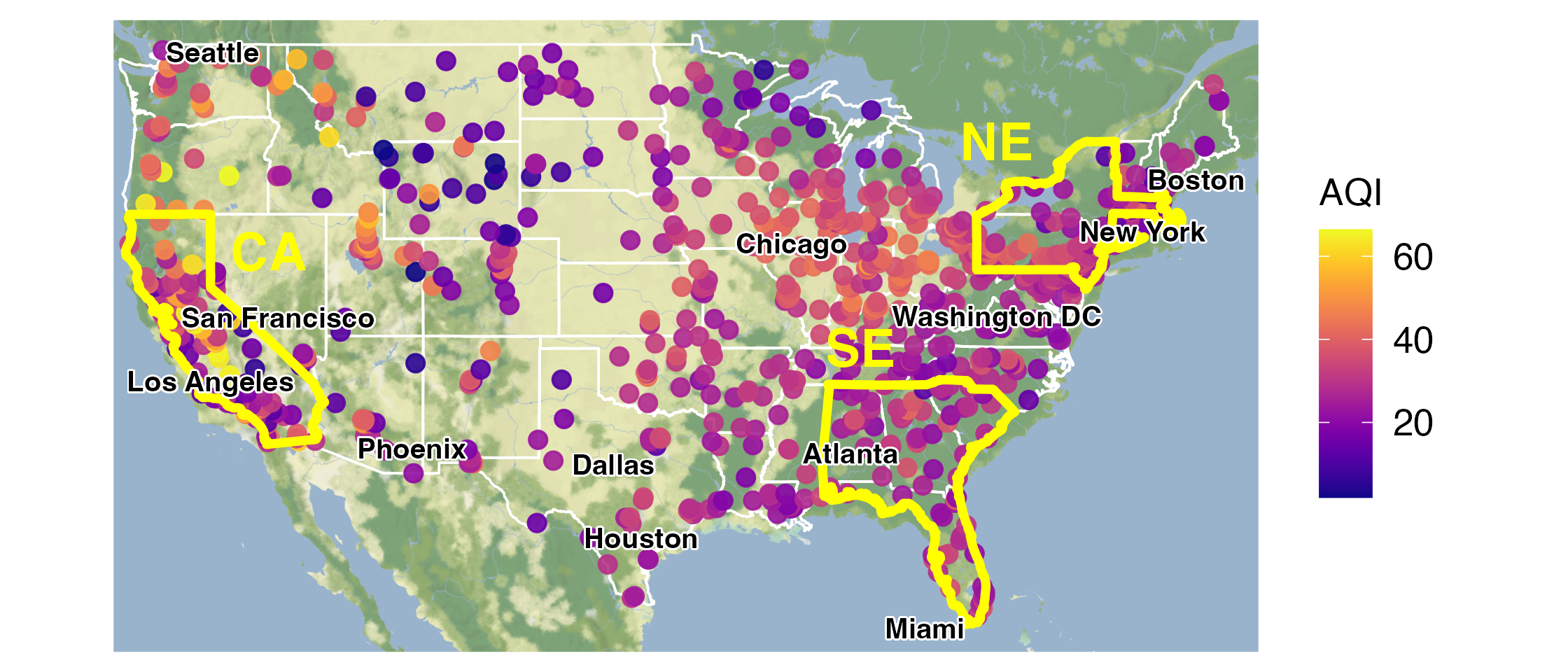} \\
\small (a) FRM AQI Observations \\[6pt]

\includegraphics[width=1\textwidth]{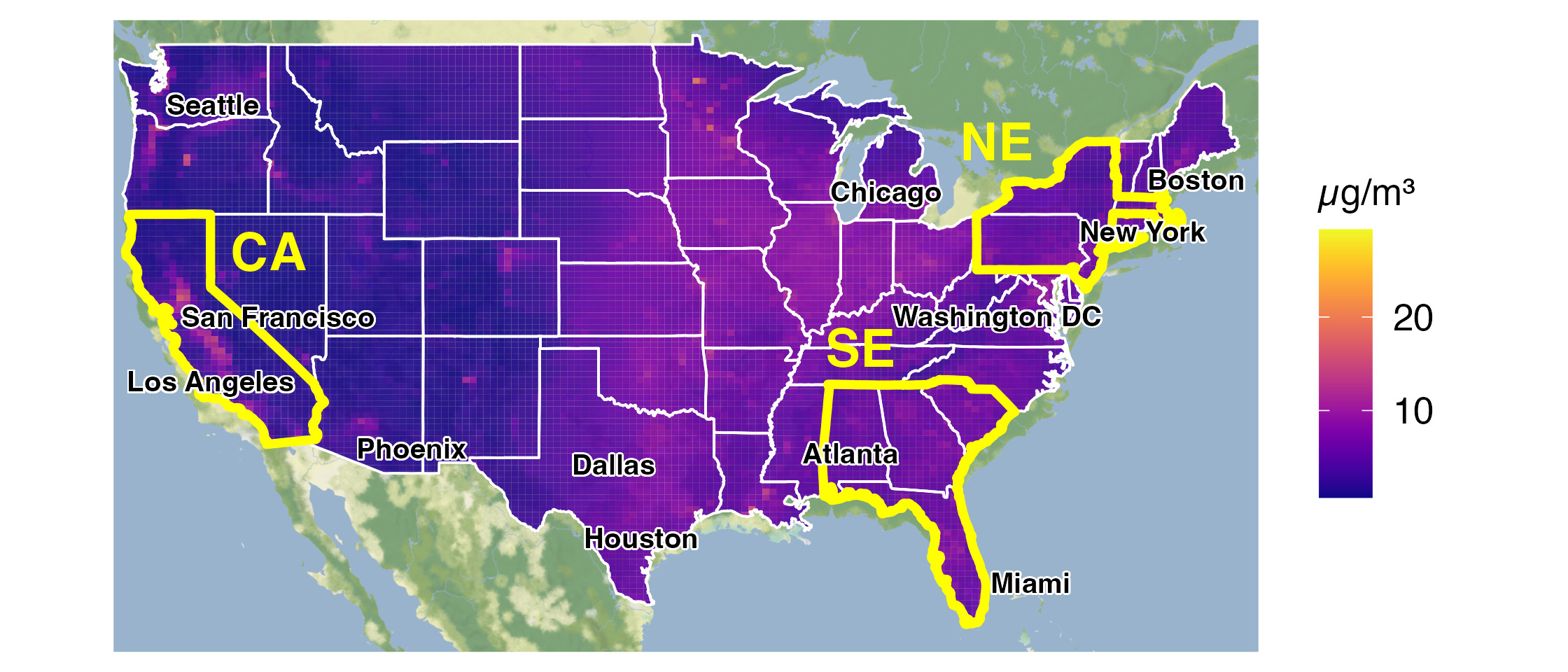} \\
\small (b) CMAQ-simulated PM$_{2.5}$ concentrations
\end{tabular}

\caption{Spatial heatmaps of (a) FRM AQI and (b) CMAQ-simulated PM$_{2.5}$ for January 2019. Regions highlighted in yellow denote California (CA), the Northeast (NE), and the Southeast (SE), which are selected for subsequent analysis.}
\label{fig:map_plot}
\end{figure}
In this section, we investigate the distributional and spatial characteristics of the FRM AQI and CMAQ-simulated PM$_{2.5}$ concentrations to motivate \textit{data fusion} and a distribution-free probabilistic joint modeling approach. The regulatory-grade FRM provides accurate but spatially sparse AQI observations, whereas the CMAQ model generates spatially dense but potentially biased PM$_{2.5}$ concentration fields on a regular grid. Figure~\ref{fig:map_plot} displays the spatial distribution of both datasets across the United States. As can be seen from the figure, AQI observations are sparsely distributed, while PM$_{2.5}$ concentrations from the CMAQ model outputs are available on a dense gridded structure. Moreover, these spatial patterns exhibit substantial heterogeneity, suggesting that the marginal distributions of both datasets may differ markedly from Gaussianity.

\begin{figure}[t!]
\centering

\begin{tabular}{c}
\includegraphics[width=0.60\textwidth]{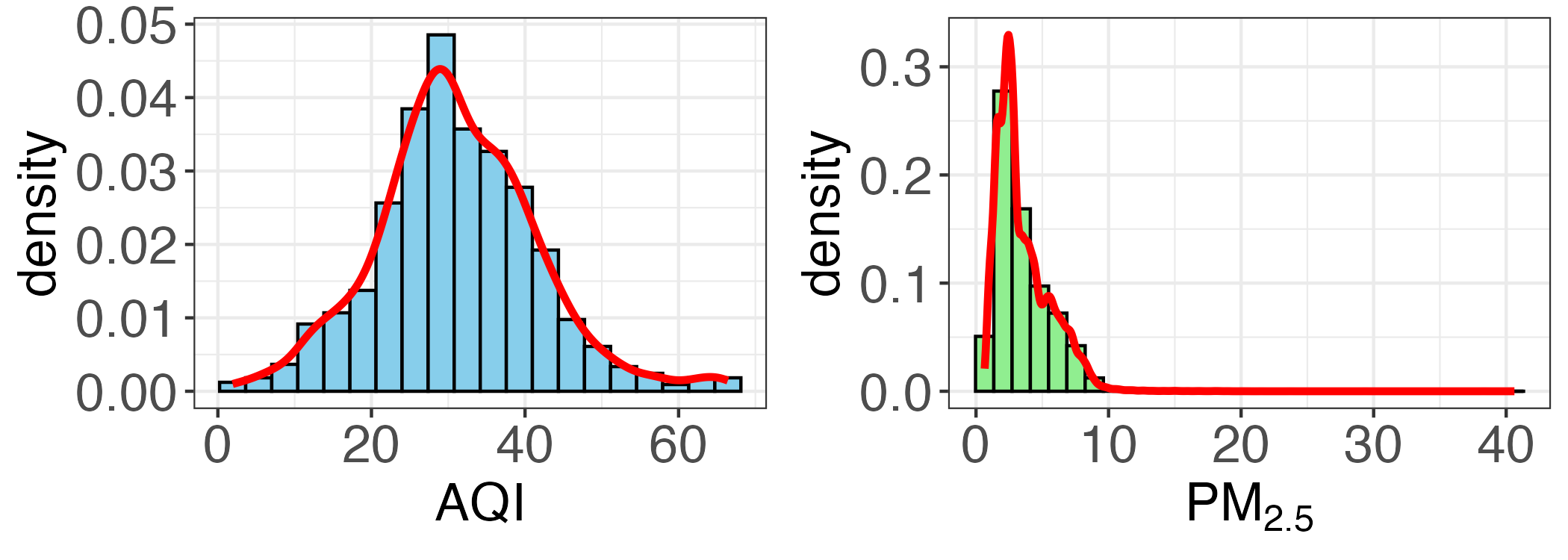} \\
\small (a) Histograms of FRM and CMAQ data \\[6pt]

\includegraphics[width=0.60\textwidth]{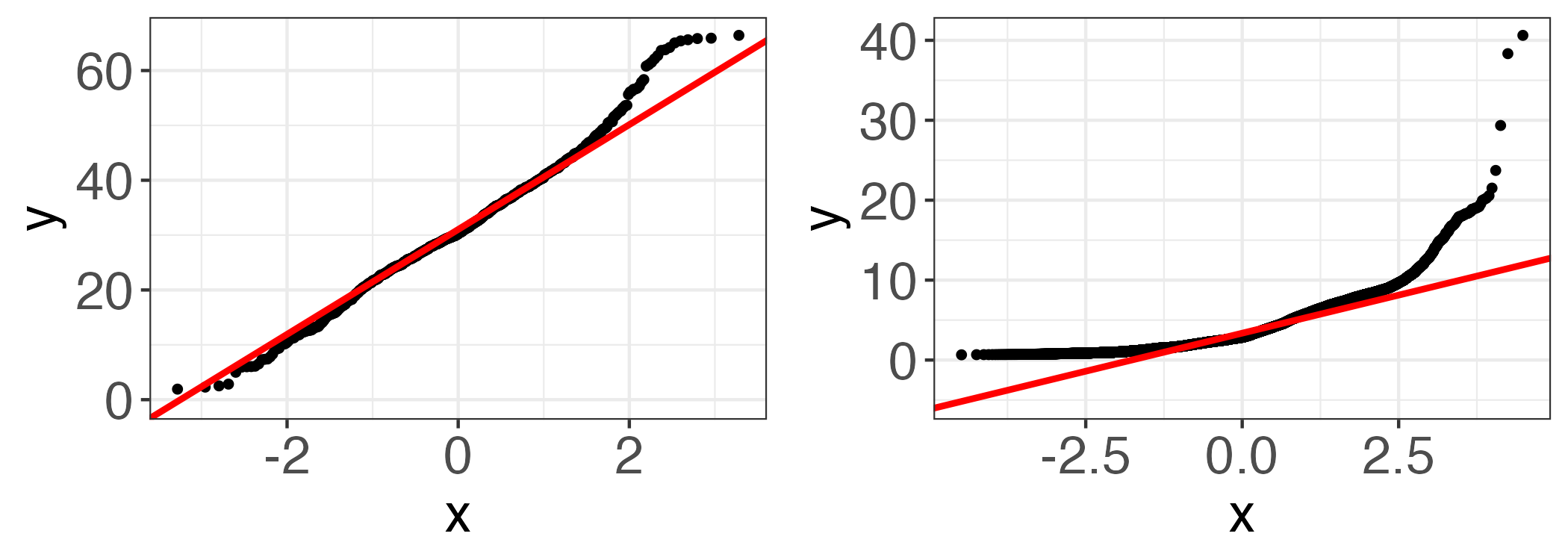} \\
\small (b) Q-Q plots of FRM and CMAQ data
\end{tabular}

\caption{Diagnostic plots for FRM AQI and CMAQ-simulated PM$_{2.5}$ concentrations. Panel (a) shows histograms illustrating the empirical distributions, while panel (b) presents Gaussian Q-Q plots assessing departures from normality.}
\label{fig:hist_qq_plot}
\end{figure}

To assess marginal behavior more formally, we examine empirical density estimates and Q-Q relationships for AQI and PM$_{2.5}$ in Figure~\ref{fig:hist_qq_plot}. Both datasets show pronounced deviations from Gaussianity. FRM AQI values present moderate right skew and a heavy upper tail, while CMAQ-simulated PM$_{2.5}$ concentrations exhibit strong right skew, with many values near zero and a long upper tail of elevated concentrations. The corresponding Q-Q plots reveal substantial curvature and clear divergence in the upper quantiles, indicating systematic departures from normality.

Summary statistics reinforce these graphical observations. AQI observations have a skewness of 0.31 and kurtosis of 3.73, whereas PM$_{2.5}$ concentrations display a more pronounced skewness of 1.32 and kurtosis of 6.88, indicating heavier tails and greater asymmetry. Anderson-Darling tests reject the null hypothesis of normality for both AQI observations ($p=1.1\times10^{-5}$) and PM$_{2.5}$ concentrations ($p<2.2\times10^{-16}$), providing strong statistical evidence against Gaussianity assumptions.

To examine regional variability, we further analyze three representative subregions: California (CA), the Northeast (NE), and the Southeast (SE), highlighted in Figure~\ref{fig:map_plot}. These areas were selected because they reflect distinct meteorological regimes and characteristic AQI-PM$_{2.5}$ relationships. AQI-PM$_{2.5}$ heatmaps for each subset reveal localized skewness, heterogeneous variance structures, and noticeable scale differences between the variables. These findings underscore the need for a data-driven fusion mechanism to combine heterogeneous data sources, together with flexible joint distributional models that can accommodate non-Gaussian behavior and complex spatial dependence.

\section{Methodology }\label{sec:method}

\subsection{Background}
Let $\mathbf{Y}(\mathbf{s}), \mathbf{s} \in D \subseteq \mathbb{R}^2$, denote a real-valued $p$-variate spatial process, where $p$ is 1 or 2, along with $k$-dimensional covariates $\mathbf{X}(\mathbf{s})$. Let $\mathbf{Z}^{(N)} \equiv \left(\mathbf{Z}(\mathbf{s}_1), \ldots, \mathbf{Z}(\mathbf{s}_N)\right)^\top$ be its realizations at $N$ spatial locations. For notational simplicity, we define the components of $\mathbf{Z}(\mathbf{s}_i)$ as collocated; however, the co-kriging formulation readily extends to non-collocated observations.  

A standard observation model assumes  
\begin{equation}
\mathbf{Z}(\mathbf{s}_i) = \mathbf{Y}(\mathbf{s}_i) + \boldsymbol{\epsilon}_i, \quad \text{for }i = 1,\dots, N,
\label{eq:data_model}
\end{equation}  
where $\boldsymbol{\epsilon}_i \sim \text{Gau}(0, \mathbb{I}_p \sigma_{\epsilon}^2)$ are independent Gaussian measurement errors. The latent process $\mathbf{Y}(\cdot)$ is modeled as  
\begin{equation}
\mathbf{Y}(\mathbf{s}_i) = \boldsymbol{\mu}(\mathbf{s}_i) + \boldsymbol{\gamma}(\mathbf{s}_i), \quad \text{for }i = 1, \dots, N,
\label{eq:process_model}
\end{equation}  
where $\boldsymbol{\mu}(\cdot)$ is the mean component and $\boldsymbol{\gamma}(\cdot) = (\gamma_1(\cdot),\dots,\gamma_p(\cdot))^\top$ is a zero-mean $p$-variate spatial process with cross-covariance function  
\begin{equation}
{C}_{uv}(\mathbf{s}_i, \mathbf{s}_j) = \text{Cov}\big(\gamma_u(\mathbf{s}_i), \gamma_v(\mathbf{s}_j)\big), \quad \text{for } u,v = 1,\dots, p, \, \text{and }i,j = 1,\dots, N.
\label{eq:covariance_gamma}
\end{equation}  

In classical co-kriging \citep{cressie1990origins}, $\boldsymbol{\mu}(\cdot)$ is typically modeled linearly as $\boldsymbol{\mu}(\mathbf{s}_i) = \mathbf{X}(\mathbf{s}_i)^\top\boldsymbol{\beta}$. Under the Gaussianity assumption of $\boldsymbol{\gamma}(\cdot)$, the conditional distribution of $\mathbf{Y}(\mathbf{s}_0)$ at a new location $\mathbf{s}_0$ given $\mathbf{X}(\mathbf{s}_0)$, $\mathbf{X}^{(N)} \equiv \left(\mathbf{X}(\mathbf{s}_1), \dots, \mathbf{X}(\mathbf{s}_N)\right)^\top$, and $\mathbf{Z}^{(N)}$ is fully defined. To model the cross-covariance structure in \eqref{eq:covariance_gamma}, a common choice is the multivariate Mat\'ern covariance function~\citep{apanasovich2012valid}:  
\begin{equation}
\mathbf{C}_{uv}(\mathbf{h}) = \sigma_{uv} M(\mathbf{h}; \nu_{uv}, \alpha_{uv}), \quad \text{for }\mathbf{h} = \|\mathbf{s}_i - \mathbf{s}_j\|, \,\text{and } u,v = 1,\dots, p,
\label{eq:matern}
\end{equation}  
where $M(\cdot)$ is the Matérn function, $\nu_{uv}$ is the smoothness parameter, and $\alpha_{uv}$ is the range parameter.  

Although co-kriging is the best linear unbiased predictor, it faces two main limitations: maximum likelihood estimation for Gaussian processes is computationally expensive for large datasets, and co-kriging may be suboptimal for non-Gaussian data such as AQI and PM$_{2.5}$. The \textit{deep classifier kriging} (DCK) approach, introduced in later sections, remedies these limitations by approximating the conditional distribution of $\mathbf{Y}(\mathbf{s}_0) \mid \mathbf{X}(\mathbf{s}_0), \mathbf{X}^{(N)}, \mathbf{Z}^{(N)}$ without any prior distributional assumption.

We define the full structure of the proposed \textit{data fusion} mechanism and DCK in the following sections. In Section~\ref{sec:data_fusion}, we introduce the proposed \textit{data fusion} mechanism and the associated algorithm. In Section~\ref{sec:discretization}, we then present the DCK framework, including an algorithm for discretizing continuous univariate or bivariate spatial processes into multivariate indicator variables representing distinct classes, a deep neural network (DNN)-based classifier for predicting the probabilities that a new location belongs to each class corresponding to the indicator variables, and a kernel smoothing approach for recovering a continuous joint conditional distribution at an unobserved location given the observed data and covariates.

\subsection{Data fusion for bivariate spatial processes}\label{sec:data_fusion}

Let us assume that the bivariate data $(Z_1(\cdot), Z_2(\cdot))^\top$, are observed at locations $\mathbf{s}_{1,i} \in \mathcal{S}_1$, \text{for }$i = 1, \ldots, N_1$, and $\mathbf{s}_{2,j} \in \mathcal{S}_2$, \text{for }$j = 1, \ldots, N_2$, where $N_2 > N_1$. We first pre-process the observed data to obtain a collocated representation of $Z_1(\cdot)$ and $Z_2(\cdot)$ by pairing $Z_1(\mathbf{s}_{1,\cdot})$ for every location in $\mathcal{S}_1$ with its $\kappa$-nearest neighbor observations of $Z_2(\mathbf{s}_{2,\cdot})$, for locations in $\mathcal{S}_2$. The resulting set is given by
\begin{equation}\label{eq:collocated_set}
  U \;\equiv\;
\left\{
\left(
Z_1(\mathbf{s}_{1,i}),\,
\frac{1}{\kappa}\sum_{j \in \mathcal{N}_\kappa(\mathbf{s}_{1,i})}
Z_2(\mathbf{s}_{2,j})
\right)^\top
\right\}_{i=1}^{N_1},
\end{equation}
where $\mathcal{N}_\kappa(\mathbf{s}_{1,i}) \subset \mathcal{S}_2$ denotes the index set of the $\kappa$ locations in $\mathcal{S}_2$ that are closest to $\mathbf{s}_{1,i}$ with respect to the Euclidean distance.

\begin{figure}[ht!]
    \centering    \includegraphics{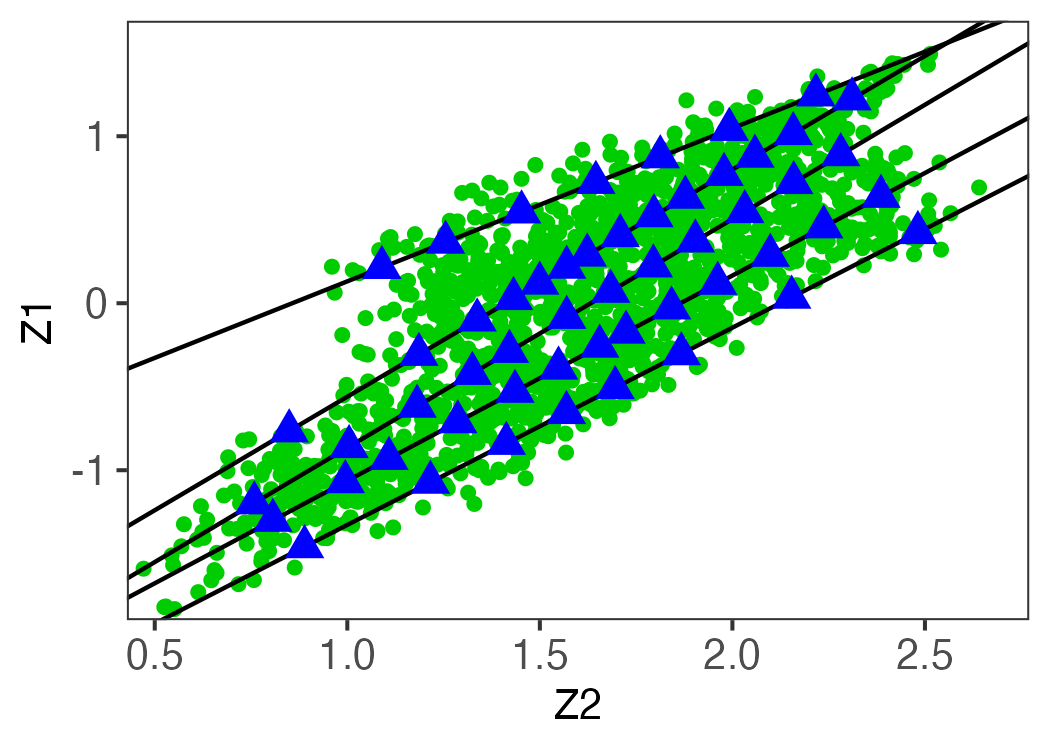}
    \caption{Scatter plot for $(Z_1(\cdot),Z_2(\cdot))^\top \in U$ (in green), the quantile lines (in black) and the middle points for each class (blue solid triangles).}
    \label{fig:reg}
\end{figure}

Now, let \(m_1\) denote the number of quantile levels. We fit $m_1$ quantile regression lines to the data in $U$, using $Z_2(\cdot)$ as the predictor and $Z_1(\cdot)$ as the response. These quantile regression lines are expressed as
$Z_1(\cdot)=b_{k} Z_2(\cdot)+a_{k}$,
for $k=1, \ldots, m_1$, and are illustrated as black lines in Figure~\ref{fig:reg}.
Then for each point in $U$, we project it onto the closest quantile regression line from which its distance is minimum yielding the set
\begin{align}
V_1 \;\equiv\; 
\Biggl\{ \; 
& \Biggl(
\frac{Z_2(\mathbf{s}_{2,.}) + b_{k^*} \left( Z_1(\mathbf{s}_{1,.}) - a_{k^*} \right)}{1 + b_{k^*}^2}, \;\;
a_{k^*} + b_{k^*} \frac{Z_2(\mathbf{s}_{2,.}) + b_{k^*} \left( Z_1(\mathbf{s}_{1,.}) - a_{k^*} \right)}{1 + b_{k^*}^2}
\Biggr) : \nonumber \\
& {k^*} = \arg\min_{k=1,\dots,m_1} 
\left| Z_1(\mathbf{s}_{1,.}) - a_k - b_k Z_2(\mathbf{s}_{2,.}) \right|, \nonumber \\
& Z_1(\mathbf{s}_{1,.}), Z_2(\mathbf{s}_{2,.}) \in U \; 
\Biggr\}.
\label{eq:projection_set_vector}
\end{align}
Note that, the $N_2 - N_1$ observations of $Z_2(\cdot)$, which are not used for projection, can be used to obtain as the following set
\begin{align}
V_2 \;\equiv\;
\left\{
\left(
Z_1^{\mathrm{exp}}(\mathbf{s}_{2,i}),\;
Z_2(\mathbf{s}_{2,i})
\right)^\top
\right\}_{i=1}^{N_2-N_1},
\label{eq:projection_set2_softNN}
\end{align}
where the expected augmented variable $Z_1^{\mathrm{exp}}(\mathbf{s}_{2,i})$ is defined as
\begin{equation}\label{eq:weighted_mean}
Z_1^{\mathrm{exp}}(\mathbf{s}_{2,i}) 
= \sum_{k=1}^{m_\mathrm{1}} p_k \, q_k,
\end{equation}
with
\begin{equation}
q_k = a_k + b_k \, Z_2(\mathbf{s}_{2,i}),
\end{equation}
\begin{equation}
r_k = \bigl| q_k - q_{k^*} \bigr|, \quad 
k^* = \arg\min_k \left\| \mathbb{Z}_{\mathrm{NN}(i)} - q_k \right\|,
\end{equation}
and
\begin{equation}\label{eq:soft_prob}
p_k = \frac{\exp\Bigl( - r_k / (\mathrm{median}(\mathbf{r}) + \varepsilon) \Bigr)}
           {\sum_{l=1}^{m_\mathrm{1}} \exp\Bigl( - r_l / (\mathrm{median}(\mathbf{r}) + \varepsilon) \Bigr)}.
\end{equation}

Here, $\mathbb{Z}_{\mathrm{NN}(i)}$ denotes the vector of the $\kappa_2$ nearest neighbor observations of $Z_1(\mathbf{s}_{1,\cdot})$, for $\mathbf{s}_{1,\cdot} \in \mathcal{S}_1$ of locations $\mathbf{s}_{2,i} \in \mathcal{S}_2$. Here, $k^*$ identifies the candidate closest to the neighbors; $p_k$ are soft probabilities used to compute the expected value; and $\varepsilon$ is a small positive constant to prevent division by zero.
We use $V = V_1 \cup V_2$ for further calculation on model building. We store the location information of $Z_2(\mathbf{s})$ for each point in $V$ in set $\mathcal{S}$. Algorithm \ref{alg:projection} provides a concise approach to the aforementioned \textit{data fusion} framework. Note that, $\mathcal{S}$ have the same elements as $\mathcal{S}_2$ but in the ordering as the set $V$. 

\begin{algorithm}[t]
\caption{Data fusion for bivariate spatial processes}
\label{alg:projection}
\begin{algorithmic}[1]
\Require Observations $\{Z_1(\mathbf{s}_{1,i})\}_{i=1}^{N_1}$ and $\{Z_2(\mathbf{s}_{2,j})\}_{j=1}^{N_2}$, number of quantile levels $m_1$
\Ensure Combined set $V = V_1 \cup V_2$ and locations $\mathcal{S}$

\State \textbf{Step 1: Collocation.} 
Form the collocated dataset $U$ as defined in Eq.~\eqref{eq:collocated_set}.

\State \textbf{Step 2: Quantile Regression.}
\Statex \quad For each quantile level $\tau_k$, $k = 1, \ldots, m_1$:
\Statex \quad \quad (a) Solve the optimization problem
\begin{equation*}
(a_k, b_k) = \arg\min_{a,b} \sum_{i = 1} ^{|U|}
\rho_{\tau_k}\!\left(Z_1(\mathbf{s}_{1,i}) - a - b\,Z_2(\mathbf{s}_{2,i})\right), \, \bigl(Z_1(\mathbf{s}_{1,i}),Z_2(\mathbf{s}_{_{2,i}})\bigr)^\top \in U
\end{equation*}
\Statex \quad \quad where $\rho_{\tau}(u) = u(\tau - \mathbb{I}\{u < 0\})$ is the quantile loss function.
\Statex \quad \quad (b) Store the estimated quantile regression line 
$ Q_k: \; Z_1(\cdot) = a_k + b_k Z_2(\cdot)$.

\State \textbf{Step 3: Projection.} 
Project each element of $U$ onto its nearest quantile regression line to obtain the set $V_1$ as in Eq.~\eqref{eq:projection_set_vector}.

\State \textbf{Step 4: Augmentation.} 
Use the remaining $N_2 - N_1$ observations of $Z_2(\cdot)$ to form the auxiliary set $V_2$ using Eq.~\eqref{eq:projection_set2_softNN}-\eqref{eq:soft_prob}.

\State \textbf{Step 5: Fusion.} 
Combine the two sets to obtain $V = V_1 \cup V_2$ and for each $Z_1(\mathbf{s}_{1,.}),Z_2(\mathbf{s}_{2,.})$ in $V$ store $\mathbf{s}_{2,.}$ in $\mathcal{S}$.
\end{algorithmic}
\end{algorithm}

\subsection{Deep classifier kriging for spatial prediction}\label{sec:discretization}

To estimate full predictive distributions, several studies have proposed transforming continuous target variables into discrete classes for probabilistic modeling. In this paradigm, approaches such as quantile classification and ordinal regression discretize continuous responses into ordered bins, allowing regression models to learn class probabilities that approximate the underlying cumulative distribution function \citep{tagasovska2019single, gustafsson2020evaluating}. Similarly, discretized likelihood or histogram-based regression methods represent the response distribution as a categorical probability mass function, enabling accurate reconstruction of continuous predictive distributions from classification outputs \citep{lakshminarayanan2017simple, mi2022training}. Compared with directly modeling continuous responses, discretization provides a more flexible framework that avoids strong distributional assumptions and allows a model to approximate complex, non-Gaussian predictive shapes. In this study, we focus on the quantile classification technique and adapt it to the spatial context to obtain an $n$-variate indicator variable for a $p$-variate continuous spatial process. Specifically, we consider the univariate ($p = 1$) and bivariate ($p = 2$) spatial processes.

\subsubsection{Discretization of spatial processes}

\textbf{Univariate scenario: }
Following upon the idea of \cite{agarwal2021copula} we obtain the multivariate indicator variables from univariate $Z(\cdot)$ by thresholding the continuous observations into $n$ non-overlapping classes. We first prepare a set of sample quantiles $\{q_1,\ldots,q_n\}$ based on some pre-defined probability values \citep{journel1983nonparametric} from the dataset; we then create $n$ sets of indicator variables as follows
\begin{equation}
z_j(\mathbf{s}_i) =
\begin{cases}
1, & \text{if } q_{j-1} \leq Z(\mathbf{s}_i) \leq q_j, \\
0, & \text{otherwise},
\end{cases}
\quad \text{for } j = 1, \ldots, n, \text{ and } i = 1, \ldots, N.
\label{eq:2}
\end{equation}
Equation \eqref{eq:2} gives us a set of binary indicator variables $\{ z_1(\mathbf{s}_{i}), z_2(\mathbf{s}_{i}), \ldots, z_n(\mathbf{s}_{i}) \}
$ for each ${Z(\mathbf{s}_{i})}$, for $ i = 1, \dots, N$. 

\noindent 
\textbf{Bivariate scenario: }
Unlike the univariate scenario, the bivariate setting extends the objective from predicting a single spatial value at a new location $\mathbf{s}_0$ to capturing the statistical dependence between $Y_1(\mathbf{s}_0)$ and $Y_2(\mathbf{s}_0)$. This dependence is characterized by their joint distribution \(F(Y_1(\mathbf{s}_0), Y_2(\mathbf{s}_0)| \mathbf{X}(\mathbf{s}_0),\mathbf{X}^{(N)}, \mathbf{Z}^{(N)} )\) and corresponding conditional distributions. Existing approaches for modeling such dependence include copula-based and quantile-based methods. Copula models directly parameterize the dependence structure between $Y_1(\cdot)$ and $Y_2(\cdot)$, offering flexibility but often incurring high computational cost \citep{nelsen2006introduction}. Alternatively, stratified quantile regression sequentially models $Y_1(\cdot)$ and $Y_2(\cdot)$ through conditional quantiles \citep{wei2008approach}, but remains simulation-intensive and relies on collocated observations. To address these challenges, we incorporate the \textit{data fusion} mechanism introduced in Section \ref{sec:data_fusion} that leverages more abundant observations of one variable to infer the less observed process, from which we derive $n$-variate indicator variables for modeling.

 \noindent \textbf{Indicator variables from bivariate continuous data :}
To create the binary indicator variables we divide the projected points in $V$ of each quantile regression $Q_k, \text{for } k = 1, \dots, m_1 $, into $m_{2,k}$ classes such that the minimum number of observations per interval is $\delta$. This results in the total number of class $n = \sum_{k=1}^{m_1}m_{2,k}$. The midpoints of these intervals are given as $\{(Z_{1,1}^{\text{node}}, Z_{2,1}^{\text{node}})^\top, \ldots , (Z_{1,n}^{\text{node}}, Z_{2,n}^{\text{node}})^\top\}$.
Similar to the univariate scenario, we define the indicator variables in the following manner:

\begin{equation}
z_j(\mathbf{s}_i) =
\begin{cases}
1, & \text{if } \bigl(Z_1(\mathbf{s}_{1,\cdot}), Z_2(\mathbf{s}_{2,\cdot})\bigr)^\top \in \mathrm{Class}_{k,c}, \\[4pt]
0, & \text{otherwise},
\end{cases}
\quad
\begin{aligned}
& c = 1, \ldots, m_{2,k}, \\
& k = 1, \ldots, m_1, \\
& \bigl(Z_1(\mathbf{s}_{1,\cdot}), Z_2(\mathbf{s}_{2,\cdot})\bigr)^\top \in V,
\end{aligned}
\end{equation}
where $\mathbf{s}_i \in \mathcal{S}$. This approach transforms the original bivariate continuous spatial process into a $n$-variate indicator spatial process, where each class represents a region in the $(Z_1, Z_2)$ space. Algorithm \ref{alg:indicator} provides a step-by-step detail of the approach.

\begin{algorithm}[ht!]
\caption{Indicator variable construction from bivariate continuous data}
\label{alg:indicator}
\begin{algorithmic}[1]
\Require Projected set $V$, minimum number of samples per class $\delta$
\Ensure Indicator variables $z_j(\mathbf{s}_i), \; j = 1, \ldots, n = \sum_k m_{2,k}$, $\mathbf{s}_{i} \in \mathcal{S}$

\State \textbf{Step 1: Interval Partitioning.} 
\For{$k = 1$ to $m_1$}
    \State Extract projected points $\{(Z_{1}^{(k)}(\mathbf{s}_{1,1}), Z_{2}^{(k)}(\mathbf{s}_{2,1}))^\top, \dots\}$ from $V$ for quantile line $Q_k$.
    \State Sort these points on the line $Q_k$ in increasing order.
    \State Divide the sorted points into $m_{2,k}$ contiguous intervals such that each interval contains at least $\delta$ samples.
    \State Store these interval boundaries as $\{\text{Class}_{k,1}, \ldots, \text{Class}_{k,m_{2,k}}\}$.
\EndFor

\State \textbf{Step 2: Midpoint Computation.}
\For{each interval $\text{Class}_{k,c}$, $c = 1,\ldots,m_{2,k}$, $k = 1,\ldots,m_1$ }
    \State Compute midpoint
    \[
    (Z_{1, kc}^{\text{node}}, Z_{2, kc}^{\text{node}}) 
    = \frac{1}{2}\big((Z_{1,\min}, Z_{2,\min}) + (Z_{1,\max}, Z_{2,\max})\big)
    \]
\EndFor
\State \textbf{Step 3: Indicator Definition.}
\For{each location $\mathbf{s}_{i} \in \mathcal{S}$ and class $j = 1,\ldots,n$, $n = \sum_k m_{2,k}$}
    \State Define
    \[
    z_j(\mathbf{s}_{i}) =
    \begin{cases}
    1 & \text{if } \bigl(Z_1(\mathbf{s}_{1,i}), Z_2(\mathbf{s}_{2,i})\bigr)^\top \in \text{Class}_j, \\
    0 & \text{otherwise.}
    \end{cases}
    \]
\EndFor

\end{algorithmic}
\end{algorithm}

\subsubsection{Estimation of class probabilities using neural networks}\label{sec:classifier}

Following the construction of indicator variables in Section \ref{sec:discretization}, the spatial prediction task can be reformulated as a classification problem,
where each indicator represents a binary event indicating whether the underlying spatial value falls within a pre-defined class.
The objective is therefore to estimate the class probabilities that serve as the basis for reconstructing the conditional cumulative distribution function (CDF).

Several classification algorithms are widely used in machine learning, including random forests~\citep{biau2016random}, XGBoost~\citep{pamina2019effective}, and deep neural network (DNN)-based classifiers~\citep{chen2014deep}.  
In this work, we adopt a DNN-based classification framework, as it provides greater flexibility in model design and enables the estimation of well-calibrated class probabilities.  


A naive choice for input features to the neural network would be the coordinate vector of spatial locations in $\mathcal{S}$. However, this approach fails to capture the underlying spatial dependence in the data. To account for spatial dependence, \citet{chen2020deepkriging} proposed using basis function embeddings for spatial coordinates.  
Motivated by this idea, we employ the multi-resolution, compactly supported Wendland radial basis functions (RBFs) introduced by \citet{nychka2015multiresolution}, defined as
\[
w(d) = \frac{(1 - d)^6}{3} \bigl(35d^2 + 18d + 3\bigr) \, \mathbbm{1}\{0 \le d \le 1\},
\]
where $d = \| \mathbf{s}_i - \mathbf{u} \| / \eta, \ \, \mathbf{s}_i \in \mathcal{S}$, $\mathbf{u} $ denotes the knot location, and $\eta$ is the bandwidth parameter controlling the spatial support. Similar to \citet{chen2020deepkriging}, 
to incorporate spatial structure, we compute $K$ RBF embeddings and combine them with the covariates $\mathbf{X}(\mathbf{s}_i)$ to obtain the vector
\[
\mathbf{X}_{\phi}(\mathbf{s}_i) = \bigl( \phi_1(\mathbf{s}_i), \phi_2(\mathbf{s}_i), \ldots, \phi_K(\mathbf{s}_i), \mathbf{X}^\top(\mathbf{s}_i)  \bigr)^\top.
\]
Thus, an $L$-layer DNN can be expressed recursively as
\begin{equation}
\begin{aligned}
h_1(\mathbf{s}_i) &= W_1 \mathbf{X}_{\phi}(\mathbf{s}_i) + b_1, 
&\quad a_1(\mathbf{s}_i) = \psi\bigl(h_1(\mathbf{s}_i)\bigr), \\
h_2(\mathbf{s}_i) &= W_2 a_1(\mathbf{s}_i) + b_2, 
&\quad a_2(\mathbf{s}_i) = \psi\bigl(h_2(\mathbf{s}_i)\bigr), \\
&\vdots \\
h_L(\mathbf{s}_i) &= W_L a_{L-1}(\mathbf{s}_i) + b_L, 
&\quad \hat{\mathbf{z}}(\mathbf{s}_i) = \sigma\bigl(h_L(\mathbf{s}_i)\bigr),
\end{aligned}
\label{eq:dnn_structure}
\end{equation}
where $\hat{\mathbf{z}}(\mathbf{s}_i) = \bigl( \hat{z}_1(\mathbf{s}_i), \hat{z}_2(\mathbf{s}_i), \ldots, \hat{z}_n(\mathbf{s}_i) \bigr)^\top$ denotes the predicted class probabilities.  
Here, $W_\ell$ and $b_\ell$ are the weight matrix and bias vector at layer $\ell$, respectively, and, $\psi(\cdot)$ and $\sigma(\cdot)$ are the activation functions.  
We use the rectified linear unit (ReLU) activation~\citep{schmidt2020nonparametric} for the hidden layers, and the softmax activation~\citep{gao2017properties} in the output layer to ensure the resulting probabilities lie in $[0, 1]$.  

Given observations $\mathbf{Z}^{(N)}$, the goal of spatial prediction is to estimate the unobserved process $\mathbf{Y}(\cdot)$ at a new location $\mathbf{s}_0$ by minimizing the expected loss
\begin{equation}\label{eq:loss_continuous}
\hat{\mathbf{Y}}^{opt}(\mathbf{s}_0 \mid \mathbf{Z}^{(N)})
= \operatorname*{argmin}_{\mathbf{a} \in \mathbb{R}^p}
\mathbb{E}\!\left\{ L\!\left(\mathbf{a}, \mathbf{Y}(\mathbf{s}_0)\right) \mid \mathbf{Z}^{(N)} \right\}.
\end{equation}
We approximate \eqref{eq:loss_continuous} in the context of binary indicator variables and choose the function $L(\cdot)$ to be the binary cross-entropy loss~\citep{buja2005loss} over the training dataset to obtain 
\begin{equation}
\begin{aligned}
R(\hat{\mathbf{z}}^{(n)},\mathbf{z}^{(n)})
&= \operatorname*{argmin}_{\boldsymbol{\theta}}
L\bigl(\hat{\mathbf{z}}^{(n)},\mathbf{z}^{(n)}; \boldsymbol{\theta}\bigr), \\
\text{where} \qquad
L\bigl(\hat{\mathbf{z}}^{(n)},\mathbf{z}^{(n)}; \boldsymbol{\theta}\bigr)
&= -\frac{1}{N} \sum_{i=1}^{N} \sum_{j=1}^{n}
z_j(\mathbf{s}_i)\,\log\!\bigl(\hat{z}_j(\mathbf{s}_i)\bigr).
\end{aligned}
\label{eq:logloss}
\end{equation}
Here $N$ is the cardinality of the set $\mathcal{S}$. In practice for faster computation we evaluate \eqref{eq:logloss} in mini-batches and optimize the loss function under an Adam learning schedule \citep{kingma2014adam}.

\subsubsection{Kernel smoothing}\label{sec:kernel_smoothing}

After obtaining the estimated class probabilities 
$\hat{\mathbf{z}}(\mathbf{s}_0)$
from the trained DCK model at location 
$\mathbf{s}_0$, 
we proceed to estimate the joint distribution of the response variables. 
A kernel-smoothing approach is employed to construct a smooth approximation of the conditional cumulative distribution function (CDF).
The estimated {bivariate conditional CDF} is defined as
\begin{equation}
    \hat{F}_{(Y_1,Y_2)}\!\left((y_1, y_2), \mathbf{s}_0 | \mathbf{X}(\mathbf{s}_0),\mathbf{X}^{(N)}, \mathbf{Z}^{(N)}\right)
    = \sum_{j=1}^{n} 
    \hat{z}_j(\mathbf{s}_0)\,
    \Phi\!\left(\frac{y_1 - Z_{1,j}^{\text{node}}}{h}\right)
    \Phi\!\left(\frac{y_2 - Z_{2,j}^{\text{node}}}{h}\right),
    \label{eq:biv_cdf}
\end{equation}
where $\Phi(\cdot)$ denotes the standard normal cumulative distribution function and 
$h$ is the kernel bandwidth parameter controlling the level of smoothing.
From \eqref{eq:biv_cdf}, the {conditional distribution} of $Y_1(\mathbf{s}_0)$ given $Y_2(\mathbf{s}_0) = \mathcal{Y}$, $\mathbf{X}(\mathbf{s}_0), \, \mathbf{X}^{(N)}, \, \mathbf{Z}^{(N)}$ can be expressed as
\begin{equation}
\begin{aligned}
     \hat{F}_{Y_1 \mid Y_2}\!\left(y_1, \mathbf{s}_0\mid \mathcal{Y}, \mathbf{X}(\mathbf{s}_0),\mathbf{X}^{(N)}, \mathbf{Z}^{(N)}\right)& 
    = \\
    \frac{1}{f_{Y_2}(\mathcal{Y}, \mathbf{s}_0|\mathbf{X}(\mathbf{s}_0),\mathbf{X}^{(N)}, \mathbf{Z}^{(N)})}
    &\sum_{j=1}^{n} 
    \hat{z}_j(\mathbf{s}_0)\,
    \Phi\!\left(\frac{y_1 - Z_{1,j}^{\text{node}}}{h}\right)
    \phi\!\left(\frac{\mathcal{Y} - Z_{2,j}^{\text{node}}}{h}\right),
\end{aligned}
    \label{eq:cond_cdf}
\end{equation}
where $\phi(\cdot)$ denotes the standard normal probability density function. 
The corresponding {marginal conditional density} of $Y_2(\mathbf{s}_0)$ is given by
\begin{equation}
    f_{Y_2}(y_2  ,\mathbf{s}_0|\mathbf{X}(\mathbf{s}_0),\mathbf{X}^{(N)}, \mathbf{Z}^{(N)})
    = 
    \sum_{j=1}^{n} 
    \hat{z}_j(\mathbf{s}_0)\,
    \phi\!\left(\frac{y_2 - Z_{2,j}^{\text{node}}}{h}\right).
    \label{eq:marginal_density}
\end{equation}

The {point prediction} and {prediction intervals} of $Y_1(\mathbf{s}_0) \mid Y_2(\mathbf{s}_0)$ can be obtained from the conditional distribution in 
\eqref{eq:cond_cdf}. 
Specifically, the $\tau$-th conditional quantile of $Y_1(\mathbf{s}_0)$ given $Y_2(\mathbf{s}_0) = \mathcal{Y}$, $\mathbf{X}(\mathbf{s}_0), \, \mathbf{X}^{(N)}, \, \mathbf{Z}^{(N)}$ at location $\mathbf{s}_0$ is
\begin{equation}\label{eq:estim_quantile}
    q_{Y_1 | Y_2}(p, \mathbf{s}_0\mid \mathcal{Y}, \mathbf{X}(\mathbf{s}_0), \, \mathbf{X}^{(N)}, \, \mathbf{Z}^{(N)})
    = 
    \hat{F}_{Y_1 \mid Y_2}^{-1}\!\left(\tau \right).
\end{equation}
This framework enables both point-wise prediction and full distributional uncertainty quantification in the bivariate spatial setting. For the univariate scenario the CDF can be easily obtained using \eqref{eq:marginal_density} by replacing $\phi(\cdot)$ with $\Phi(\cdot)$. Note that in practical applications, $Y_2(\mathbf{s}_0)$ is typically unobserved. Therefore, the conditional density of $Y_1(\mathbf{s}_0)$ is evaluated by treating $Z_2(\mathbf{s}_0)$ as the conditioning variable in place of $Y_2(\mathbf{s}_0)$. 

Choice of $h$ in kernel smoothing to estimate the conditional CDF is critical.
Based on our experiments, we adopt the following rule-of-thumb:
$$
h=p \cdot \frac{C}{3} \cdot \sigma_h \cdot N^{-\alpha}
$$
where $p$ is the dimensionality of the response variable. The constant $C$ is a user-specified positive integer that controls the overall scale of the bandwidth. The parameter $\sigma_h$ represents the scale of the response variable, estimated using the median absolute deviation (MAD). The training sample size is denoted by $N$, and the exponent $\alpha = \frac{1}{3}$ is a commonly used value in nonparametric smoothing. The division by 3 serves as an additional regularization to moderate the effective bandwidth. A detailed study on the choice of these hyper-parameters is given in Section~\ref{S-sec:sensitivity} of the Supplementary Material.

\section{Simulation experiments}\label{sec:simulation}

In this section, we evaluate the performance of our proposed DCK model against classical co-kriging (CK) \citep{cressie2015statistics}, assuming Gaussianity in the bivariate setting. We also compare our model's performance with the univariate DeepKriging (DK) \citep{chen2020deepkriging} along with uncertainty quantification as discussed in \cite{nag2025bivariate}. We compare competing models based on mean absolute error ({MAE}), prediction interval coverage probability (PICP), and average interval length (AL); further details on these metrics are provided in Section~\ref{S-sec:metrics} of the Supplementary Material. We have also included a separate simulation experiment for the univariate scenario in Section~\ref{S-sec:univariate} of the Supplementary Material, along with a sensitivity analysis of the hyperparameters in Section~\ref{S-sec:sensitivity}. 



\subsection{Bivariate scenario}

We simulate the bivariate random field $\mathbf{Y}(\cdot) \equiv (Y_1(\cdot), Y_2(\cdot))^\top$ over the domain $[0,1] \times [0,1]$ with $3600$ irregularly spaced spatial locations, using a zero-mean Gaussian process with a bivariate Matérn covariance function as defined in \eqref{eq:matern}. We adopt a quasi-parsimonious specification \citep{gneiting2010matern}, in which the cross-covariance parameters satisfy
\[
\nu_{12} = \frac{\nu_{11} + \nu_{22}}{2}, 
\qquad 
\alpha_{12} = \frac{\alpha_{11} + \alpha_{22}}{2},
\]
ensuring a valid and interpretable covariance structure.
For the first variable $Y_1(\cdot)$, we set $\alpha_{11} = 0.2$, $\nu_{11} = 0.8$, and $\sigma_{11}^2 = 0.89$. For the second variable $Y_2(\cdot)$, we choose $\alpha_{22} = 0.4$, $\nu_{22} = 0.8$, and $\sigma_{22}^2 = 1.3$. The cross-covariance is determined by $\alpha_{12} = 0.3$, $\nu_{12} = 0.8$, and $\rho_{12} = 0.8$, producing a full $2N \times 2N$ covariance matrix from which joint realizations are sampled. 

For the non-Gaussian scenario, we apply the Tukey $g$-$h$ transformation \citep{xu2017tukey} to each field, using parameters $(g = 0.5, h = 0.5)$ to induce skewness and heavy tails. The simulations here as well are performed using $3600$ irregularly spaced locations.

To better align this simulation with our data application, we add Gaussian noise with standard deviation $\sigma_{\epsilon} = 0.1$ to obtain $\mathbf{Z}(\cdot) \equiv (Z_1(\cdot), Z_2(\cdot))^\top$, retain $100$ locations for testing, and designate the remaining $N = 3500$ as training locations. Among these, we randomly select $500$ observations for $Z_1(\cdot)$ while using all $3500$ observations of $Z_2(\cdot)$. The goal is to predict $Y_1(\cdot)$ at the unobserved test locations using the $500$ available observations of $Z_1(\cdot)$ and the full set of $Z_2(\cdot)$ observations. This design introduces non-collocation between the two variables and creates a challenging interpolation task for $Y_1(\cdot)$. This entire process is repeated for 100 simulations using the same sets of spatial training and test locations. 

For the DCK model, we first apply the \textit{data fusion} of Algorithm~\ref{alg:projection} to obtain a collocated projected set of observations. For each simulation, we fit $m_1 = 5$ quantile regression lines modeling $Z_1(\cdot)$ as a function of $Z_2(\cdot)$ at quantile levels $\tau = \{0.05, 0.275, 0.5, 0.725, 0.95\}$. Each data point is projected onto its nearest quantile line and stratified based on its projected distance from the origin. To ensure adequate sample support, each class contains at least $\delta = 15$ observations. This adaptive procedure yields $n = 132$ classes used for classification. We fix $C = 12$ in the DCK framework. For the DK model framework we only use the $500$ observations of $Z_1(\cdot)$ to do univariate spatial interpolation without borrowing any information from $Z_2(\cdot)$.

\subsection{Simulation results}

\begin{table}[t!]
\centering
\caption {MAE, PICP for (95\% nominal), and AL with corresponding standard errors (in parentheses as relative percentages), together with computation time (in seconds), for CK and DCK methods applied to different bivariate simulation experiments.}
\label{tab:2d}
\begin{tabular}{||c| c c c c c||}
\hline
Scenario & Model & MAE (SE) & PICP (SE) & AL (SE) & Time (s) \\
\hline\hline
\multirow{2}{*}{Gaussian} 
  & CK  & 0.14 (0.2) & 97.4 (0.2) & 0.8 (1.7) & 18678.1 \\
 & DCK & 0.16 (0.9) & 97.2 (0.3) & 1.1 (3.3) &   23.1 \\
 & DK & 0.23 (1.3) & 91.2 (0.4) & 1.2 (3.5) &   379.2 \\

\hline
\multirow{2}{*}{non-Gaussian}

 & CK  & 0.48 (3.3) & 93.1 (0.9) & 1.9 (5.2) & 19749.2 \\
 & DCK & 0.30 (2.5) & 94.9 (0.4) & 1.7 (10.4) &   21.8 \\
 & DK & 0.49 (1.7) & 77.2 (1.9) & 1.4 (4.5) &   397.3 \\

\hline
\end{tabular}

\end{table}

Table~\ref{tab:2d} summarizes the simulation results. It can be seen that DCK performs comparably to CK in the Gaussian setting while outperforming CK in the non-Gaussian setting. Notably, DCK reduces computation time by approximately 800-fold faster, demonstrating exceptional scalability. Moreover, DCK outperforms DK across all evaluation metrics.

These results highlight that the proposed \textit{data fusion} mechanism in Algorithm~\ref{alg:projection} enables DCK to outperform univariate DK by effectively fusing information from heterogeneous data sources. Furthermore, without imposing any distributional assumptions, DCK provides accurate predictions and well-calibrated uncertainty estimates while offering substantial computational advantages.

\section{Conditional modeling of AQI given \texorpdfstring{PM$_{2.5}$}{PM2.5} concentrations over the United States}
\label{sec:application}

In this section, we apply the proposed DCK framework to model the AQI measurements from FRM monitoring stations conditional on PM$_{2.5}$ concentrations from the CMAQ numerical model over the United States. A joint analysis of FRM AQI and CMAQ-simulated PM$_{2.5}$ concentrations along with subsequent comparisons have been provided in Section~\ref{S-sec:joint} of the Supplementary Material. 

\subsection{Data pre-processing}

Before model fitting, several preprocessing steps are undertaken to ensure consistency between the two data sets. As introduced in Section~\ref{sec:exploratory_analysis}, AQI observations are denoted by $Z_1(\cdot)$ and PM$_{2.5}$ concentrations by $Z_2(\cdot)$. Because the two data sets are spatially misaligned, we first harmonize them using the \textit{data fusion} mechanism of Algorithm~\ref{alg:projection}, which projects each AQI monitoring site to its nearest PM$_{2.5}$ grid cell to form collocated pairs $(Z_1, Z_2)$. These collocated samples, together with PM$_{2.5}$-only locations, serve as inputs to the DCK model.

We extract January~2019 monthly mean AQI and PM$_{2.5}$ concentrations, yielding approximately $N_1 \approx 1{,}000$ AQI observations and $N_2 \approx 130{,}000$ PM$_{2.5}$ grid values. The projection step produces a collocated dataset $U$, while the remaining PM$_{2.5}$-only locations constitute $U_{N_2 - N_1}$. To maintain computational feasibility for GP-based benchmarks such as co-kriging, we randomly sample 10{,}000 PM$_{2.5}$-only locations from within the convex hull of the AQI network, preserving the spatial footprint of the study area.

Indicator variables are then generated using Algorithm~\ref{alg:indicator}, and Equation~\eqref{eq:estim_quantile} is used to estimate conditional quantiles and predict the latent AQI field $Y_1(\cdot)$ over the domain. To capture conditional dependence, we fit $m_1 = 5$ quantile regression functions, resulting in approximately 50{,}000 effective samples combining collocated and PM$_{2.5}$-only points. Finally, the collocated data are split into 90\% training and 10\% testing sets using a fixed random seed to ensure reproducibility.

\subsection{Comparison with Gaussian process kriging}\label{sec:comparison_ck}

Because AQI measurements, which are considered the regulatory gold standard, are spatially sparse, our goal is to predict AQI concentrations by conditioning on the denser PM$_{2.5}$ fields. Similar to Section \ref{sec:simulation} we compare the predictive performance of DCK to CK using four metrics: MAE, PICP, AL, and computation time. Results are summarized in Table~\ref{tab:pm25_results}.

\begin{table}[t!]
\centering
\caption{MAE, PICP (95\% nominal), AL, and computation time (seconds) for CK and DCK applied to the FRM AQI and the CMAQ-simulated PM$_{2.5}$ data sets.}
\label{tab:pm25_results}
\begin{tabular}{||c c c c c||}
\hline
Method & MAE  & PICP  & AL & Time (s) \\
\hline\hline
CK  & 9.2 & 36.1 & 10.9 & 114726.7 \\
DCK & 4.8 & 98.9 & 39.3 & 491.1 \\
\hline
\end{tabular}
\end{table}

The results in the table clearly demonstrate that DCK delivers substantially higher predictive accuracy, reducing the mean absolute error (MAE) by nearly 50\% compared to classical kriging (CK). In terms of uncertainty quantification, DCK achieves a prediction interval coverage probability (PICP) close to the nominal 95\% target, whereas CK markedly under-covers, indicating overly narrow and overconfident intervals. The wider intervals produced by DCK are therefore more realistic and better calibrated. Additional details and visual comparisons are provided in Section~\ref{S-sec:visual_ck} of the Supplementary Material.

A notable practical advantage of DCK is its computational scalability. Inference with DCK completes in approximately 491 seconds, whereas CK requires over 114,000 seconds under the same conditions, corresponding to a speedup of more than 200-fold. Moreover, the DCK model can be trained offline, enabling near-instantaneous interpolation at new locations, whereas CK still requires substantial computation during the prediction stage even after parameter estimation.

Overall, these findings illustrate that DCK not only provides more accurate predictions and well-calibrated uncertainty estimates, but also dramatically improves computational efficiency, making it highly suitable for large-scale spatial datasets.

\subsection{Spatial prediction and hotspot analysis across subregions}

To further examine regional variability in model performance, we focus on three representative subregions of the United States: California (CA), the Northeast (NE), and the Southeast (SE). These regions capture diverse pollution regimes, ranging from strong inland-coastal contrasts in California to dense urban corridors in the Northeast and comparatively diffuse patterns in the Southeast. The subsets contain 108 AQI and 2,686 PM$_{2.5}$ locations in CA, 95 AQI and 1,873 PM$_{2.5}$ locations in NE, and 77 AQI and 3,499 PM$_{2.5}$ locations in SE. This provides a balanced combination of high-quality FRM AQI observations and spatially dense CMAQ-simulated PM$_{2.5}$ concentrations.

\begin{figure}[t!]
\centering

\begin{tabular}{cc}
\includegraphics[width=0.5\textwidth]{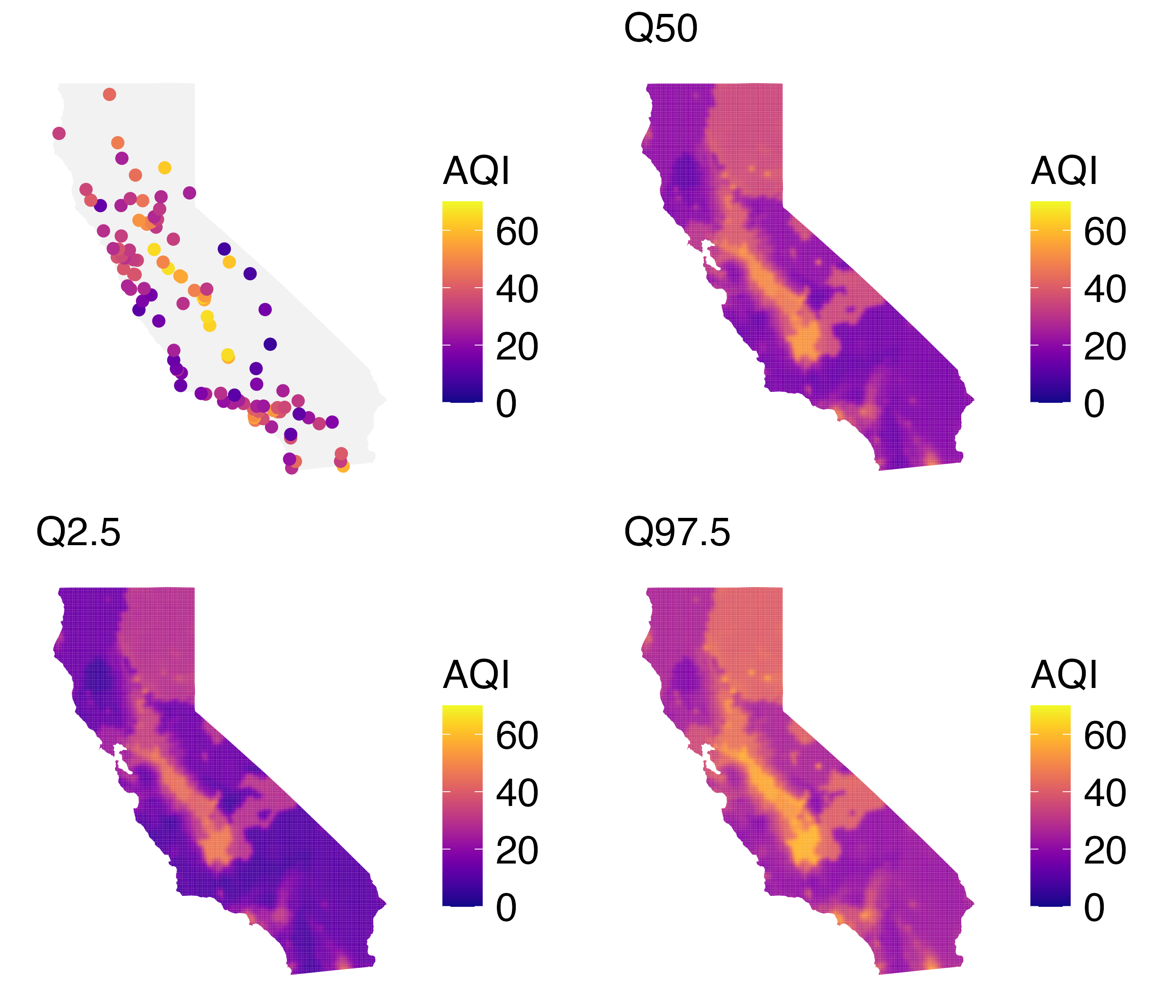} &
\includegraphics[width=0.5\textwidth]{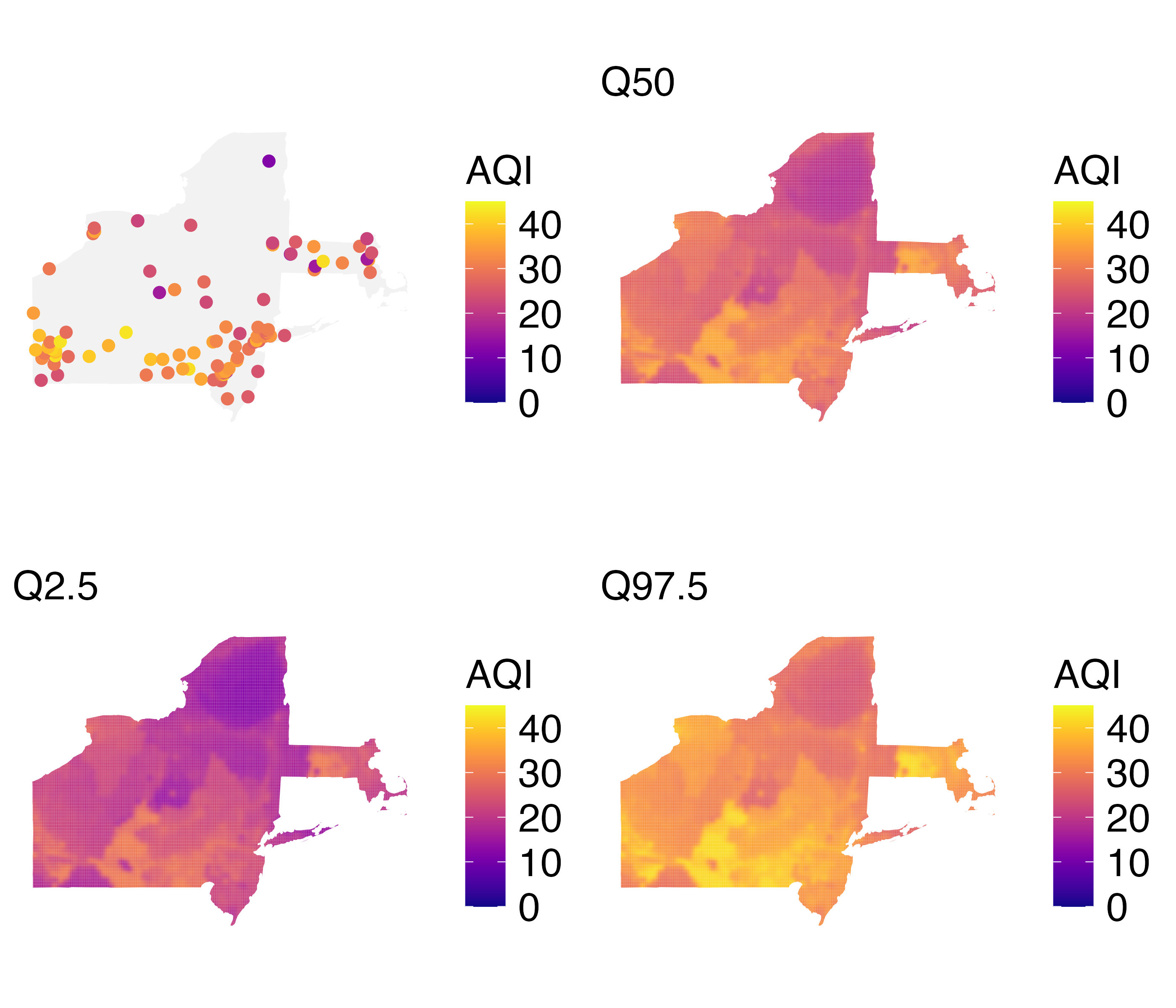} \\
\small (a) California & \small (b) Northeast
\end{tabular}

\vspace{0.4cm}

\begin{tabular}{c}
\includegraphics[width=0.5\textwidth]{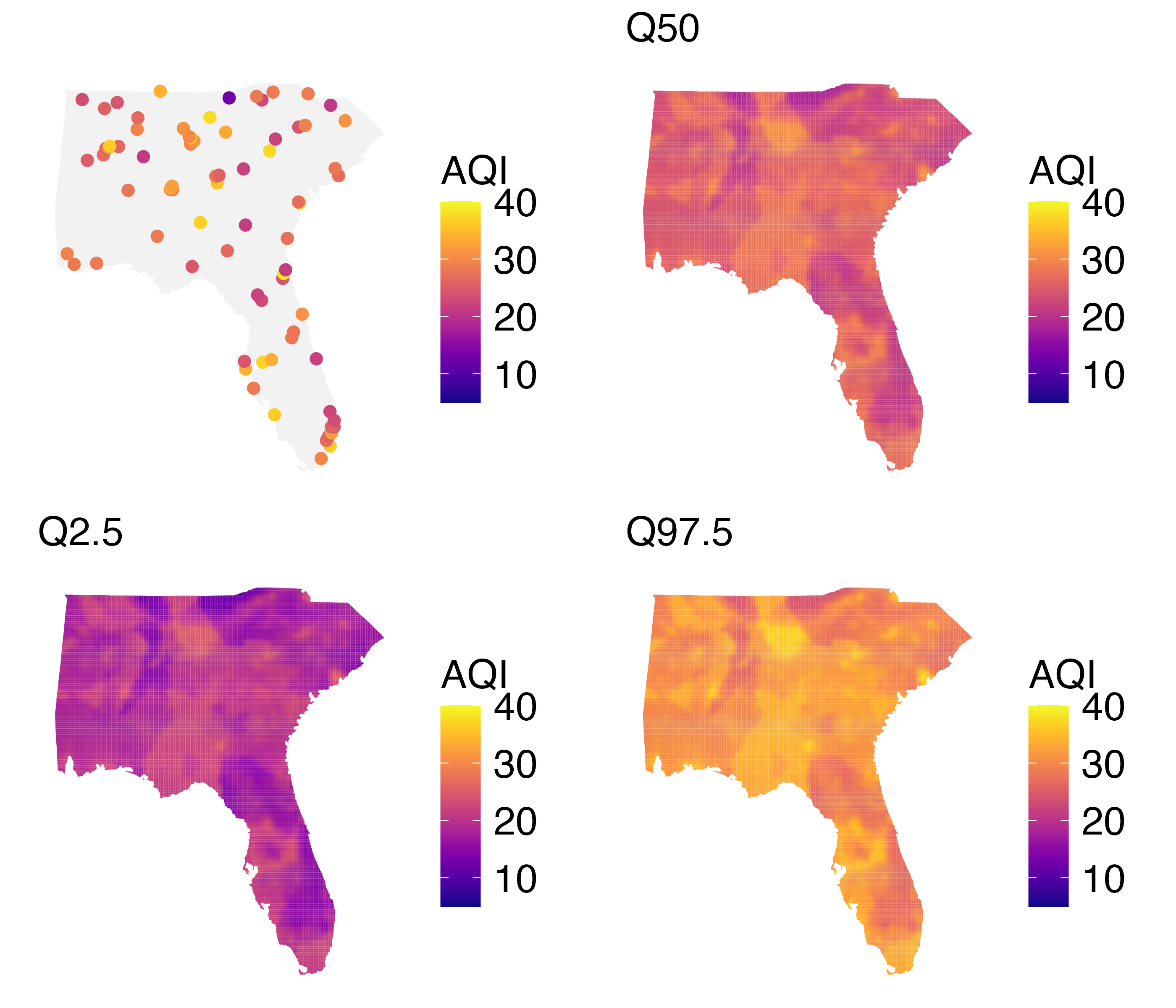} \\
\small (c) Southeast
\end{tabular}

\caption{Observed FRM AQI measurements and DCK-predicted conditional quantiles (Q2.5, Q50, and Q97.5) for the (a) California, (b) Northeast, and (c) Southeast subregions.}
\label{fig:heat_CA_NE_SE}
\end{figure}

Figure~\ref{fig:heat_CA_NE_SE} presents the observed FRM AQI values together with the DCK-based conditional predictive quantiles at the 2.5, 50, and 97.5 percent levels for CA, the NE, and the SE. Pronounced regional contrasts are evident. In California, the AQI observations exhibit a clear inland hotspot extending across the Central Valley and adjacent interior regions, in sharp contrast to substantially lower AQI levels along the coastal corridor. The DCK median surface (Q50) accurately reproduces this inland-coastal gradient, while the upper quantile (Q97.5) both intensifies and spatially expands the hotspot, reflecting elevated uncertainty and heightened risk in these interior areas. This behavior is consistent with well-documented accumulation of particulate matter under stagnant meteorological conditions and complex topography in inland California.

In the Northeast, elevated AQI values are concentrated along the densely urbanized corridor surrounding the New York metropolitan area, with a gradual attenuation toward inland and northern regions. The DCK predictions closely track this spatial structure, and the Q97.5 surface broadens the region of elevated AQI relative to the median, indicating increased uncertainty and potential for extreme events in major urban centers. By contrast, the Southeast displays comparatively weak spatial gradients, with AQI observations appearing more homogeneous across the region. The corresponding DCK surfaces exhibit smooth spatial variation and relatively narrow separation between quantiles, suggesting more stable air quality conditions and lower spatial variability.

\begin{figure}[t!]
\centering

\begin{tabular}{cc}
\includegraphics[width=0.5\textwidth]{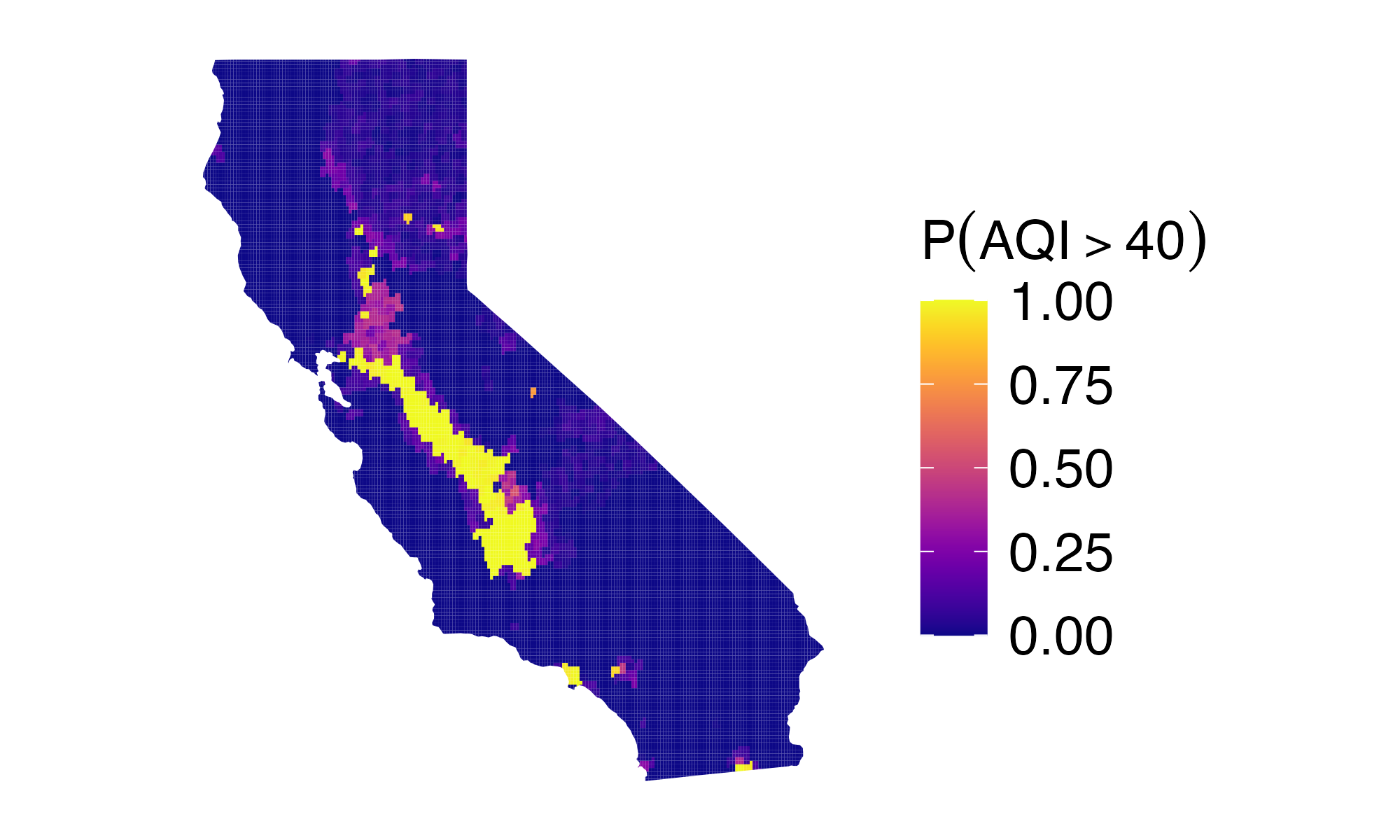} &
\includegraphics[width=0.5\textwidth]{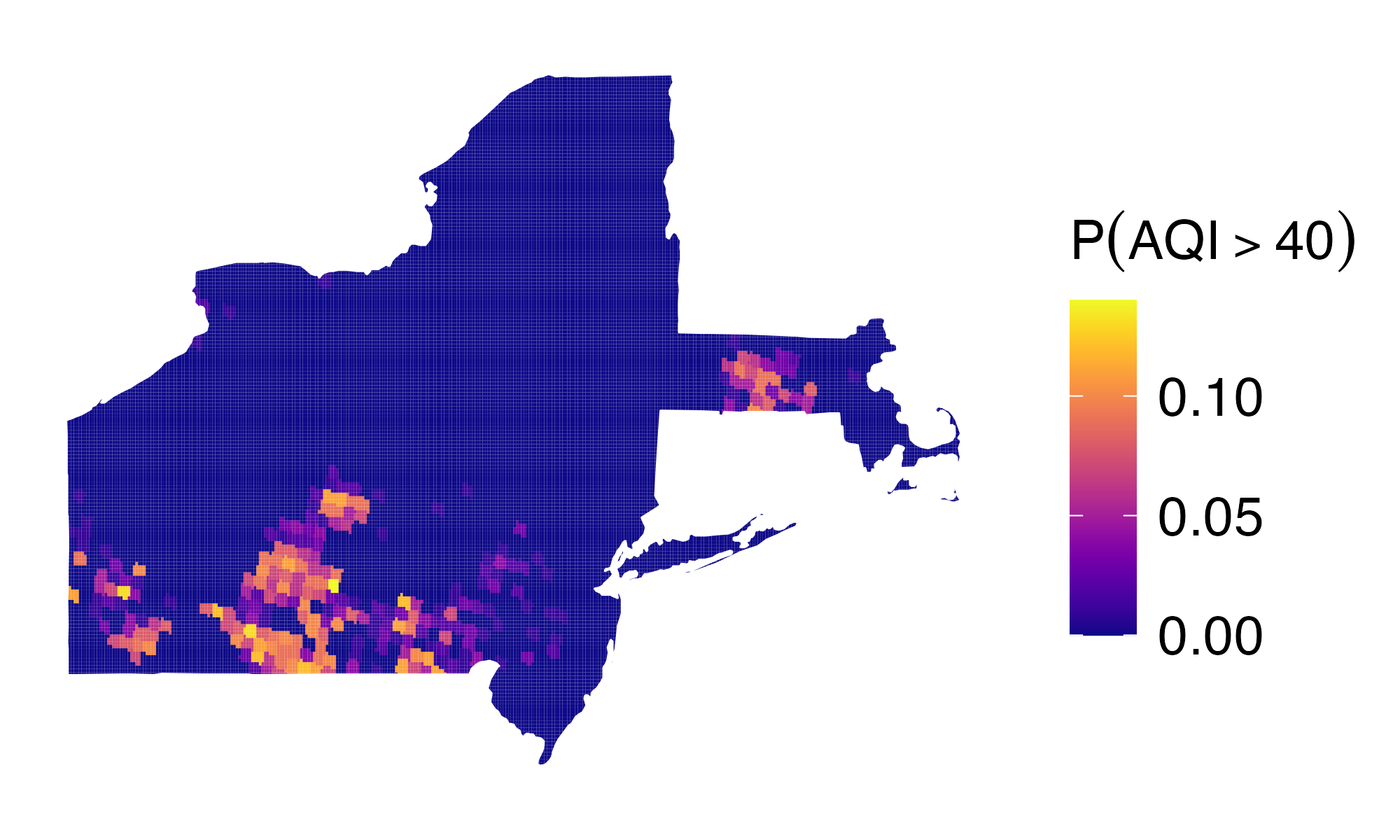} \\
\small (a) California & \small (b) Northeast
\end{tabular}

\vspace{0.4cm}

\begin{tabular}{c}
\includegraphics[width=0.5\textwidth]{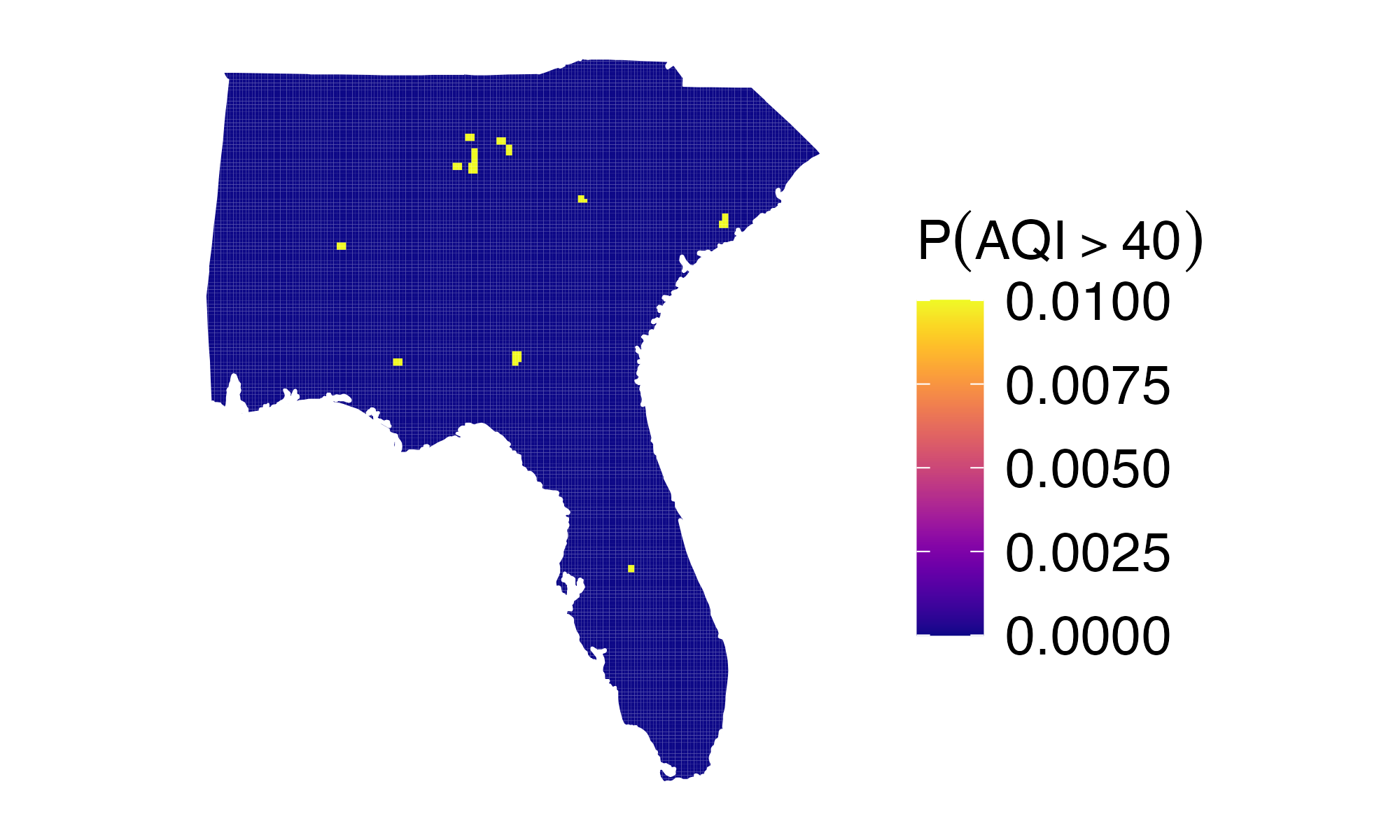} \\
\small (c) Southeast
\end{tabular}

\caption{Estimated probability of AQI exceeding 40 for (a) California, (b) the Northeast, and (c) the Southeast, based on 100 posterior predictive samples from the DCK model.}
\label{fig:extreme_CA_NE_SE}
\end{figure}

To further assess extreme pollution risk, Figure~\ref{fig:extreme_CA_NE_SE} shows the estimated probability of exceeding an AQI threshold of 40, computed as the proportion of posterior predictive samples exceeding the threshold at each location. California exhibits the most pronounced exceedance patterns, with a contiguous high-risk inland region that aligns with the observed hotspot in the predictive quantile maps. The Northeast shows more localized and moderate exceedance probabilities, primarily concentrated near major urban areas. In the Southeast, exceedance probabilities are near zero across the domain, indicating a low likelihood of high-AQI episodes during the study period.

Collectively, these results highlight the ability of DCK to adapt to diverse spatial pollution regimes. The framework captures sharp inland-coastal contrasts, urban-driven hotspots, and smoothly varying regional patterns, while simultaneously providing spatially coherent and interpretable uncertainty quantification. Such probabilistic outputs are particularly valuable for identifying regions at elevated risk of extreme AQI events and for supporting region-specific air quality management and policy decisions.

\section{Discussion}\label{sec:discussion}

This study introduces a nonparametric \textit{data fusion} mechanism for non-collocated bivariate spatial processes, enabling the principled integration of heterogeneous data sources such as FRM AQI observations and CMAQ model outputs. Our proposed DCK framework provides a flexible deep learning approach for spatial probabilistic modeling that addresses the key limitations of classical kriging and existing deep learning methods. By directly learning conditional distribution functions in a data-driven manner, DCK delivers accurate point predictions alongside well-calibrated uncertainty estimates without imposing Gaussian assumptions.  In the national-scale application, DCK yields bias-adjusted AQI prediction surfaces with improved accuracy, reliable coverage, and competitive computational efficiency relative to conventional methods.

From an applied perspective, the AQI application highlights the practical relevance of the proposed methodology for regulatory and public health decision-making. By modeling AQI directly, rather than pollutant concentrations alone, DCK aligns naturally with the metric used in air quality communication and policy. The framework leverages spatially sparse but high-fidelity FRM AQI observations together with spatially complete CMAQ-simulated PM$_{2.5}$ outputs to produce probabilistic AQI surfaces that capture regional heterogeneity, urban hotspots, and inland gradients. Importantly, the resulting predictive distributions enable the estimation of exceedance and extreme-event probabilities, which are critical for risk-based assessments and targeted intervention strategies. To our knowledge, this is the first statistical framework that performs distribution-free probabilistic interpolation of AQI by fusing regulatory AQI observations with model-based PM$_{2.5}$ fields, representing a novel contribution to the spatial statistics and environmental modeling literature.

Beyond its applied contributions, DCK offers a conceptual advance at the interface of geostatistics and deep learning by casting spatial prediction as a classification-based estimation of conditional distribution functions. This perspective provides a unifying framework that bridges classical kriging and modern neural approximators while retaining interpretability through probabilistic outputs.

Several directions for future work remain. Extending DCK to explicitly account for temporal dynamics would facilitate spatio-temporal AQI forecasting. Incorporating physical or chemical constraints from atmospheric processes may further enhance robustness and interpretability. Finally, evaluation of the framework across additional environmental and geophysical data sets would help assess its generality and scalability.

\section*{Statements and Declarations}

\if0\blind
\paragraph{Acknowledgements}
The authors thank Professor Noel Cressie and Professor Tatiyana Apanasovich
for their valuable advice and constructive feedback.

\paragraph{Funding}
The research of H.J.-W. and J.C. was partially supported by a Sustainability
Research Institute (SRI) Pilot Award from the George Washington University.
P.N. and Y.S. acknowledge support from King Abdullah University of Science
and Technology (KAUST).
\fi

\paragraph{Competing Interests}
The authors have no relevant financial or non-financial interests to disclose.

\if0\blind
\paragraph{Author Contributions}
J.C. and P.N. contributed to the methodology, software development, formal
analysis, simulation studies, data analysis, visualization, and preparation
of the original manuscript. H.J.-W. and Y.S. contributed to the methodology,
supervision, and review and editing of the manuscript. All authors read and
approved the final manuscript.
\fi

\paragraph{Data Availability}
The data used in this study were obtained from publicly available sources
maintained by the U.S. Environmental Protection Agency. The regulatory
monitoring data and numerical model outputs used in the air-quality
application are available from the data sources cited in the manuscript.
Additional information on data processing and analysis is provided in the
Supplementary Material.

\section*{Supplementary Material}
\begin{description}
\item[Supplement:] Performance metrics, univariate simulation experiments, sensitivity analyses of the tuning parameters, computational considerations, and further analyses of the AQI and PM$_{2.5}$ data. The Supplementary Material follows the References of this article.
\item[Code:] The code supporting the simulation studies and data application is available at \if0\blind\url{https://github.com/junyuchen929/Deep-classifier-kriging-for-probabilistic-spatial-prediction-of-air-quality-index}\else\texttt{<github link>}\fi.
\end{description}

\begin{singlespace}
\bibliographystyle{dcu}
\bibliography{ref}

@article{scheuerer2015variogram,
  title={Variogram-based proper scoring rules for probabilistic forecasts of multivariate quantities},
  author={Scheuerer, Michael and Hamill, Thomas M},
  journal={Monthly Weather Review},
  volume={143},
  number={4},
  pages={1321--1334},
  year={2015}
}

@article{gneiting2007strictly,
  title={Strictly proper scoring rules, prediction, and estimation},
  author={Gneiting, Tilmann and Raftery, Adrian E},
  journal={Journal of the American statistical Association},
  volume={102},
  number={477},
  pages={359--378},
  year={2007},
  publisher={Taylor \& Francis}
}

@book{nelsen2006introduction,
  author    = {Nelsen, Roger B.},
  title     = {An Introduction to Copulas},
  year      = {2006},
  publisher = {Springer},
  address   = {New York}
}

@article{wei2008approach,
  title={An approach to multivariate covariate-dependent quantile contours with application to bivariate conditional growth charts},
  author={Wei, Ying},
  journal={Journal of the American Statistical Association},
  volume={103},
  number={481},
  pages={397--409},
  year={2008},
  publisher={Taylor \& Francis}
}

@book{cressie2015statistics,
  author    = {Cressie, Noel A. C.},
  title     = {Statistics for Spatial Data},
  edition   = {Revised Edition},
  year      = {2015},
  publisher = {John Wiley \& Sons},
  address   = {Hoboken, NJ}
}

@article{berrocal2010space,
  title={Space-time data fusion under error in computer model output: An application to modeling air quality},
  author={Berrocal, Veronica J and Gelfand, Alan E and Holland, David M},
  journal={Biometrics},
  volume={66},
  number={3},
  pages={1234--1242},
  year={2010}
}

@article{zhang2018deep,
  title={Deep air learning: Interpolation, prediction, and feature analysis of fine-grained air quality},
  author={Zhang, Jingyuan and Zheng, Yu and Qi, Di and Li, Ruiyuan and Yi, Xiaohui and Li, Tianrui},
  journal={IEEE Transactions on Knowledge and Data Engineering},
  volume={30},
  number={12},
  pages={2285--2297},
  year={2018}
}

@article{cheng2021deep,
  title={Deep learning for {PM}$_{2.5}$ forecasting: A review and framework},
  author={Cheng, Jing and Wei, Yunqi and Chen, Yong},
  journal={Environmental Modelling \& Software},
  volume={145},
  pages={105206},
  year={2021}
}

@article{NAG2023100773,
title = {Spatio-temporal DeepKriging for interpolation and probabilistic forecasting},
journal = {Spatial Statistics},
volume = {57},
pages = {100773},
year = {2023},
issn = {2211-6753},
doi = {https://doi.org/10.1016/j.spasta.2023.100773},
url = {https://www.sciencedirect.com/science/article/pii/S2211675323000489},
author = {Nag, Pratik and Sun, Ying and Reich, Brian J.}
}

@inproceedings{tagasovska2019single,
  title={Single-model uncertainties for deep learning},
  author={Tagasovska, Natasa and Lopez-Paz, David},
  booktitle={Advances in Neural Information Processing Systems},
  volume={32},
  year={2019}
}

@inproceedings{gustafsson2020evaluating,
  title={Evaluating scalable bayesian deep learning methods for robust computer vision},
  author={Gustafsson, Fredrik K and Danelljan, Martin and Schon, Thomas B},
  booktitle={IEEE/CVF Conference on Computer Vision and Pattern Recognition Workshops (CVPRW)},
  pages={318--319},
  year={2020}
}

@article{mi2022training,
  title={Training deep histogram-based models for uncertainty quantification in regression tasks},
  author={Mi, Lei and Zhang, Han and Gong, Yuxuan and Lin, Weiyu},
  journal={Pattern Recognition},
  volume={127},
  pages={108616},
  year={2022}
}

@article{cressie2008fixed,
  title={Fixed rank kriging for very large spatial data sets},
  author={Cressie, Noel and Johannesson, Gardar},
  journal={Journal of the Royal Statistical Society Series B: Statistical Methodology},
  volume={70},
  number={1},
  pages={209--226},
  year={2008},
  publisher={Oxford University Press}
}

@article{gneiting2010matern,
  title={Mat{\'e}rn cross-covariance functions for multivariate random fields},
  author={Gneiting, Tilmann and Kleiber, William and Schlather, Martin},
  journal={Journal of the American Statistical Association},
  volume={105},
  number={491},
  pages={1167--1177},
  year={2010},
  publisher={Taylor \& Francis},
  doi={10.1198/jasa.2010.tm09420}
}

@article{Guinness02102018,
author = {Joseph Guinness},
title = {Permutation and Grouping Methods for Sharpening Gaussian Process Approximations},
journal = {Technometrics},
volume = {60},
number = {4},
pages = {415--429},
year = {2018},
publisher = {ASA Website},
doi = {10.1080/00401706.2018.1437476},

    note ={PMID: 31447491},


URL = { 
    
        https://doi.org/10.1080/00401706.2018.1437476
    
    

},
eprint = { 
    
        https://doi.org/10.1080/00401706.2018.1437476
    
    

}

}

@article{datta2016nearest,
  title={On nearest-neighbor {G}aussian process models for massive spatial data},
  author={Datta, Abhirup and Banerjee, Sudipto and Finley, Andrew O and Gelfand, Alan E},
  journal={Wiley Interdisciplinary Reviews: Computational Statistics},
  volume={8},
  number={5},
  pages={162--171},
  year={2016},
  publisher={Wiley Online Library}
}

@inproceedings{lakshminarayanan2017simple,
  title={Simple and scalable predictive uncertainty estimation using deep ensembles},
  author={Lakshminarayanan, Balaji and Pritzel, Alexander and Blundell, Charles},
  booktitle={Advances in Neural Information Processing Systems},
  volume={30},
  year={2017}
}

@article{chen2020deepkriging,
  author = {Chen, Wanfang and Li, Yuxiao and Reich, Brian J and Sun, Ying},
  title = {DeepKriging: Spatially Dependent Deep Neural Networks for Spatial Prediction},
  journal = {Statistica Sinica},
  volume = {34},
  number = {1},
  pages = {291--311},
  year = {2024}
 }

@article{agarwal2021copula,
  title={Copula-based multiple indicator kriging for non-{G}aussian random fields},
  author={Agarwal, Gaurav and Sun, Ying and Wang, Huixia J},
  journal={Spatial Statistics},
  volume={44},
  pages={100524},
  year={2021},
  publisher={Elsevier}
}

@article{lung2025peaks,
  title={Peaks, sources, and immediate health impacts of PM2. 5 and PM1 exposure in Indonesia and Taiwan with microsensors},
  author={Lung, Shih-Chun Candice and Tsou, Ming-Chien Mark and Cheng, Chih-Hui Chloe and Setyawati, Wiwiek},
  journal={Journal of Exposure Science \& Environmental Epidemiology},
  volume={35},
  number={2},
  pages={264--277},
  year={2025},
  publisher={Nature Publishing Group US New York}
}

@misc{board2024inhalable,
  title={Inhalable Particulate Matter and Health (PM2. 5 and PM10},
  author={Board, CAR},
  year={2024}
}

@article{appel2020community,
  title={The Community Multiscale Air Quality ({CMAQ}) model versions 5.3 and 5.3. 1: system updates and evaluation},
  author={Appel, K Wyat and Bash, Jesse O and Fahey, Kathleen M and Foley, Kristen M and Gilliam, Robert C and Hogrefe, Christian and Hutzell, William T and Kang, Daiwen and Mathur, Rohit and Murphy, Benjamin N and others},
  journal={Geoscientific Model Development Discussions},
  volume={2020},
  pages={1--41},
  year={2020},
  publisher={G{\"o}ttingen, Germany}
}

@article{nag2025bivariate,
  title={Bivariate DeepKriging for large-scale spatial interpolation of wind fields},
  author={Nag, Pratik and Sun, Ying and Reich, Brian J},
  journal={Technometrics},
  volume = {67},
number = {3},
pages = {397--408},
year = {2025},
  publisher={Taylor \& Francis}
}

@article{cressie1990origins,
  title={The origins of kriging},
  author={Cressie, Noel},
  journal={Mathematical geology},
  volume={22},
  number={3},
  pages={239--252},
  year={1990},
  publisher={Springer}
}

@article{nychka2015multiresolution,
  title={A multiresolution Gaussian process model for the analysis of large spatial datasets},
  author={Nychka, Douglas and Bandyopadhyay, Soutir and Hammerling, Dorit and Lindgren, Finn and Sain, Stephan},
  journal={Journal of Computational and Graphical Statistics},
  volume={24},
  number={2},
  pages={579--599},
  year={2015},
  publisher={Taylor \& Francis}
}

@article{journel1983nonparametric,
  title={Nonparametric estimation of spatial distributions},
  author={Journel, Andr{\'e} G},
  journal={Journal of the International Association for Mathematical Geology},
  volume={15},
  number={3},
  pages={445--468},
  year={1983},
  publisher={Springer}
}

@article{biau2016random,
  title={A random forest guided tour},
  author={Biau, G{\'e}rard and Scornet, Erwan},
  journal={Test},
  volume={25},
  number={2},
  pages={197--227},
  year={2016},
  publisher={Springer}
}

@article{fuentes2005model,
  title={Model evaluation and spatial interpolation by Bayesian combination of observations with outputs from numerical models},
  author={Fuentes, Montserrat and Raftery, Adrian E},
  journal={Biometrics},
  volume={61},
  number={1},
  pages={36--45},
  year={2005},
  publisher={Oxford University Press}
}

@misc{epaAQS,
  author       = {{U.S. EPA.}},
  title        = {Air Quality System Federal Reference Method monitoring data},
  year         = {2025},
  note         = { (downloaded on 3 January 2025)},
  howpublished = {\url{https://www.epa.gov/aqs}},
}

@misc{epaCMAQdata,
  author       = {{U.S. EPA.}},
  title        = {Community Multiscale Air Quality model output and related data sets },
  year         = {2025},
  note         = {(downloaded on 3 January 2025)},
  howpublished = {\url{https://www.epa.gov/cmaq/forms/download-cmaq-data}},
}

@article{pamina2019effective,
  title={An effective classifier for predicting churn in telecommunication},
  author={Pamina, Jeyakumar and Raja, Beschi and SathyaBama, S and Sruthi, MS and VJ, Aiswaryadevi and others},
  journal={Jour of Adv Research in Dynamical \& Control Systems},
  volume={11},
  year={2019}
}

@article{chen2014deep,
  title={Deep learning-based classification of hyperspectral data},
  author={Chen, Yushi and Lin, Zhouhan and Zhao, Xing and Wang, Gang and Gu, Yanfeng},
  journal={IEEE Journal of Selected topics in applied earth observations and remote sensing},
  volume={7},
  number={6},
  pages={2094--2107},
  year={2014},
  publisher={IEEE}
}

@article{apanasovich2012valid,
  title={A valid {M}at{\'e}rn class of cross-covariance functions for multivariate random fields with any number of components},
  author={Apanasovich, Tatiyana V and Genton, Marc G and Sun, Ying},
  journal={Journal of the American Statistical Association},
  volume={107},
  number={497},
  pages={180--193},
  year={2012},
  publisher={Taylor \& Francis}
}

@article{kingma2014adam,
  title={Adam: A method for stochastic optimization},
  author={Kingma, Diederik P},
  journal={arXiv preprint arXiv:1412.6980},
  year={2014}
}

@article{xu2017tukey,
  title={Tukey g-and-h random fields},
  author={Xu, Ganggang and Genton, Marc G},
  journal={Journal of the American Statistical Association},
  volume={112},
  number={519},
  pages={1236--1249},
  year={2017},
  publisher={Taylor \& Francis}
}

@article{schmidt2020nonparametric,
  title={Nonparametric regression using deep neural networks with ReLU activation function},
  author={Schmidt-Hieber, Johannes},
  journal={The Annals of Statistics},
  volume={48},
  number={4},
  pages={1875--1897},
  year={2020},
  publisher={Institute of Mathematical Statistics}
}

@article{gao2017properties,
  title={On the properties of the softmax function with application in game theory and reinforcement learning},
  author={Gao, Bolin and Pavel, Lacra},
  journal={arXiv preprint arXiv:1704.00805},
  year={2017}
}

@article{buja2005loss,
  title={Loss functions for binary class probability estimation and classification: Structure and applications},
  author={Buja, Andreas and Stuetzle, Werner and Shen, Yi},
  journal={Working draft, November},
  volume={3},
  year={2005},
  publisher={Citeseer}
}
\end{singlespace}
\end{document}


\begin{center} {\Large{\bf Supplementary Material for ``Deep Classifier Kriging for Probabilistic Spatial Prediction of Air Quality Index''}}
\end{center}

\baselineskip=12pt
\vskip 5mm

\setcounter{page}{1}
\setcounter{section}{0}
\setcounter{equation}{0}
\setcounter{table}{0}
\setcounter{figure}{0}

This Supplementary Material provides additional simulation results, sensitivity analyses and further supporting analyses on FRM AQI and CMAQ-simulated PM$_{2.5}$ concentrations that complement the main text. Section~\ref{sec:fsim} presents further simulation experiments for the univariate setting, together with detailed descriptions of the evaluation metrics. Section~\ref{sec:sensitivity} reports sensitivity analyses with respect to key tuning parameters. Section~\ref{sec:computation} discusses computational considerations and scalability of the proposed DCK framework. Finally, Section~\ref{sec:analysis} presents additional analyses for AQI prediction, including residual diagnostics, predictive interval assessment, and joint distribution diagnostics.

\section{Further simulation details }\label{sec:fsim}

\subsection{Performance metrics}\label{sec:metrics}

We assess the predictive performance of each model using evaluation metrics computed on a set of test locations namely $D_0$. Specifically, we use Median Absolute Error (MAE) to quantify the accuracy of point predictions for locations in $D_0$:
\begin{equation}
  \text{MAE} =\frac{1}{|D_0|} \frac{1}{T} \sum_{\mathbf{s}_0 \in D_0}\sum_{b=1}^{T} \left| \hat{Y}_p^{(b)}(\mathbf{s}_0) - Y_p^{(b)}(\mathbf{s}_0) \right|, \quad, k = 1,2.
  \label{eq:5}
\end{equation}
where $T$ is the total number of simulations, and $b=1, \ldots, T$ denotes the simulation index. $\hat{Y}_p^{(b)}(\cdot)$ represents the estimated median of the $p$-th spatial variable in the $b$-th simulation.
To evaluate the quality of the prediction intervals, we use Prediction Interval Coverage Probability (PICP)
and Average Length (AL) given as
\begin{equation}
\begin{aligned}
  \mathrm{PICP} &= \frac{1}{|D_0|} \frac{1}{T} \sum_{\mathbf{s}_0 \in D_0}\sum_{b=1}^{T} \mathbbm{1} \left\{ \hat{Y}_p^{(b)}(\mathbf{s}_0) \in \left[ L_p^{(b)}(\mathbf{s}_0),\; U_p^{(b)}(\mathbf{s}_0) \right] \right\}, \\
  \mathrm{AL} &= \frac{1}{|D_0|} \frac{1}{T} \sum_{\mathbf{s}_0 \in D_0}\sum_{b=1}^{T} \left( U_p^{(b)}(\mathbf{s}_0) - L_p^{(b)}(\mathbf{s}_0) \right),
\end{aligned}
\label{eq:6}
\end{equation}
where $L_p^{(b)}\left(\mathbf{s}_0\right)=q_{\mathbf{s}_0}(\alpha / 2), \, U_p^{(b)}\left(\mathbf{s}_0\right)=q_{\mathbf{s}_0}(1-\alpha / 2), \, b = 1, \ldots, T$, are the lower and upper prediction bounds for location $\mathbf{s}_0$ at confidence level $\alpha$.

\subsection{Simulation of univariate scenario}\label{sec:univariate}

To assess the performance of the proposed DCK framework under different distributional assumptions, we consider two scenarios: a Gaussian scenario with exponential covariance function and a non-Gaussian scenario where a Gaussian process is simulated and transformed using the Tukey g-and-h transformation to obtain a non-Gaussian process.
In each of these, we consider both $N^* = 1600$ and $N^* = 3600$ training samples. We fix the number of classes $n=30$ and the constant $C=12$ in both scenarios for the consistency. The results are averaged over 100 simulation replicates.

    For Gaussian scenario, we simulate 100 replicates from the Gaussian processes $Y(\cdot)$ with exponential covariance function given as
    

\[
\mathrm{Cov}\{Y(\mathbf{s}_1),Y(\mathbf{s}_2)\}
= C(h_d)
= \exp\!\left(-\frac{h}{\gamma}\right),
\]
    where $h_d = \left\|\mathbf{s}_1-\mathbf{s}_2\right\|$ is the distance between the locations $\boldsymbol{s}_1$ and $\boldsymbol{s}_2$, and $\gamma$ is the range parameter.    
    We take $\gamma=0.5$ and simulate $N^*$ spatial observations at irregularly spaced locations over the region $[0,1] \times[0,1]$. The irregular spatial locations are generated by the following procedure: First, $40 \times 40$ spatial locations are generated on the regular grid $[0,1] \times[0,1]$; then a random perturbation uniformly distributed on $[-0.4,0.4]$ is added to each location. 
    
    We randomly divide $90 \%$ data for training and leave $10 \%$ data for validation. 
We compute empirical quantiles at evenly spaced levels from 0.01 to 0.99 and use them to define $n$ thresholds. Each observation is assigned a class label based on the threshold it falls into. We incorporate two competing methods for comparison;  
The first one is the classical Kriging (CK) \citep{cressie2015statistics}, with exponential covariance; the second one is the univariate DeepKriging (DK) method with ensembles to produce prediction uncertainties \citep{nag2025bivariate}. To compute the  prediction interval of DK, an ensemble of 20 neural network models are considered along with a neighbourhood size of 40 to compute the mis-specification variance.

In addition to the Gaussian scenario, we also consider a non-Gaussian spatial process generated through a nonlinear transformation of a latent Gaussian process. Following the structure of \eqref{M-eq:process_model} in the main text, we simulate a stationary Gaussian process with mean function  $\mu(\mathbf{s})$, given as
    $$
    \begin{aligned}
    \mu(\mathbf{s})= & x_1(\mathbf{s})^2-x_2(\mathbf{s})^2+x_3(\mathbf{s})^2-x_4(\mathbf{s})^2-x_5(\mathbf{s})^2+2 x_1(\mathbf{s}) x_2(\mathbf{s})+3 x_2(\mathbf{s}) x_3(\mathbf{s}) \\
    & -2 x_3(\mathbf{s}) x_5(\mathbf{s})+10 x_1(\mathbf{s}) x_4(\mathbf{s})+\sin \left(x_1(\mathbf{s})\right) x_2(\mathbf{s}) x_3(\mathbf{s})+\cos \left(x_2(\mathbf{s})\right) x_3(\mathbf{s}) x_5(\mathbf{s}) \\
    & +x_1(\mathbf{s}) x_2(\mathbf{s}) x_4(\mathbf{s}) x_5(\mathbf{s}),
    \end{aligned}
    $$
    where each covariate $x_i(\mathbf{s}), i=1, \ldots, 5$ were simulated independently from a stationary Gaussian process with Mat\'ern covariance function, with standard deviation $\sigma=0.9$, smoothness $\nu=0.5$, and range $\alpha=0.1$, over a set of $N^*$ irregularly sampled locations on the unit square $[0,1] \times[0,1]$, similar to the Gaussian experiments. The spatial random effect $\gamma(\mathbf{s})$ was generated from a zero-mean Gaussian process again with a Mat\'ern covariance function, with standard deviation $\sigma=0.7$, smoothness $\nu=0.5$, and range $\alpha=0.2$.
To introduce non-Gaussianity we transform $\gamma(\mathbf{s})$ using the Tukey $g$-and-$h$ transformation (\cite{xu2017tukey}) with $g=0.8, h=0.5$. We also incorporate two competing methods for comparison.
The first one is classical kriging (CK), implemented with a Mat\'ern covariance function, which is optimal under the Gaussian assumption but generally mis-specified in the non-Gaussian setting.
The second one is the univariate DeepKriging (DK) method.

\begin{table}[ht]
\centering
\caption{MAE, PICP for 95\% coverage probability, and AL with corresponding standard errors (shown in parentheses as relative percentages), together with computation time (in seconds), for CK, DCK and DK methods applied different univariate simulation experiments.}
\label{tab:non_gaussian_gk_dnn_dk}
\begin{tabular}{||c|c|c c c c c||}
\hline
Scenario & $N$ & Method & MAE (SE) & PICP (SE) & AL (SE) & Time (s) \\
\hline\hline
\multirow{6}{*}{Gaussian} 
 & 1600 & CK  & 0.15 (0.1) & 95.0 (0.2) & 0.75 (0.2) & 199.9 \\
 &      & DCK & 0.18 (0.1) & 93.7 (0.5) & 0.93 (1.8) &  12.3 \\
 &      & DK  & 0.18 (0.1) & 95.0 (0.2) & 0.93 (0.3) & 141.5 \\
\cline{2-7}
 & 3600 & CK  & 0.13 (0.1) & 94.9 (0.1) & 0.62 (0.1) & 1776.7 \\
 &      & DCK & 0.15 (0.1) & 94.2 (0.4) & 0.75 (1.3) &   23.9 \\
 &      & DK  & 0.15 (0.1) & 94.9 (0.1) & 0.76 (0.3) &  209.0 \\
\hline
\multirow{6}{*}{non-Gaussian}
 & 1600 & CK  & 5.0 (8.3) & 94.6 (0.2) & 28.2 (61.5) &  570.5 \\
 &      & DCK & 2.8 (9.5) & 97.1 (0.2) & 21.3 (76.6) &   13.2 \\
 &      & DK  & 2.5 (7.6) & 96.9 (0.1) & 16.5 (58.0) &  206.0 \\
\cline{2-7}
 & 3600 & CK  & 4.4 (7.2) & 94.9 (0.2) & 25.3 (72.0) & 6944.1 \\
 &      & DCK & 2.1 (7.2) & 97.1 (0.2) & 16.4 (57.9) &   25.0 \\
 &      & DK  & 1.7 (6.6) & 97.2 (0.1) & 11.5 (57.9) &  281.0 \\

\hline
\end{tabular}

\end{table}

Table \ref{tab:non_gaussian_gk_dnn_dk} summarizes the experimental results, showing that in Gaussian scenario, our method, DCK performs comparably to CK in both point and interval prediction, achieving similar MAE and PICP with slightly larger AL, while reducing computation time by an order of magnitude
. Compared to DK, DCK delivers comparable accuracy and uncertainty calibration but is over ten times faster since it estimates prediction intervals directly through conditional CDF approximation rather than ensemble-based uncertainty modeling. In non-Gaussian scenario, DCK substantially outperforms CK across all metrics, achieving lower MAE, higher PICP and shorter AL, while significantly reducing runtime. The DCK model also demonstrates a favorable trade-off compared to DK, which, while achieving strong and often best-in-class predictive performance across various non-Gaussian experiments and outperforming other state-of-the-art low-rank approximation methods such as \textit{FRK} \citep{cressie2008fixed}, \textit{NNGP} \citep{datta2016nearest}, and \textit{GpGp} \citep{Guinness02102018}, incurs significantly higher computational cost. In contrast, DCK attains comparable uncertainty calibration with much faster inference.
From the above comparison it is evident that DCK maintains accuracy and reliability while providing superior scalability and efficiency, making it a practical alternative to classical kriging for large-scale or time-sensitive tasks.

\begin{figure}
    \centering
    \includegraphics[width=1\linewidth]{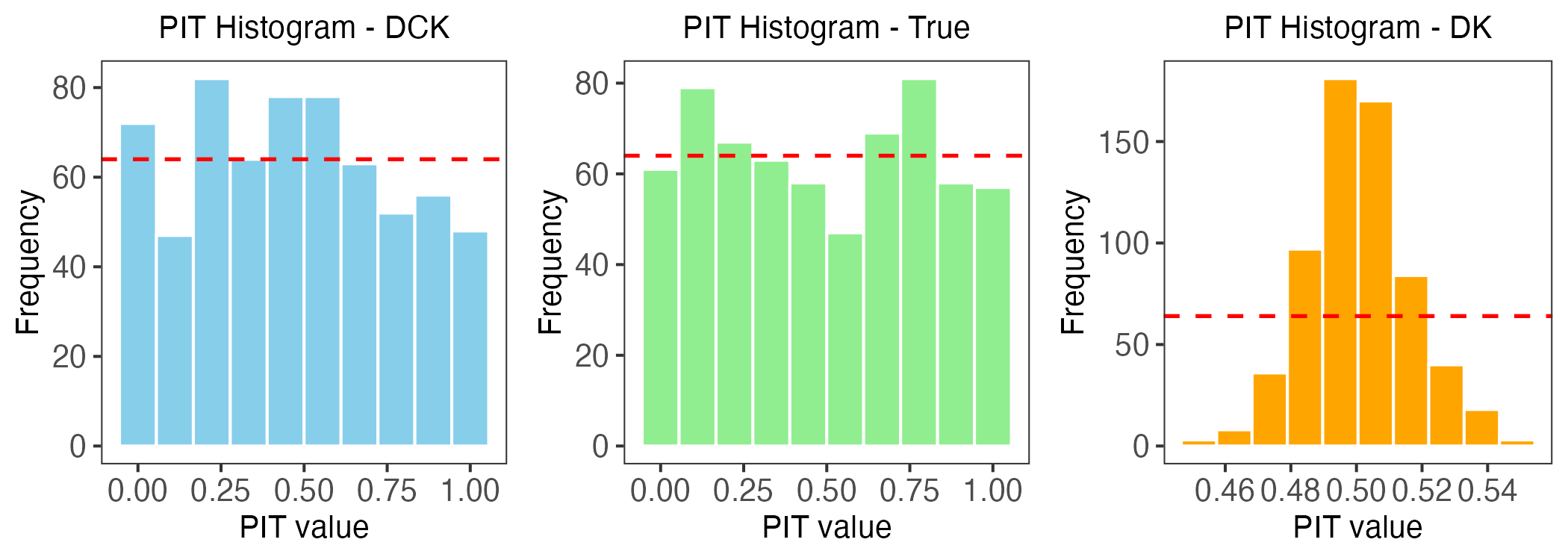}
    \caption{PIT histograms comparing the DCK, DK, and the true field in univariate scenario.}
    \label{fig:pit1}
    
\end{figure}

To further verify the advantages of DCK from posterior distribution estimation compared with DK, we compared the Probability Integral Transform (PIT) histograms of the DCK, DK, and true field models for the non-Gaussian experiments. The results are depicted in Figure~\ref{fig:pit1}. Under a well-calibrated model, the PIT values should follow a uniform distribution on ([0, 1]). 
To get a larger test set and clearer visualization, the plots were generated using training data with 6,400 locations.
From the plots it can be seen that DK exhibits a concentrated shape around the center, indicating under-dispersion and overconfident predictive intervals. While DK consistently demonstrates strong predictive performance across different settings, it is important to note that its formulation inherently assumes Gaussianity of the predictive distribution. This assumption, although effective for achieving accurate point predictions and reasonable interval coverage, may limit its ability to fully capture the underlying distributional characteristics in non-Gaussian settings. As a result, the PIT histogram of DK can deviate from uniformity, reflecting a mismatch between the assumed and actual data distributions. In contrast, the DCK model relaxes this Gaussianity assumption by directly estimating the empirical conditional CDF, allowing it to better represent non-Gaussian behaviors while maintaining reliable uncertainty quantification.

\section{Sensitivity analysis}\label{sec:sensitivity}

To further examine the robustness and stability of the proposed DCK model, we conduct a sensitivity analysis on several key hyper-parameters as discussed in the last paragraph of Section~\ref{M-sec:kernel_smoothing} of the main text.

We first examine the number of classes $n$, which is specified directly in the univariate
setting and controlled by the minimum sample size per class $\delta$ in the bivariate setting, as well as the bandwidth $h$, which controls the degree of spatial smoothing through a scaling parameter $C$ in the kernel function. These two parameters jointly determine the resolution and locality of the conditional distribution estimation: $n$ influences the granularity of the conditional CDF approximation, while $h$ (via $C$) adjusts the effective neighborhood size in spatial interpolation.
In addition, we investigate the number of quantile regressions $m_1$ in the bivariate scenario, which determines the lines used to model the conditional dependence between $Y_1(\cdot)$ and $Y_2(\cdot)$ and influence how finely the cross-variable relationship is captured.

\subsection{Number of classes n}

In the univariate scenario, the number of classes $n$, which is also the number of quantile thresholds, controls how finely we divide the range of the response variable. A larger $n$ allows the model to represent more detailed variation in the conditional distribution, potentially improving prediction quality. However, increasing $n$ also leads to more class labels in the classification task, causing longer computation time. 
\begin{figure}[ht]
    \centering
    \includegraphics[width=\textwidth]
    {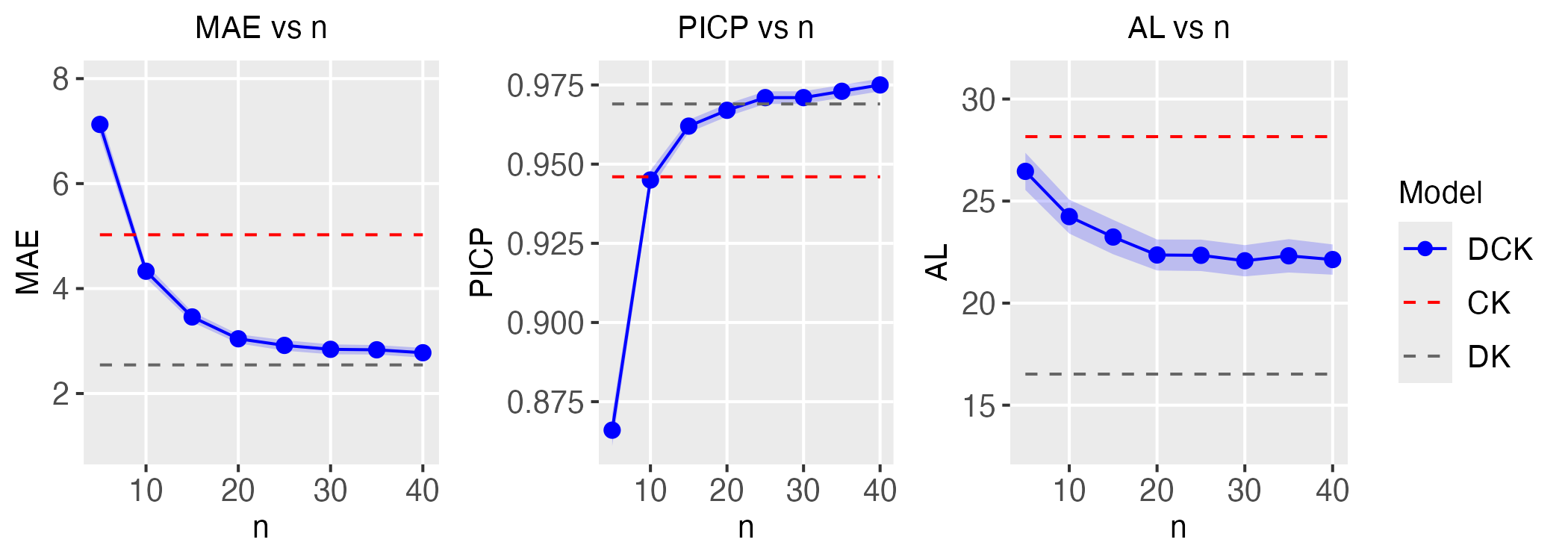}
    \caption{Performance with different $n$ in the univariate scenario (non-Gaussian data, $N^* = 1600$).}
    \label{ng1600_k}
\end{figure}
Figure \ref{ng1600_k} demonstrates the effect of $n$ on the performance of the model. Overall, the results show that increasing $n$ improves the model performance in terms of MAE, PICP, and AL.

\begin{figure}[t!]
\centering

\begin{tabular}{cc}
\includegraphics[width=0.4\textwidth]{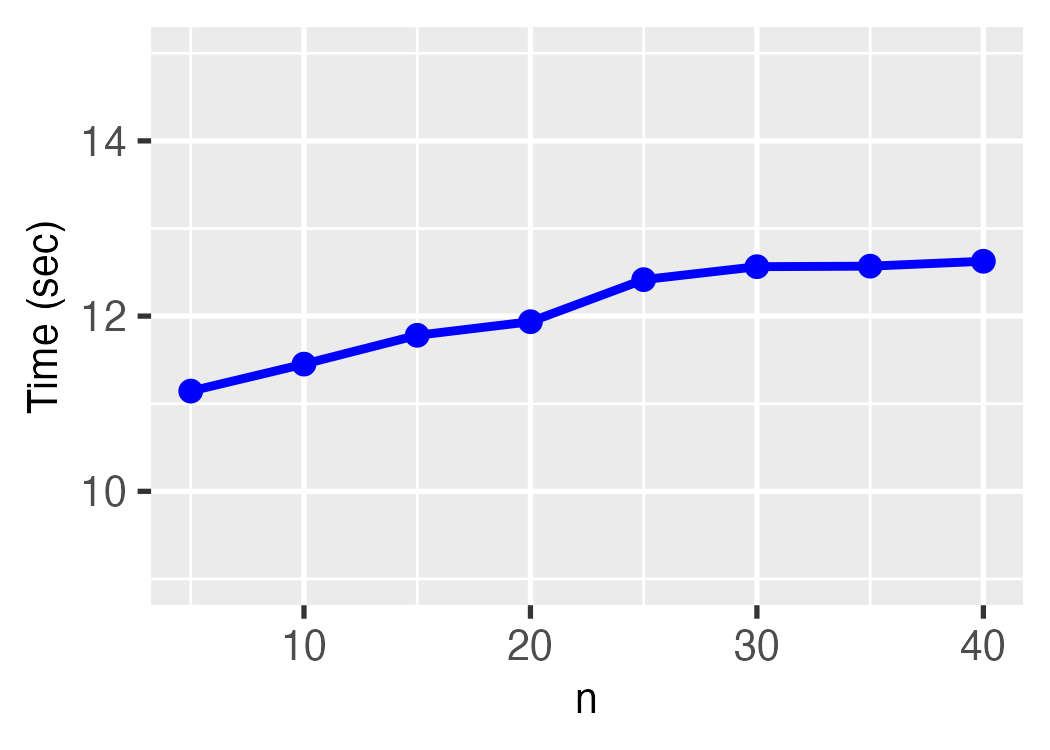} &
\includegraphics[width=0.4\textwidth]{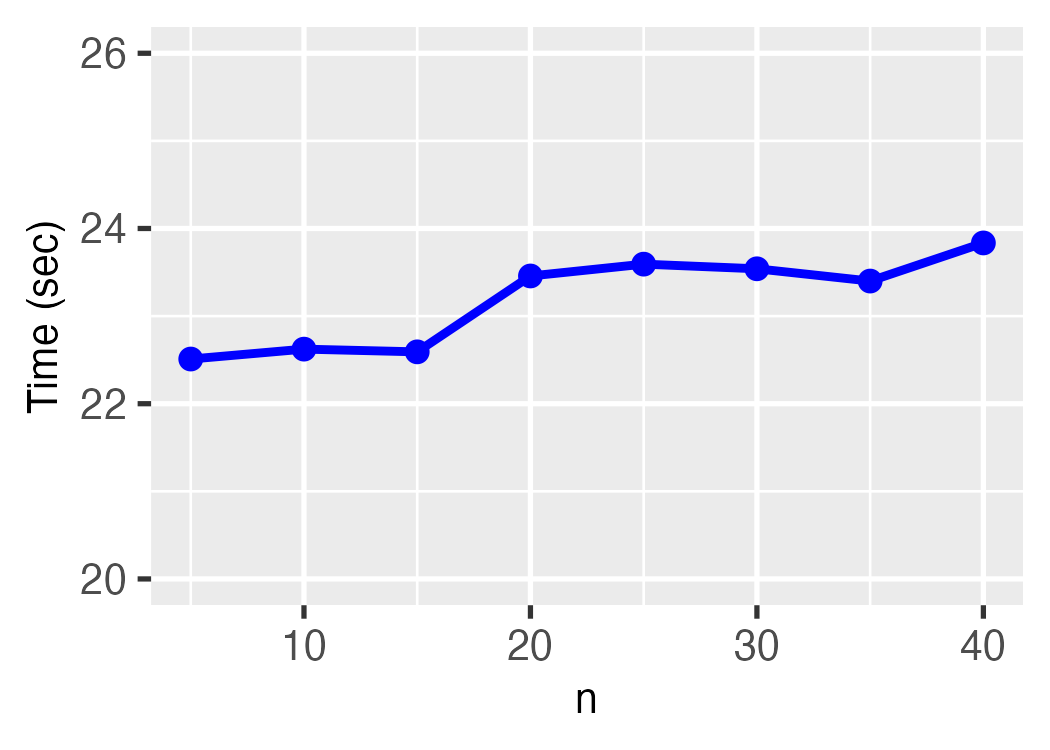} \\
\small (a) Gaussian data, $N^*=1600$ &
\small (b) Gaussian data, $N^*=3600$
\end{tabular}

\vspace{0.35cm}

\begin{tabular}{cc}
\includegraphics[width=0.4\textwidth]{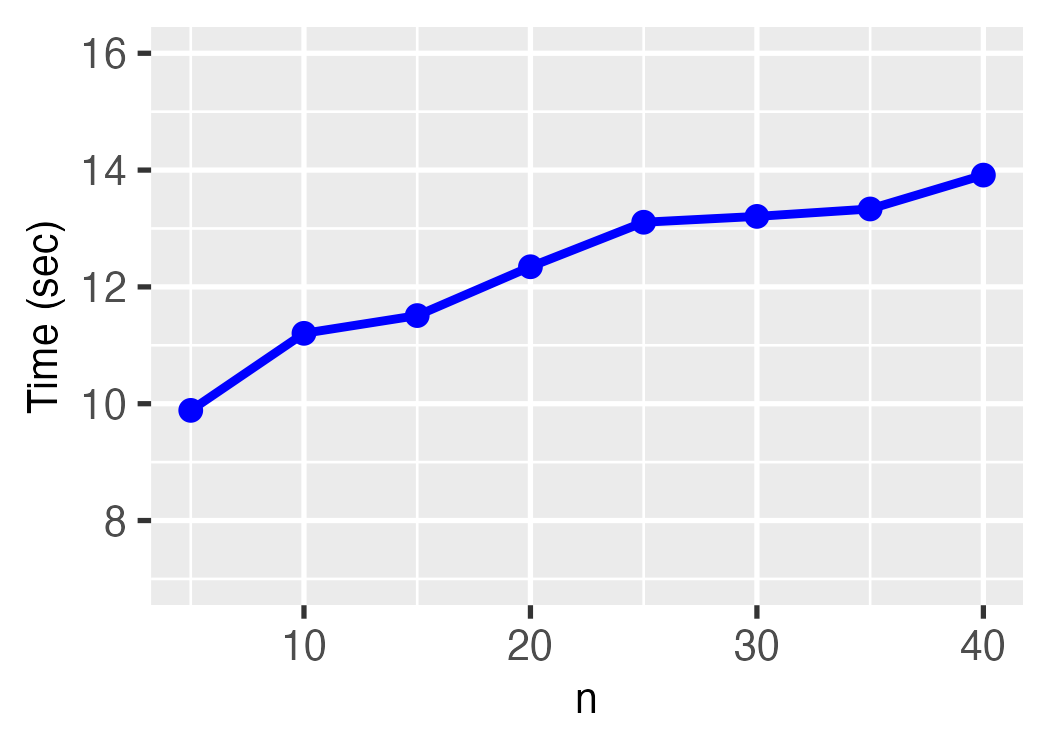} &
\includegraphics[width=0.4\textwidth]{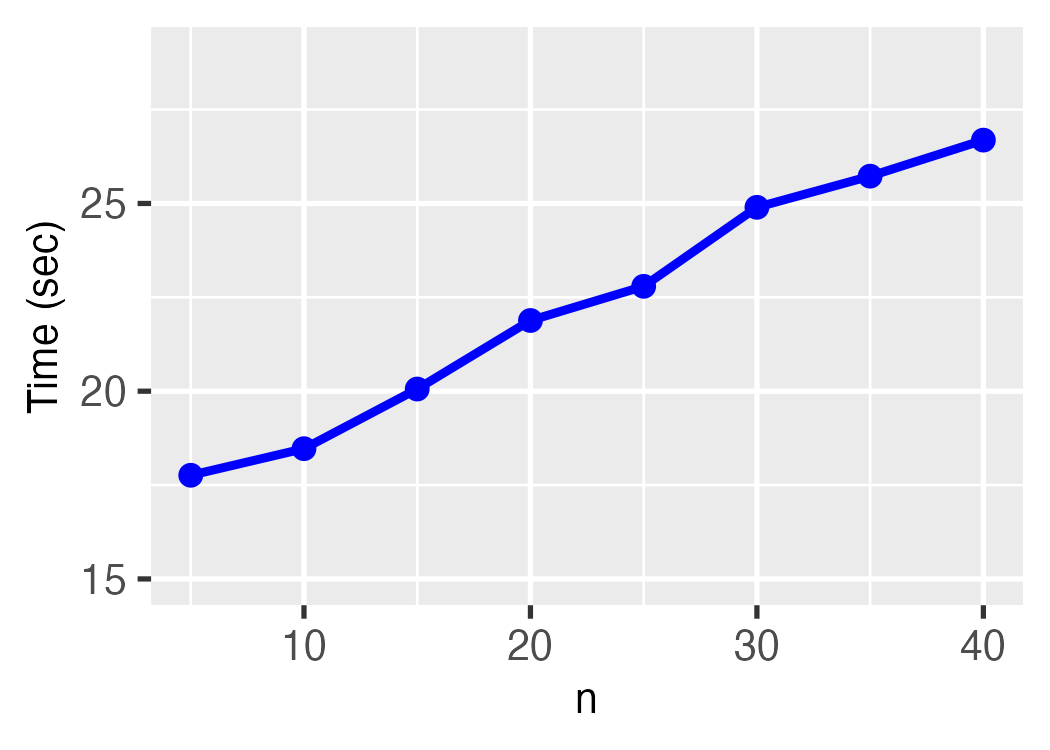} \\
\small (c) non-Gaussian data, $N^*=1600$ &
\small (d) non-Gaussian data, $N^*=3600$
\end{tabular}

\caption{Computation time v.s. $n$ under the univariate Gaussian and non-Gaussian scenario.}
\label{fig:ktime_all}
\end{figure}

In addition to improved predictive performance, increasing $n$ also leads to longer computation time, where the model must handle a larger number of threshold-based classification tasks under greater data complexity, as shown in Figure \ref{fig:ktime_all}.

A practical trade-off between predictive accuracy and computational cost across all scenarios indicates that choosing $n \geq 15$ is appropriate. This conclusion is supported by additional analyses conducted in the univariate setting under three scenarios: a Gaussian case with 1,600 samples, a Gaussian case with 3,600 samples, and a non-Gaussian case with 3,600 samples. In the bivariate setting, the minimum number of samples per class, denoted by $\delta$, also affects the total number of classes and exhibits an inverse relationship with $n$, since larger values of $\delta$ result in fewer classes. Taken together, these results suggest that a robust and practical choice across a wide range of settings is $n \geq 15$ and $\delta \leq 50$. We further examine computational efficiency and find that the runtime of DCK increases with $n$ and decreases with $\delta$, reflecting the corresponding changes in model complexity.

\begin{figure}[H]
    \centering    \includegraphics[width = \textwidth]{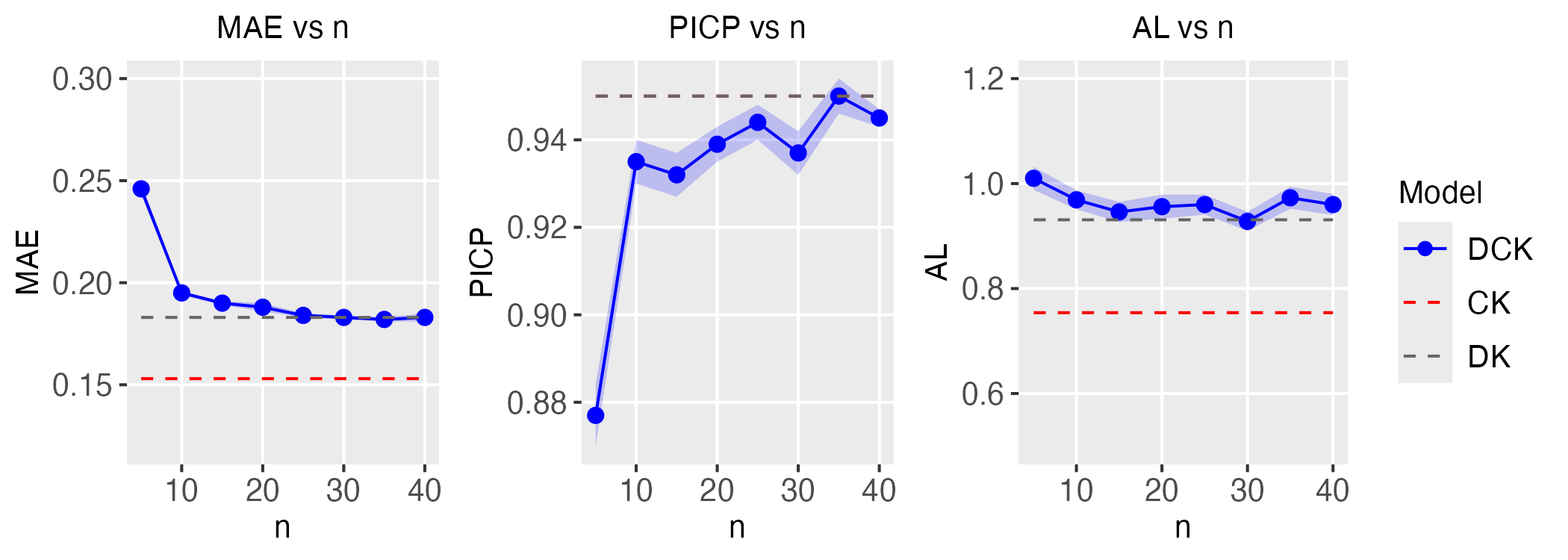}
    \caption{Performance with different $n$ in the univariate scenario (Gaussian, $N^* = 1600$).}
\end{figure}

\begin{figure}[H]
    \centering    \includegraphics[width = \textwidth]{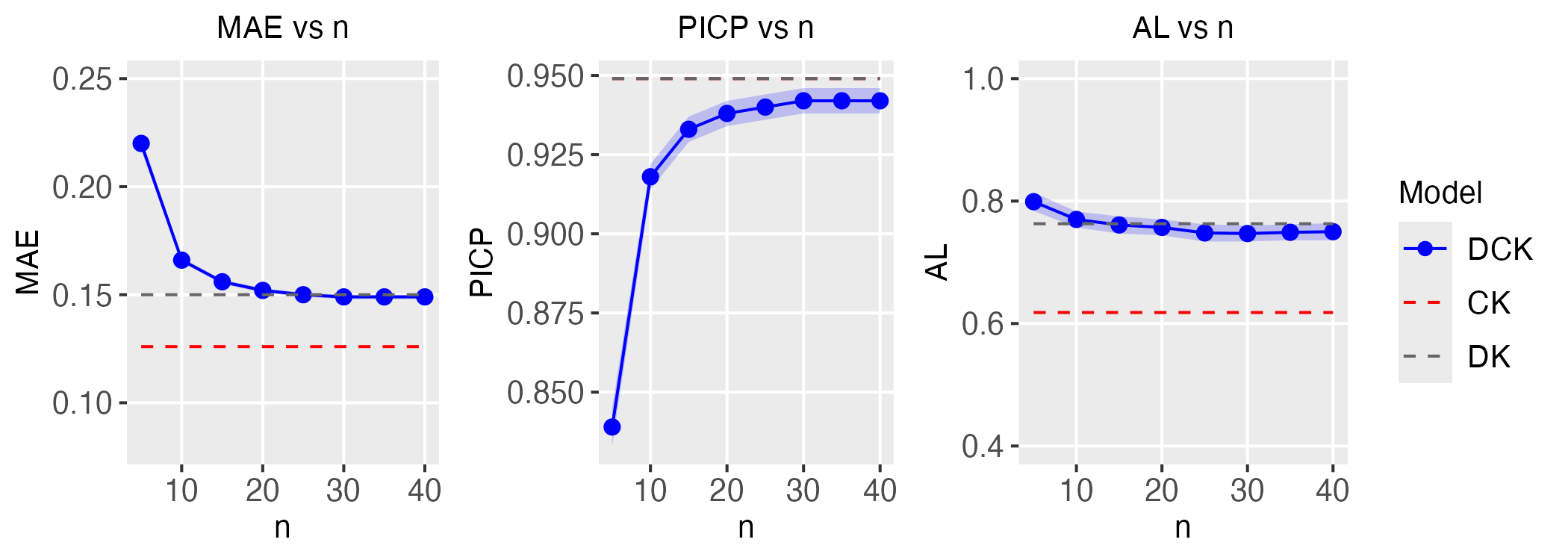}
    \caption{Performance with different $n$ in the univariate scenario (Gaussian, $N^* = 3600$).}
\end{figure}

\begin{figure}[H]
    \centering    \includegraphics[width = \textwidth]{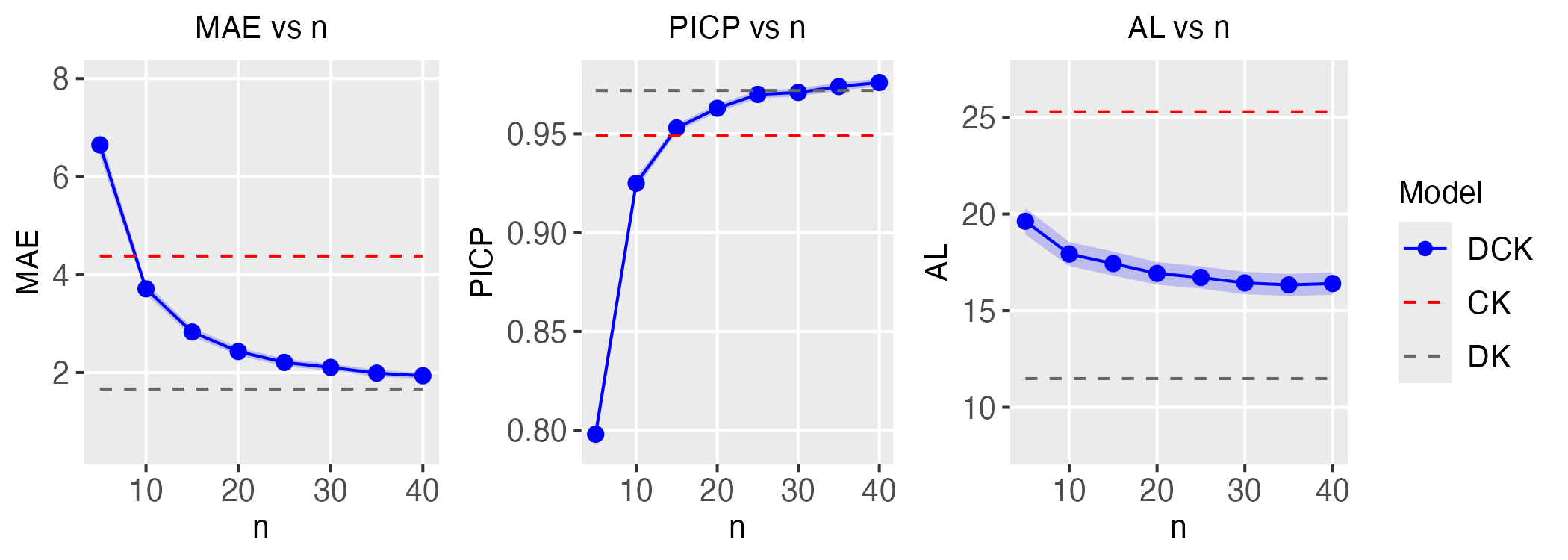}
    \caption{Performance with different $n$ in the univariate scenario (non-Gaussian, $N^* = 3600$).}
\end{figure}

\begin{figure}[H]
    \centering    \includegraphics[width = \textwidth]{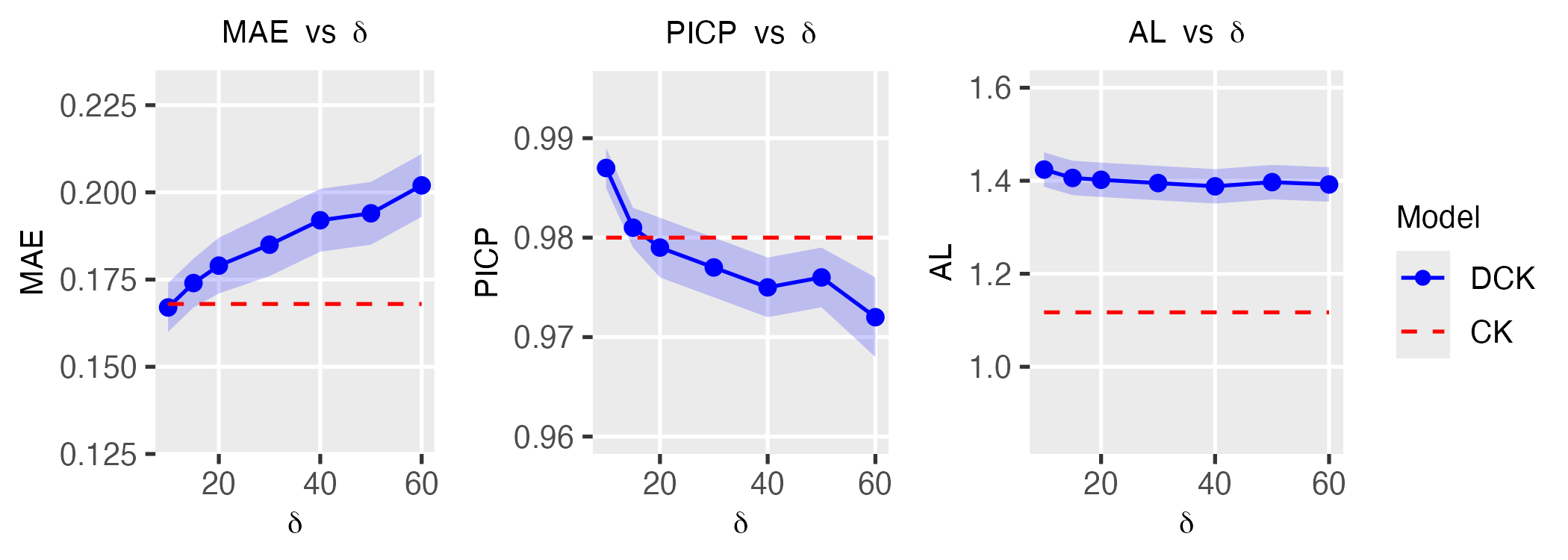}
    \caption{Performance with different $\delta$ in the bivariate scenario (Gaussian, $N^* = 1600$).}
\end{figure}

\begin{figure}[H]
    \centering    \includegraphics[width = \textwidth]{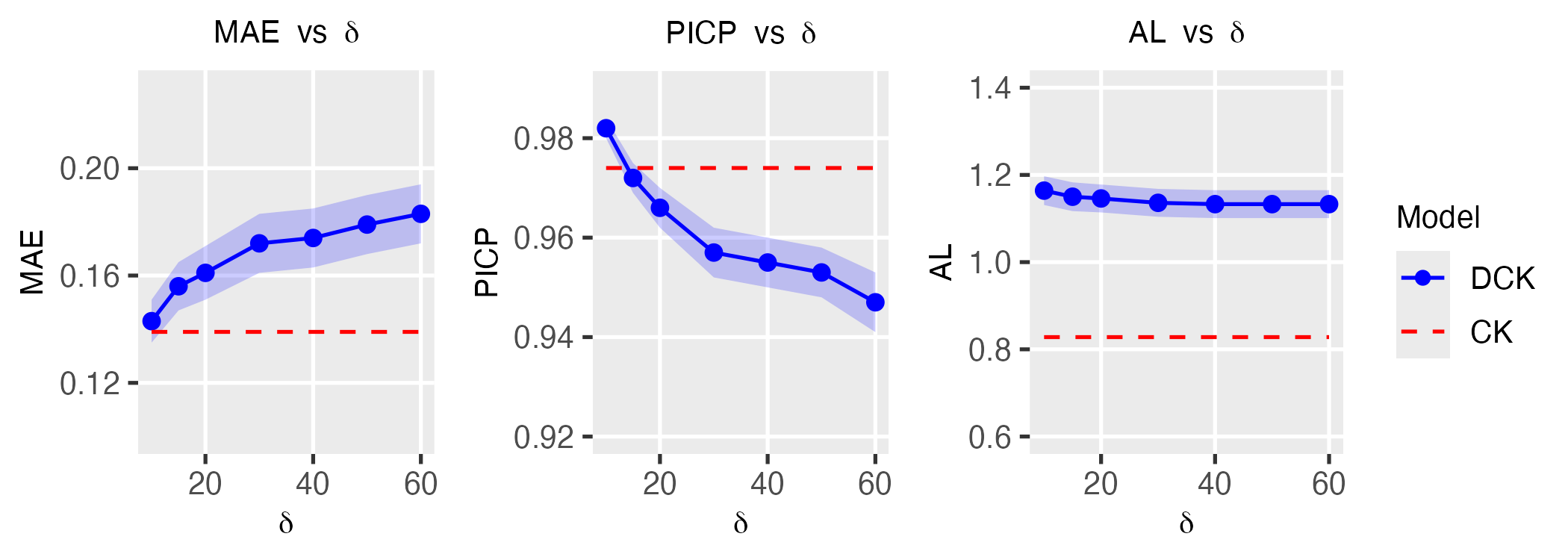}
    \caption{Performance with different $\delta$ in the bivariate scenario (Gaussian, $N^* = 3600$).}
\end{figure}

\begin{figure}[H]
    \centering    \includegraphics[width = \textwidth]{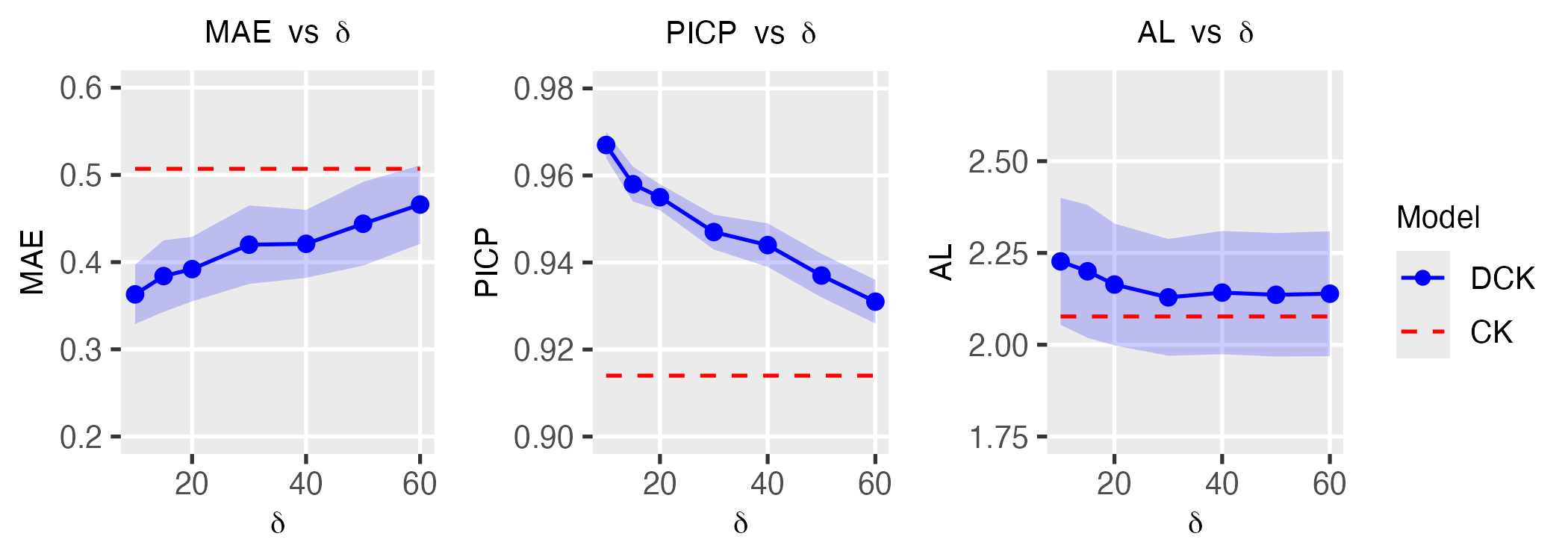}
    \caption{Performance with different $\delta$ in the bivariate scenario (non-Gaussian, $N^* = 1600$).}
\end{figure}

\begin{figure}[H]
    \centering    \includegraphics[width = \textwidth]{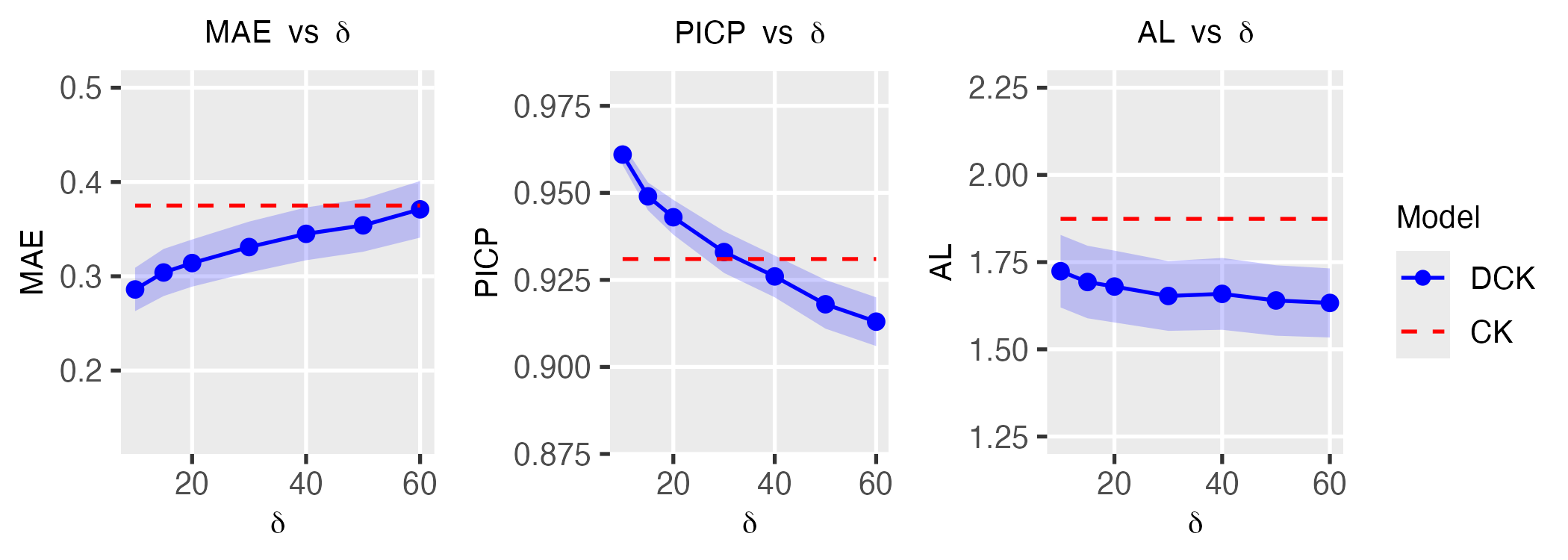}
    \caption{Performance with different $\delta$ in the bivariate scenario (non-Gaussian, $N^* = 3600$).}
\end{figure}

\begin{figure}[t!]
\centering

\begin{tabular}{cc}
\includegraphics[width=0.4\textwidth]{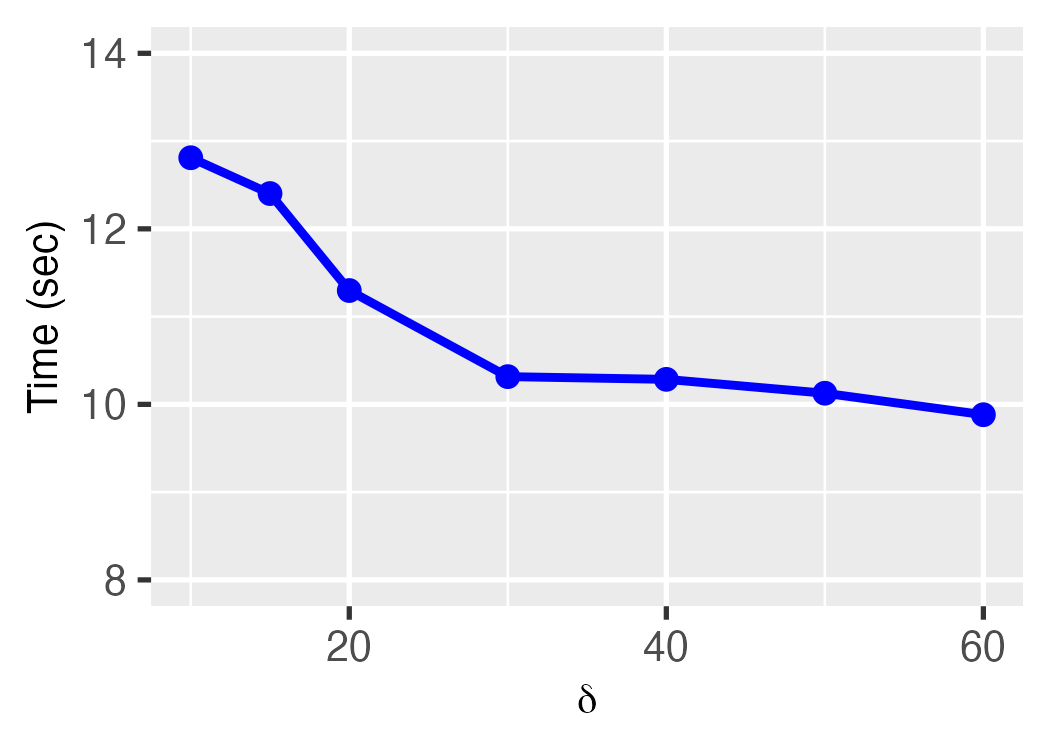} &
\includegraphics[width=0.4\textwidth]{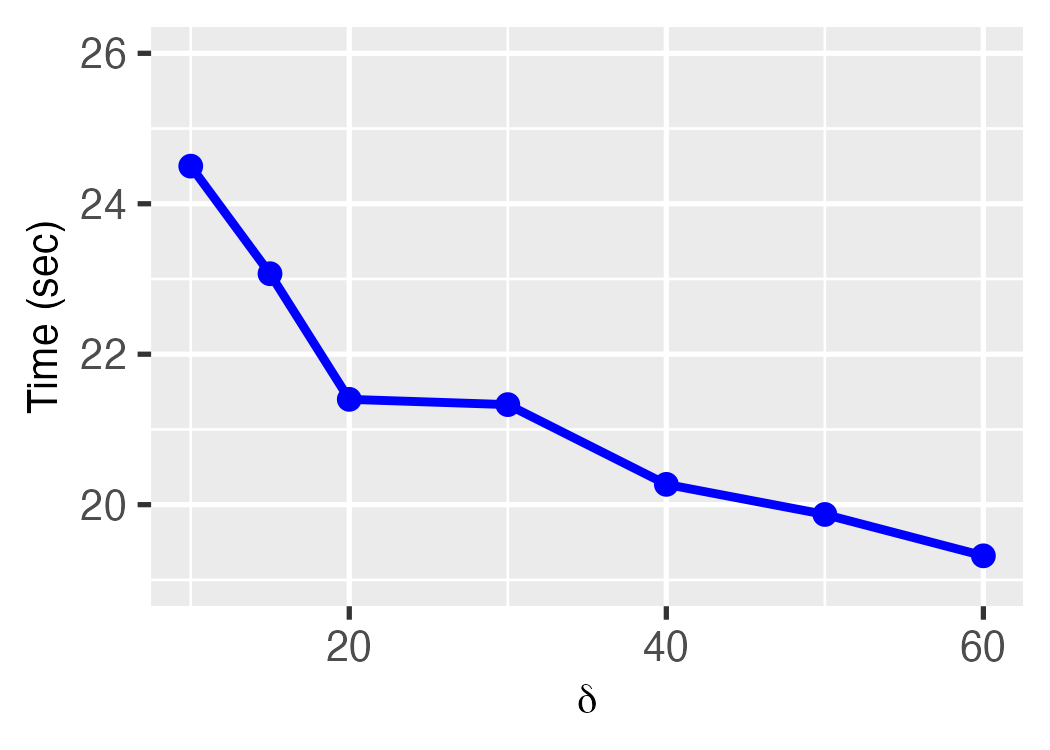} \\
\small (a) Gaussian data, $N^*=1600$ &
\small (b) Gaussian data, $N^*=3600$
\end{tabular}

\vspace{0.35cm}

\begin{tabular}{cc}
\includegraphics[width=0.4\textwidth]{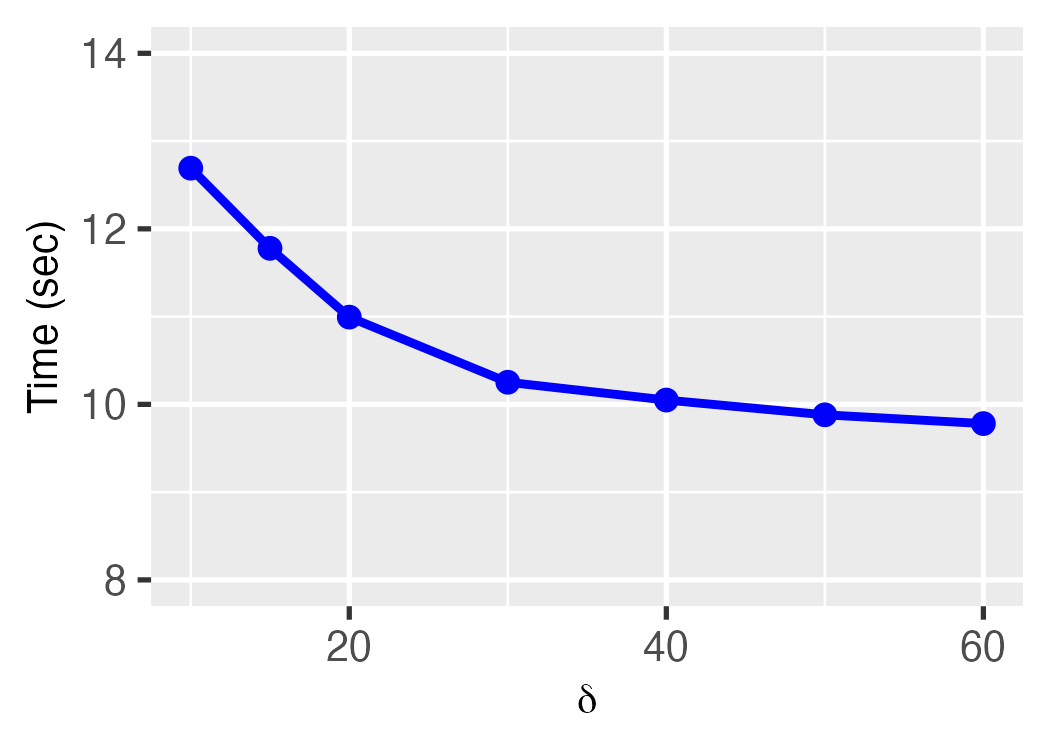} &
\includegraphics[width=0.4\textwidth]{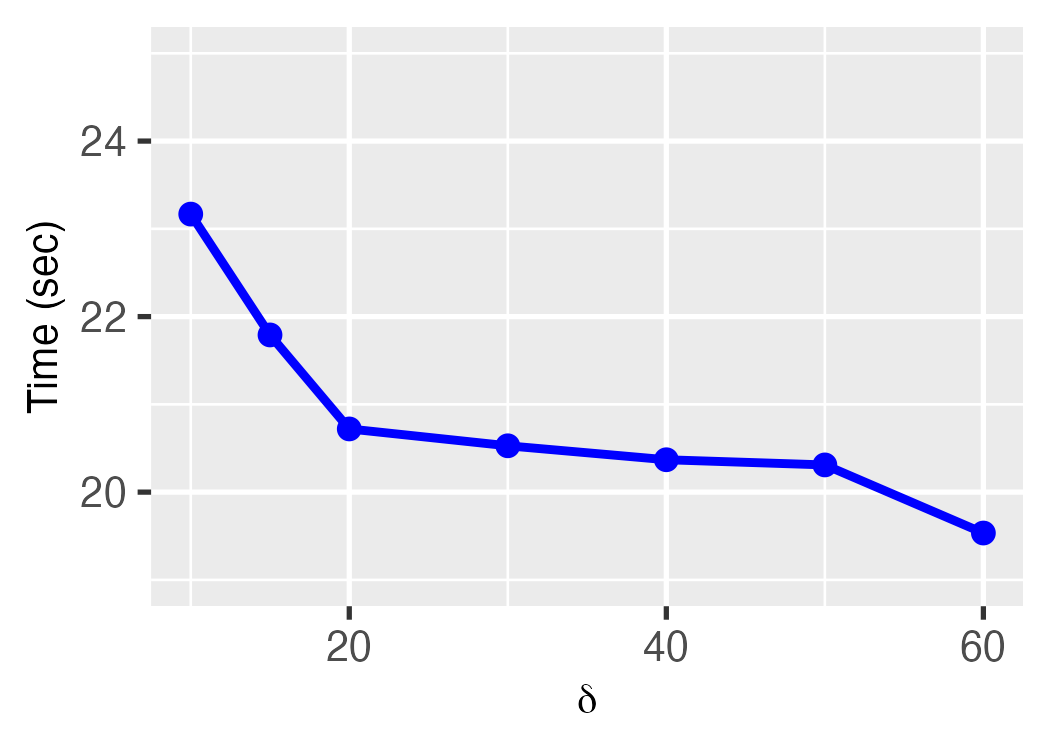} \\
\small (c) non-Gaussian data, $N^*=1600$ &
\small (d) non-Gaussian data, $N^*=3600$
\end{tabular}

\caption{Computation time v.s. $\delta$ under bivariate Gaussian and non-Gaussian scenarios.}
\label{fig:deltatime_all}
\end{figure}

\subsection{\texorpdfstring{Bandwidth $h$}{Bandwidth h}}

Since the method exhibits relatively low sensitivity to $n$, we fix $n = 30$ throughout the univariate simulation studies. This allows us to focus on the sensitivity of the results to the choice of the kernel smoothing bandwidth $h$.
$h$ controls the degree of smoothing in the conditional CDF estimation. By adjusting $h$ via the scaling parameter $C$, one can modulate the smoothness of the DNN-based conditional CDF, thereby affecting both the coverage and width of the resulting prediction intervals.

\begin{figure}[H]
    \centering    \includegraphics[width = \textwidth]{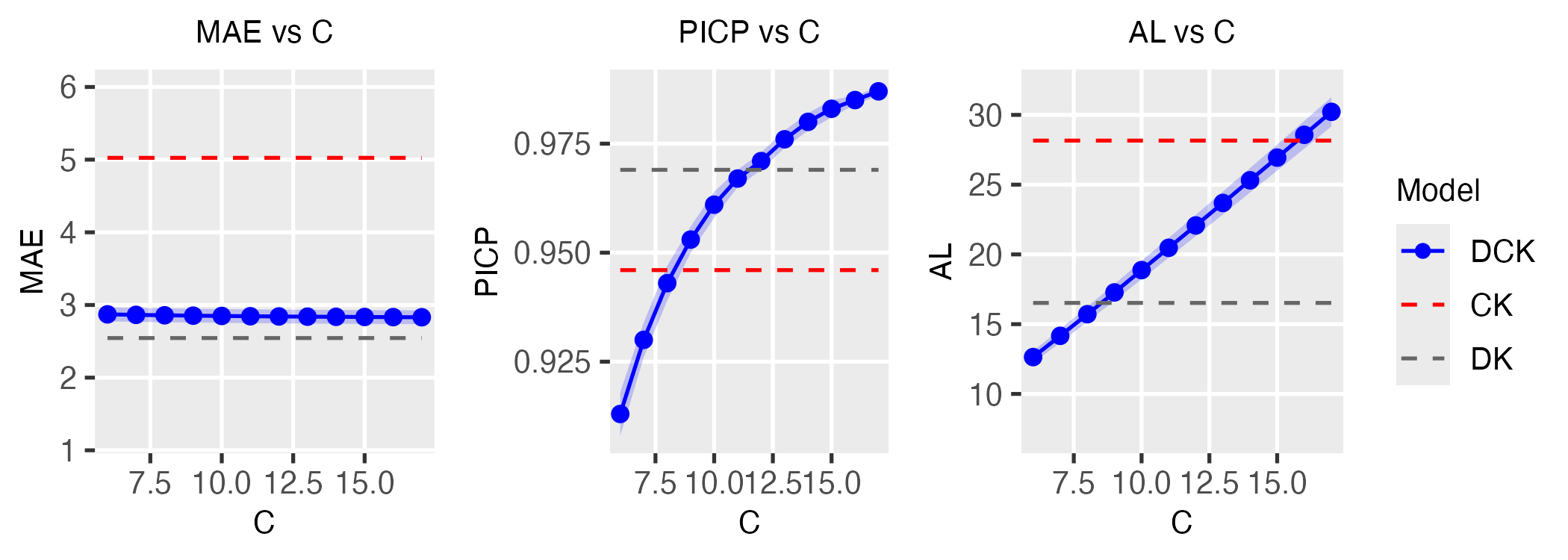}
    \caption{Performance with different $C$ in the univariate scenario (non-Gaussian, $N^* = 1600$).}
    \label{ng1600_c}
\end{figure}

Figure \ref{ng1600_c} illustrates that MAE remains relatively stable across varying $C$, while PICP consistently increases and AL grows monotonically. When $C$ is too small, the PICP falls below the nominal level due to under-smoothed CDF estimates, while overly large $C$ leads to excessively wide intervals and potential over-coverage. This reflects the trade-off between predictive uncertainty and interval coverage: larger $C$ yields wider prediction intervals and improved coverage. 

We further conduct sensitivity analyses for the parameter $C$ across additional scenarios, including univariate Gaussian settings with 1,600 and 3,600 samples, a univariate non-Gaussian setting with 3,600 samples, as well as the bivariate scenario. The bivariate results exhibit trends similar to those observed in the univariate cases. Balancing predictive performance and stability across all settings, a practical range of $C \in [10, 15]$ emerges as robust and reliable. Additional details regarding the selection of this parameter are provided in Section~\ref{sec:computation}.

\begin{figure}[H]
    \centering    \includegraphics[width = \textwidth]{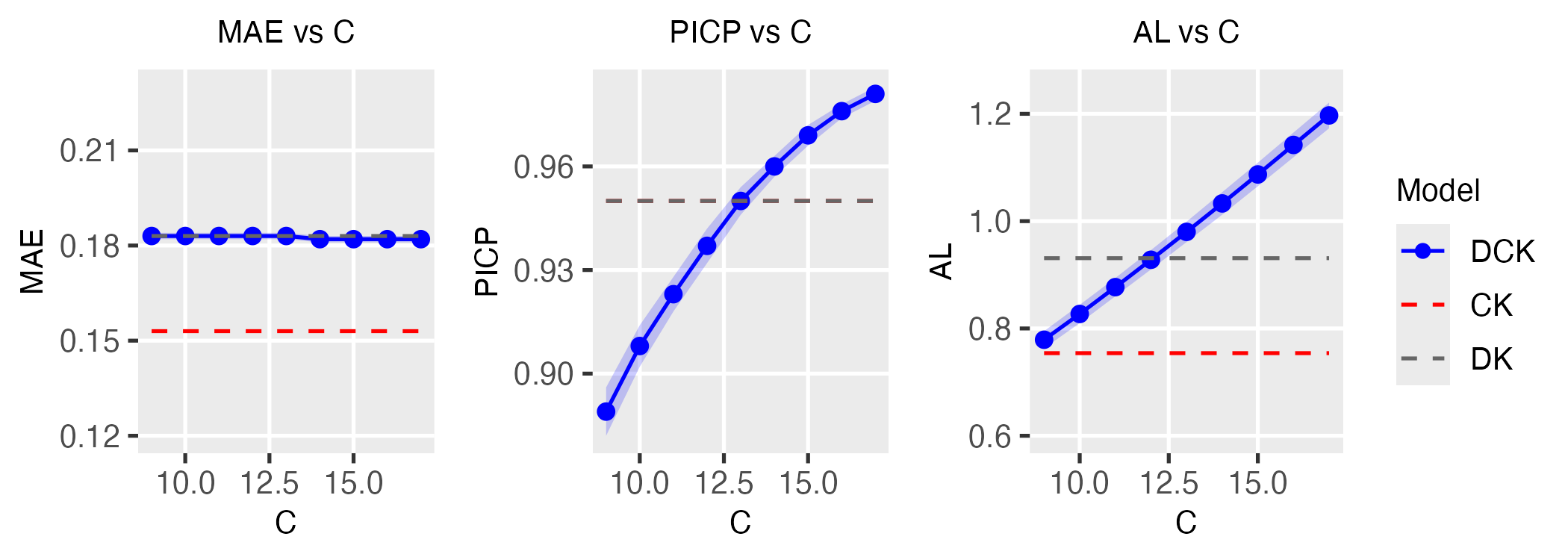}
    \caption{Performance with different $C$ in the univariate scenario (Gaussian, $N^* = 1600$).}
\end{figure}

\begin{figure}[H]
    \centering    \includegraphics[width = \textwidth]{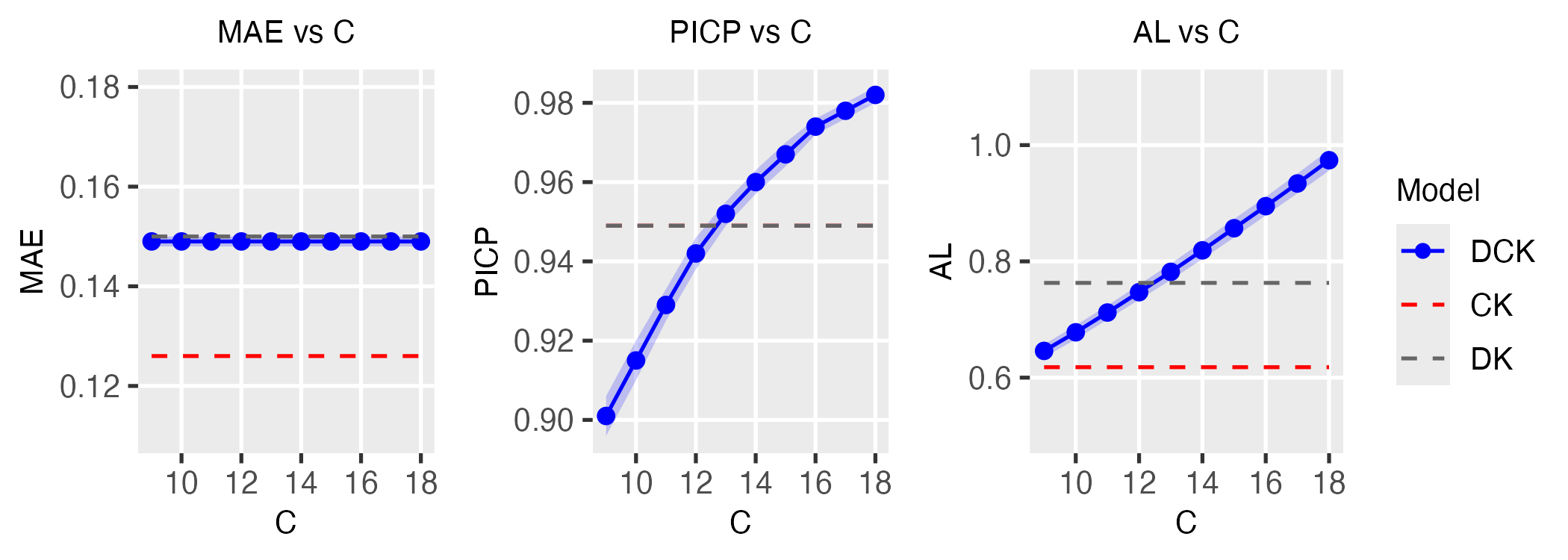}
    \caption{Performance with different $C$ in the univariate scenario  (Gaussian, $N^* = 3600$).}
\end{figure}

\begin{figure}[H]
    \centering    \includegraphics[width = \textwidth]{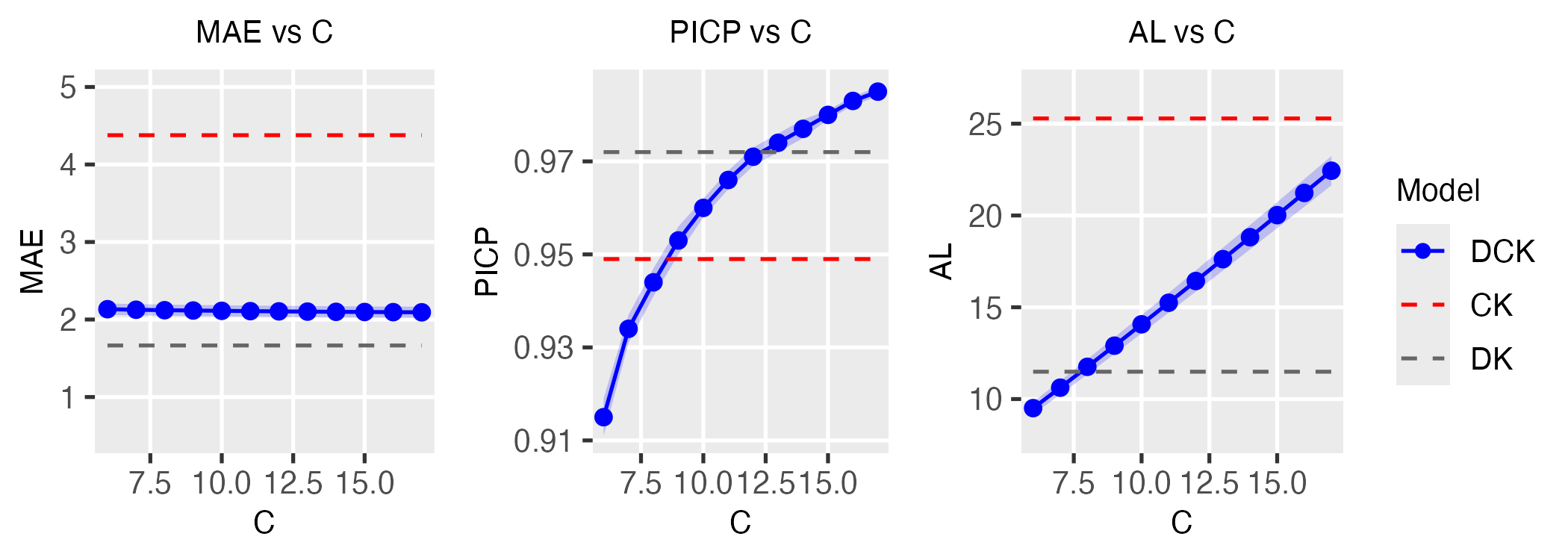}
    \caption{Performance with different $C$ in the univariate scenario  (non-Gaussian, $N^* = 3600$).}
\end{figure}

\begin{figure}[H]
    \centering    \includegraphics[width = \textwidth]{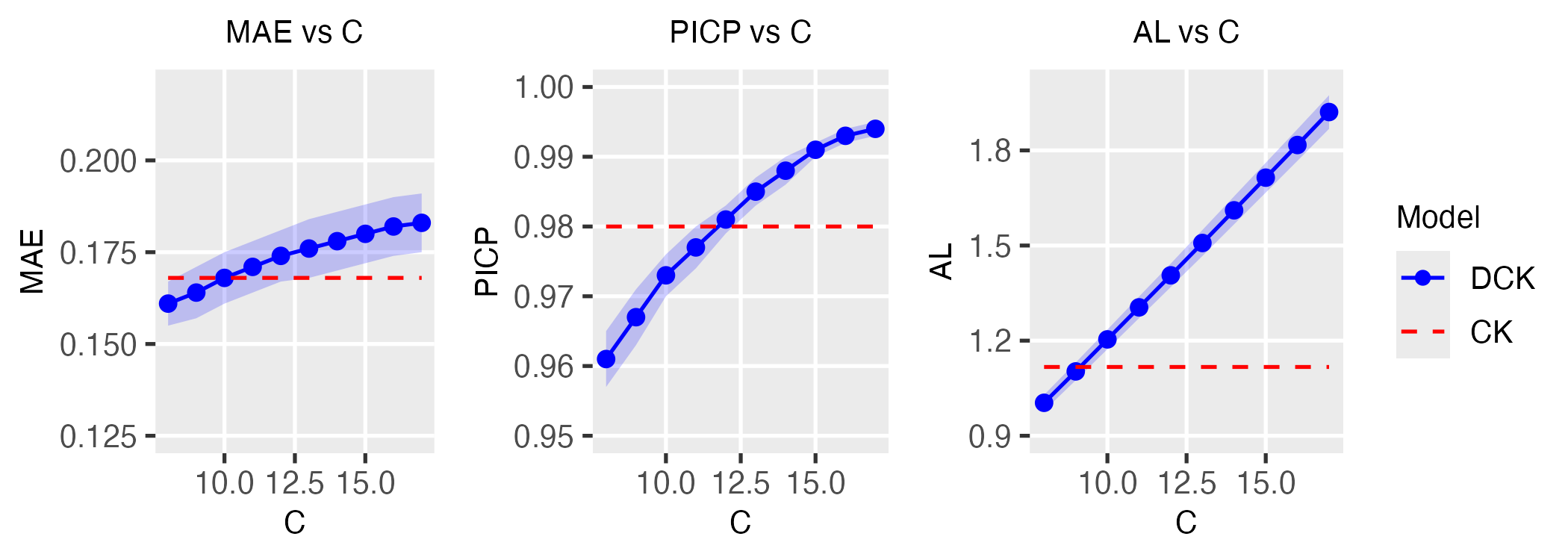}  \caption{ Performance with different $C$ in the bivariate scenario (Gaussian, $N^* = 1600$).}
\end{figure}

\begin{figure}[H]
    \centering    \includegraphics[width = \textwidth]{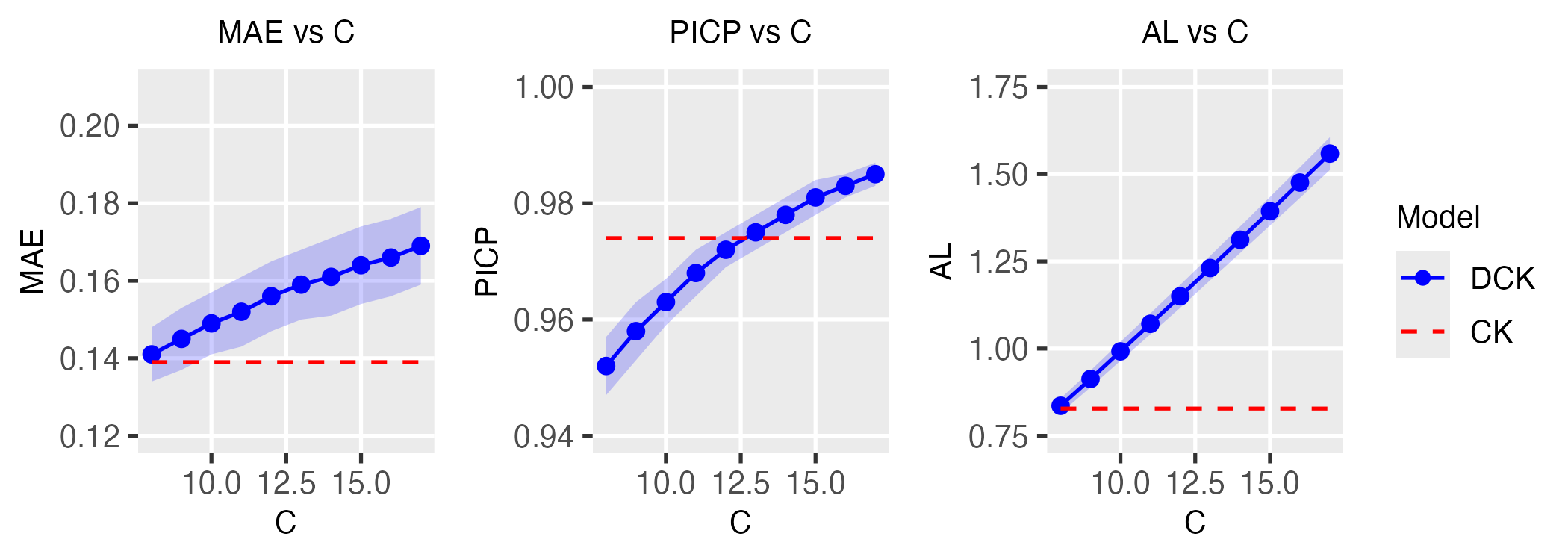}  \caption{Performance with different $C$ in the bivariate scenario (Gaussian, $N^* = 3600$).}
\end{figure}

\begin{figure}[H]
    \centering    \includegraphics[width = \textwidth]{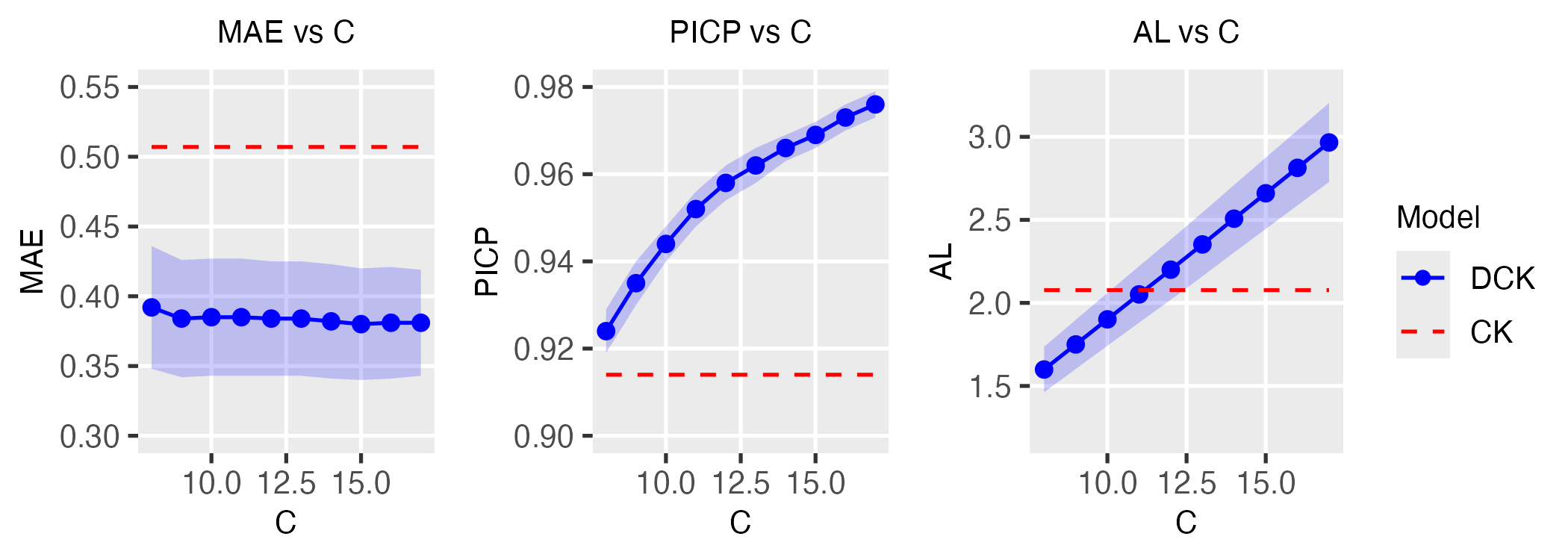}  \caption{Performance with different $C$ in the bivariate scenario (non-Gaussian, $N^* = 1600$).}
\end{figure}

\begin{figure}[H]
    \centering    \includegraphics[width = \textwidth]{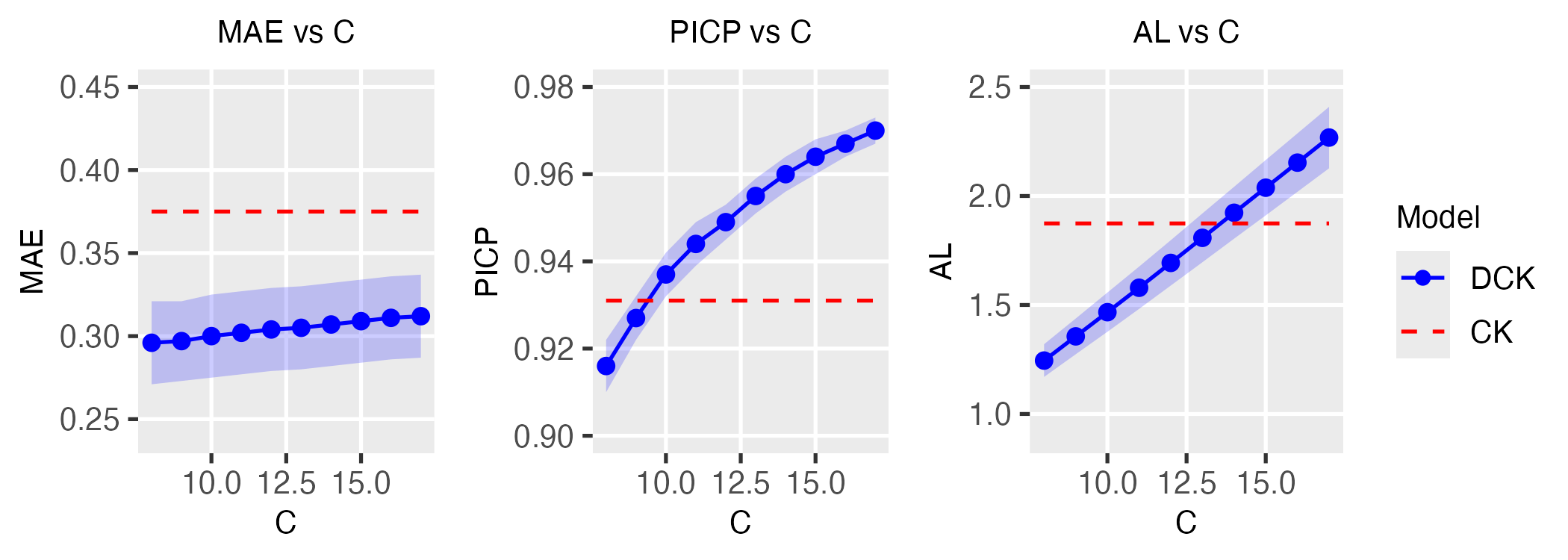}  \caption{Performance with different $C$ in the biivariate scenario (non-Gaussian, $N^* = 3600$).}
\end{figure}

\subsection{\texorpdfstring{Number of regressions $m_1$}{Number of regressions m1}}

In the bivarite scenario, the parameter $m_1$ controls how many quantile regression lines are used to capture the conditional relationship between $Y_1(\cdot)$ and $Y_2(\cdot)$. A larger $m_1$ allows for a finer and more expressive representation of the conditional structure, enabling the model to better capture the patterns in the bivariate distribution. However, increasing $m_1$ also results in fewer projected points per line when the total sample size is fixed, which may lead to sparsity in downstream interval division, thus reducing prediction accuracy.

\begin{figure}[H]
    \centering    \includegraphics[width = \textwidth]{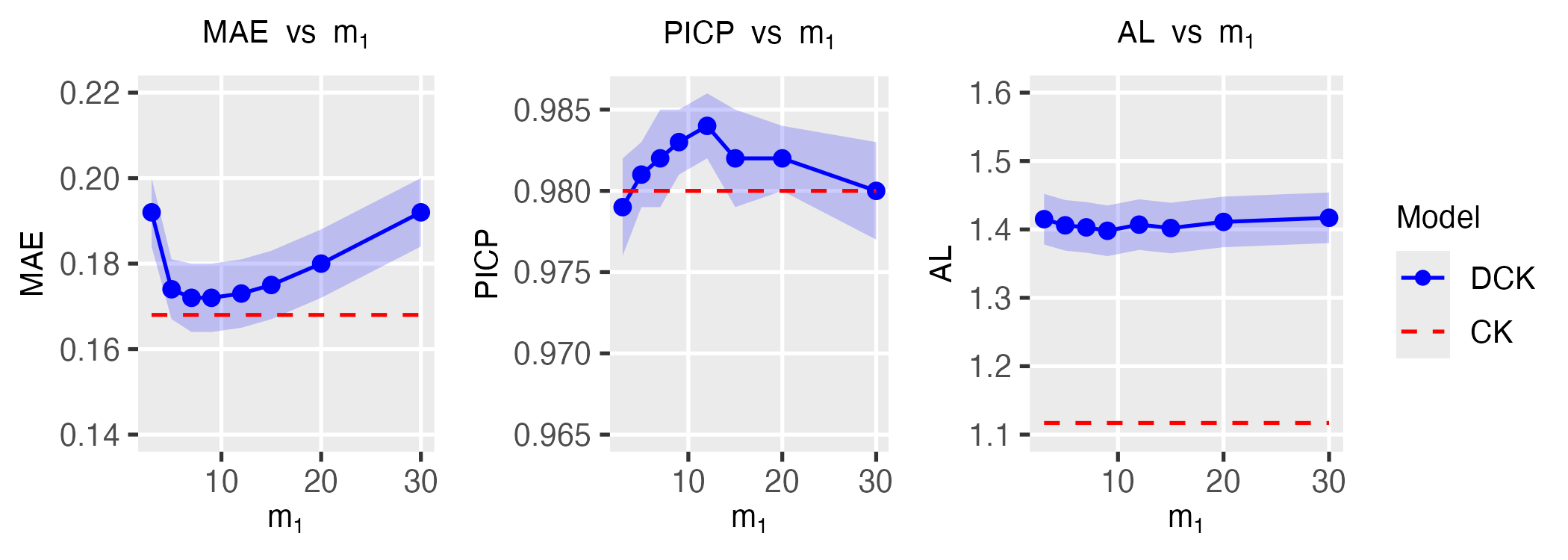}
    \caption{Performance with different $m_1$ in the bivariate scenario (Gaussian data, $N^* = 1600$).}
\end{figure}

\begin{figure}[H]
    \centering    \includegraphics[width = \textwidth]{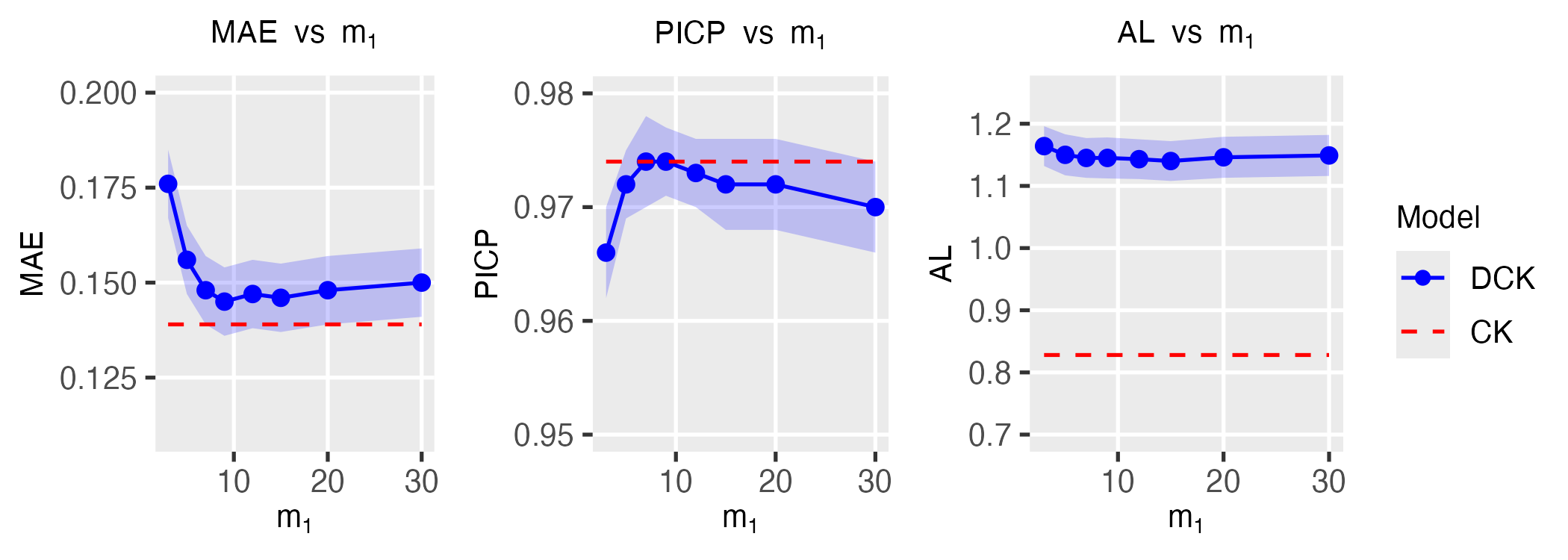}
    \caption{Performance with different $m_1$ in the bivariate scenario (Gaussian data, $N^* = 3600$).}
\end{figure}

\begin{figure}[H]
    \centering    \includegraphics[width = \textwidth]{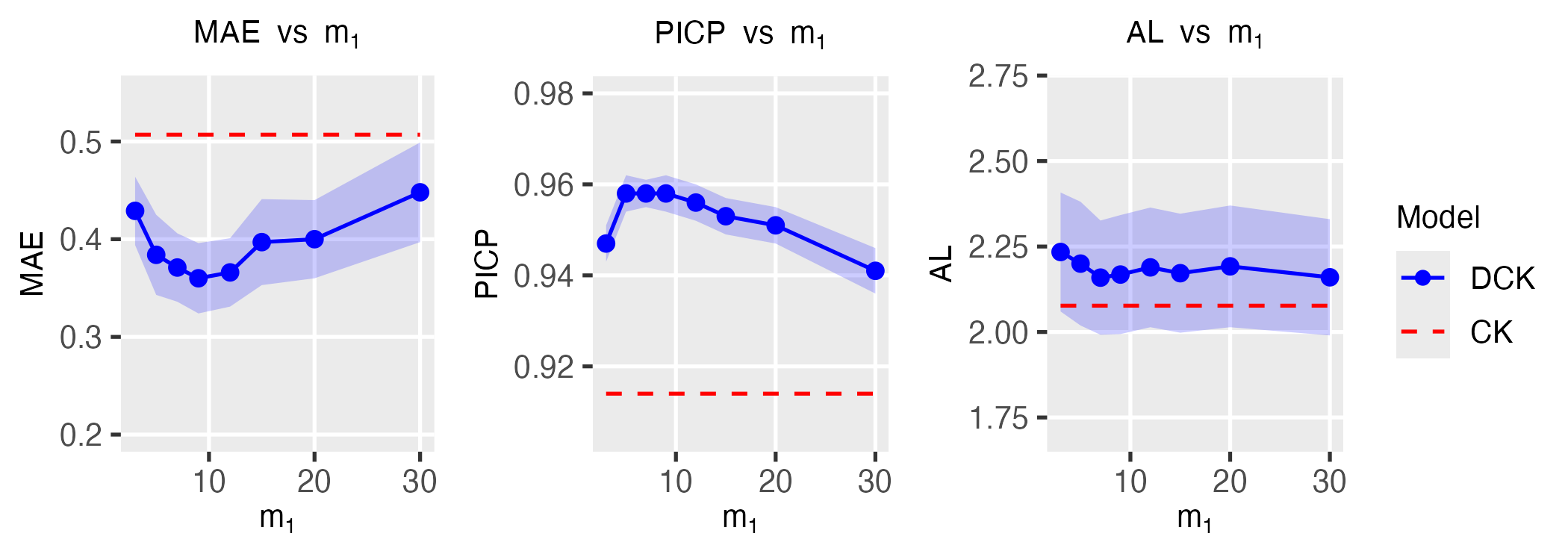}
    \caption{Performance with different $m_1$ in the bivariate scenario (non-Gaussian data, $N^* = 1600$).}
\end{figure}

\begin{figure}[H]
    \centering    \includegraphics[width = \textwidth]{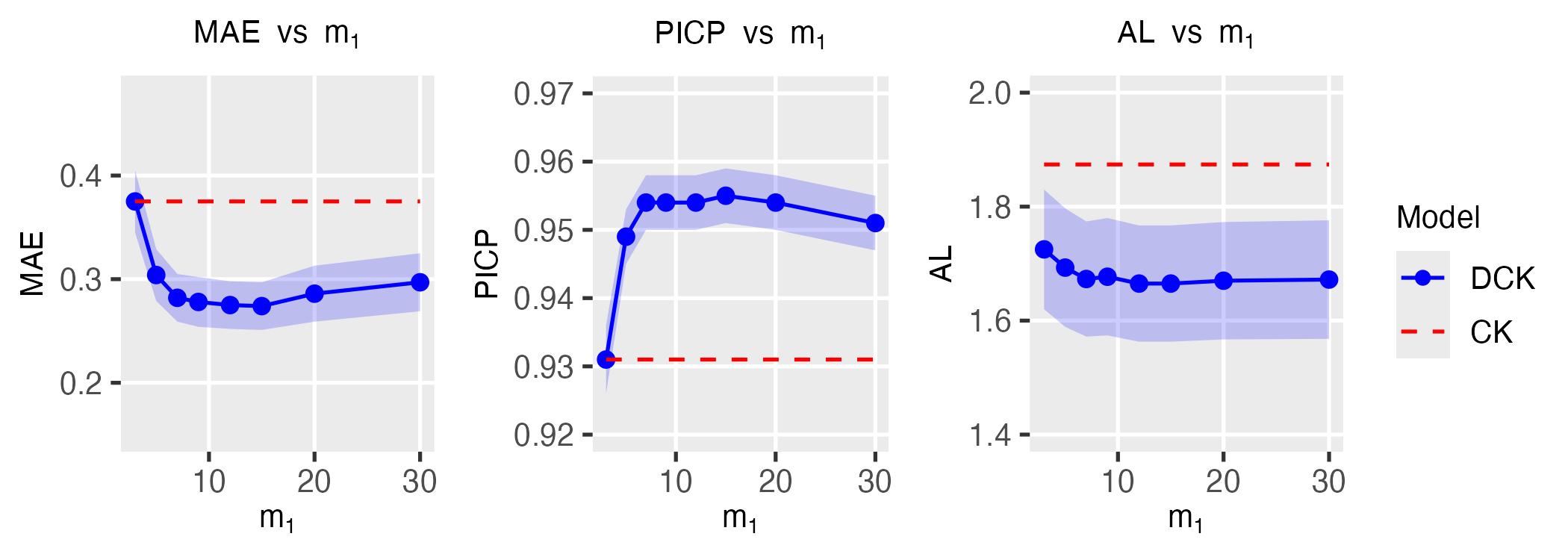}
    \caption{Performance with different $m_1$ in the bivariate scenario (non-Gaussian data, $N^* = 3600$).}
\end{figure}

We can see increasing $m_1$ initially improves the model performance in terms of MAE and PICP. 
As $m_1$ increases, the model gains a better understanding of the conditional relationship, yielding more accurate point predictions and more reliable uncertainty quantification. However, when $m_1$ becomes too large (e.g., $m_1 \geq 30$), the number of samples per regression line becomes insufficient, which negatively affects the stability of the classification task and leads to degraded performance.
We recommend selecting $m_1 \in[5,15]$, with optimal performance typically observed around $5 \sim 10$.

\section{Computational considerations and scalability}\label{sec:computation}

Throughout the sensitivity analysis across all the simulation scenarios, we summarize the following practical guidelines: A practical range for the kernel smoothing parameter is $C \in [10, 15]$.
In univariate scenario, a reasonable choice for the number of quantile thresholds is $n \geq 15$; In bivariate scenario, an effective setting for the number of quantile regression lines is $m_1 \in [5, 15]$, and a recommended upper bound for the minimum number of samples per class is $\delta \leq 50$.

Beyond parameter tuning, it is also important to understand the computational scalability of our method. In particular, we compare the computational complexity of deep neural networks (DNNs) with that of traditional kriging to highlight the efficiency advantage of DNNs in large-scale settings. 
Deep neural networks involve matrix multiplication in several layers to give us the resulting output. Whereas, kriging involves a $N \times N$ matrix inversion. Note that, time complexity of multiplication of one $m \times n$ and $n \times p$ matrix is $O(mnp)$. So  single layer $i$ of the neural network with minibatch size of $b$, $n_i$ input nodes and $m_i$ output nodes will have the time complexity of $O(bn_im_i)$. 
A neural network with $L$ layers will have time complexity $O(\sum_{i=1}^{L}bn_im_i)$. On the other hand, time complexity of Kriging is $O(N^3)$. Hence it can be seen that for large $N$ DNNs with adequate number of layers and nodes is more computationally efficient than classical kriging.  

\section{Further analysis on AQI and \texorpdfstring{PM$_{2.5}$}{PM2.5} concentrations}\label{sec:analysis}

\subsection{Visual comparisons with Gaussian process kriging}\label{sec:visual_ck}

To further assess the spatial behavior of prediction errors and uncertainty quantification of the conditional AQI prediction in Section~\ref{M-sec:comparison_ck} of the main text, we conducted a region-specific comparison between DCK and classical Gaussian process kriging (CK) using localized spatial visualizations.
Focusing on the U.S. Gulf Coast region, spanning Texas, Louisiana, Alabama, and Florida, we examined both pointwise prediction residuals and predictive intervals.
Through residual maps, coverage diagnostics, and representative interval visualizations, we demonstrate that DCK substantially reduces localized bias and provides more reliable and spatially consistent uncertainty estimates than CK.

\begin{figure}
    \centering
    \includegraphics[width = \textwidth]{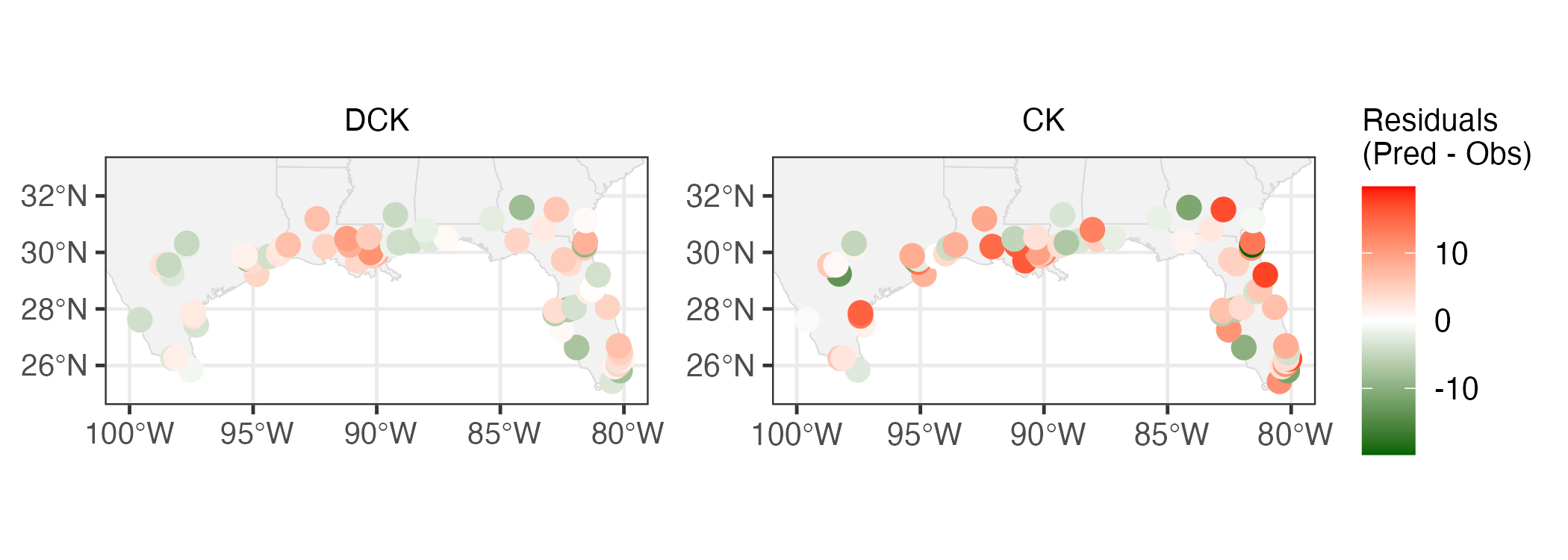}
    \caption{Spatial distribution of prediction residuals for DCK and CK models over the U.S. Gulf Coast region.}
    \label{fig:map}
\end{figure}

Firstly we plotted the prediction residuals (Predicted - Observed) for both the DCK and CK models in this area (Figure \ref{fig:map}).
The color gradient represents the direction and magnitude of prediction error, with red indicating overprediction and green indicating underprediction.
As shown, the CK model exhibits both severe over-predictions and localized under-predictions, with large positive residuals clustered around Louisiana and the Florida Panhandle, suggesting pronounced spatial bias and overconfident estimates.
In contrast, the DCK model produces more balanced and spatially homogeneous residuals, with smaller deviations around zero, indicating improved calibration and reduced local bias.
This regional visualization confirms that DCK more effectively captures spatial heterogeneity and mitigates systematic prediction errors in coastal environments where Gaussian assumptions are less appropriate.

\begin{figure}
    \centering
    \includegraphics[]{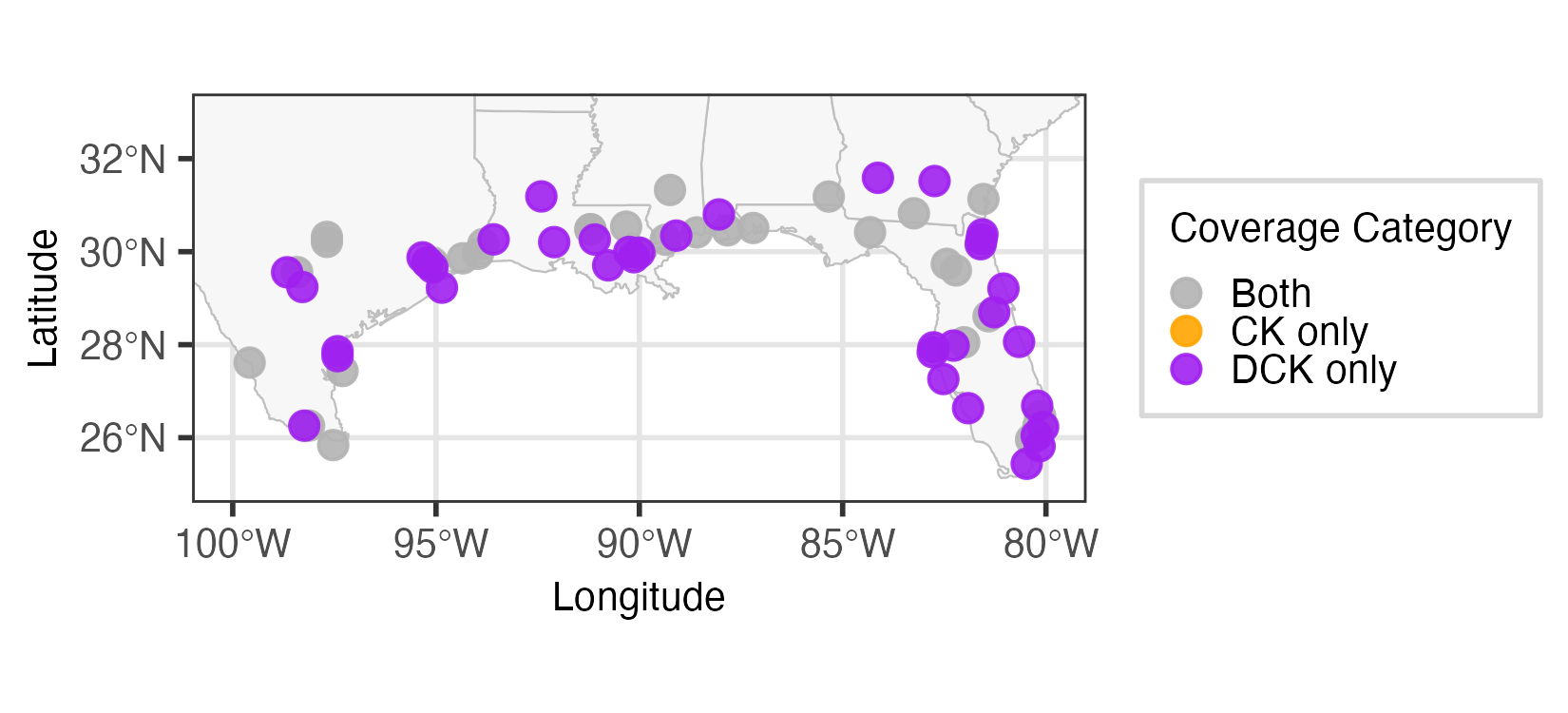}
    \caption{Comparison of predictive interval coverage between DCK and CK over the Gulf Coast region.}
    \label{fig:map2}
\end{figure}

Besides point estimates for the residuals, we further examined whether the predictive intervals were properly calibrated across the same region. As shown in Figure \ref{fig:map2}, we first visualized which test locations were successfully covered by each model’s 95\% predictive intervals. The purple points indicate sites where DCK’s intervals contained the observed AQI values while CK’s did not, whereas gray points represent locations covered by both models. Most of the CK failures are concentrated along coastal Louisiana and the Florida Panhandle, regions characterized by strong spatial gradients and non-Gaussian behavior. This visual evidence confirms that DCK not only improves local calibration but also provides more spatially consistent uncertainty representation.

\begin{figure}
    \centering
    \includegraphics{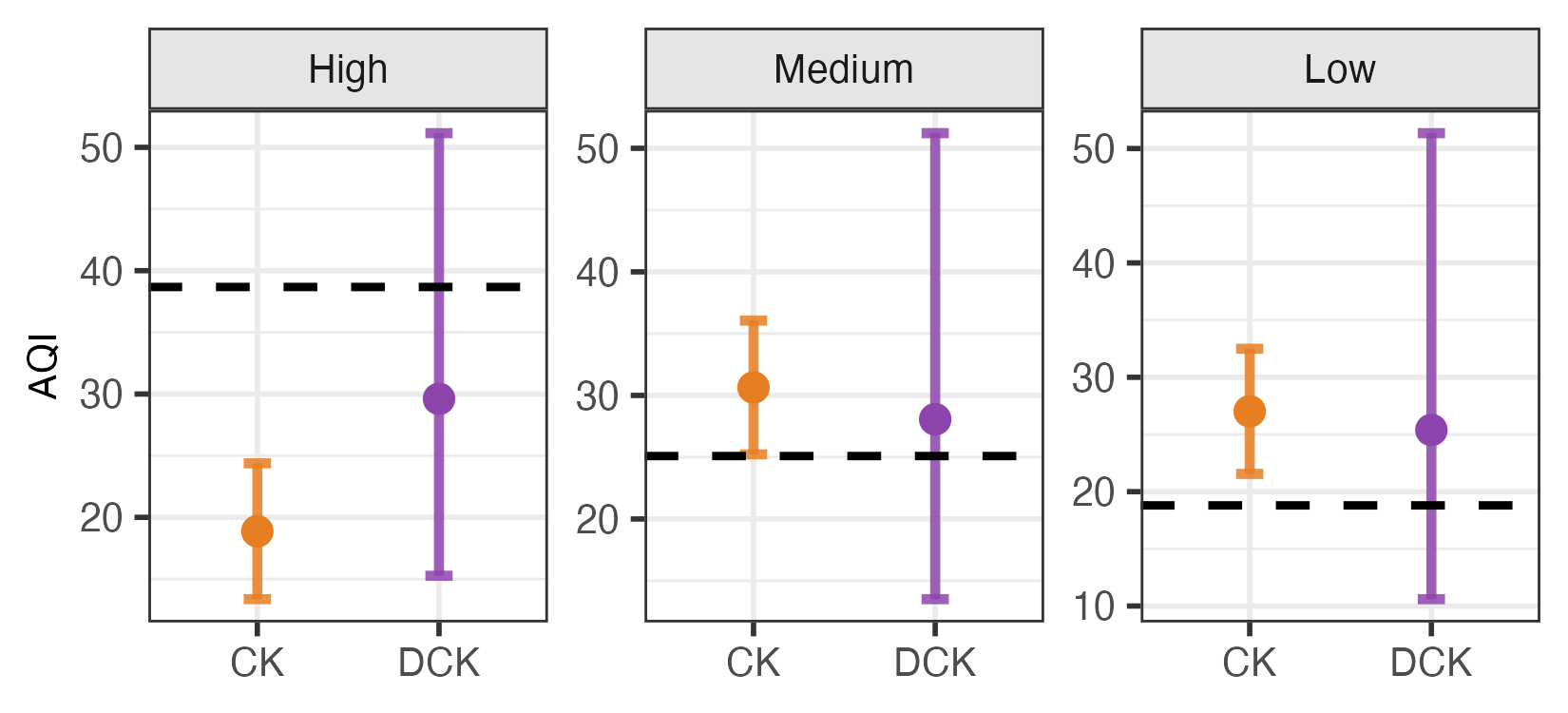}
    \caption{Representative test points illustrating interval calibration under low, medium, and high AQI conditions (DCK-only coverage).}
    \label{fig:interval}
\end{figure}

As mentioned earlier, the PICP was 36.1\% for CK and 98.9\% for DCK, highlighting a substantial gap in interval reliability. To illustrate this discrepancy, Figure \ref{fig:interval} presents representative test points from low, medium, and high PM$_{2.5}$ conditions within the Gulf Coast. The DCK intervals (purple) adaptively expand with concentration levels and successfully include the ground truth (black dashed lines), while the CK intervals (orange) remain overly narrow and miss the observations. Together, these results highlight that DCK produces better-calibrated, more adaptive, and spatially stable predictive intervals, capturing heterogeneity and uncertainty that conventional Gaussian assumptions fail to represent.

\subsection{Joint modeling of AQI and \texorpdfstring{PM$_{2.5}$}{PM2.5}}\label{sec:joint}

Following \eqref{M-eq:biv_cdf} in the main text, we compute the statistical dependence between
$Y_1(\mathbf{s}_0)$ and $Y_2(\mathbf{s}_0)$ through their conditional joint distribution.
In the air quality application, $Z_1(\cdot)$ denotes FRM
observations and $Z_2(\cdot)$ represents CMAQ-simulated PM$_{2.5}$ numerical model outputs. This joint analysis is of significant scientific interest, as it enables a more fundamental characterization of the complete dependency structure between the regulatory-grade FRM observations and the CMAQ model outputs.
A robust estimate of this joint distribution is essential for in-depth model diagnostics, revealing complex, non-linear biases in the CMAQ model (e.g., state-dependent over- or under-prediction). Furthermore, it facilitates more sophisticated probabilistic risk assessments, such as quantifying the joint probability of the scenarios where $Z_1(\cdot)$ attains hazardous levels despite concurrently low values of $Z_2(\cdot)$, which is of particular relevance to environmental risk management and public health policy.

To quantitatively compare the competing models DCK, DK, and CK, two standard multivariate proper scoring rules are considered: the energy score (ES, \cite{gneiting2007strictly}) and the variogram score (VS, \cite{scheuerer2015variogram}), the definitions of these are as follows,
\begin{align*}
   \text{ES}= &\frac{1}{|D_0|}\frac{1}{M} \sum_{\mathbf{s}_0 \in D_0}\sum_{m=1}^M \sqrt{\left(Y_1^{(m)}(\mathbf{s}_0)-Z_1(\mathbf{s}_0)\right)^2+\left(Y_2^{(m)}(\mathbf{s}_0)-Z_2(\mathbf{s}_0)\right)^2}\\
   &-\frac{1}{|D_0|}\frac{1}{2 M^2} \sum_{\mathbf{s}_0 \in D_0}\sum_{m=1}^M \sum_{m^{\prime}=1}^M \sqrt{\left(Y_1^{(m)}-Y_1^{(m^{\prime})}\right)^2+\left(Y_2^{(m)}-Y_2^{(m^{\prime})}\right)^2}, \\
   \text{VS} =& \frac{1}{|D_0|}\sum_{\mathbf{s}_0 \in D_0}\left(\left|Z_1(\mathbf{s}_0)-Z_2(\mathbf{s}_0)\right|^\beta-\frac{1}{M} \sum_{m=1}^M\left|Y_1^{(m)}-Y_2^{(m)}\right|^\beta\right)^2,
\end{align*}
where $D_0$ denotes the set of held-out test locations, $\mathbf{Z}(\cdot)=(Z_1(\cdot), Z_2(\cdot))^\top$ denotes the observed bivariate outcome, and $\mathbf{Y}^{(m)}(\cdot)=(Y_1^{(m)}(\cdot),$ $ Y_2^{(m)}(\cdot))^\top$, $m = 1, \dots, M$, represents the set of $M$ Monte Carlo predictive samples drawn from the estimated joint distribution, $\beta$ is the variogram order parameter, set to
$0.5$ in our analysis. For ES the first term measures the average Euclidean distance between the predictive samples and the observation, while the second term accounts for the average pairwise distance among predictive samples. Their difference quantifies the energy distance between the predictive and true distributions.
A smaller ES therefore indicates that the predictive distribution is closer to the true observation.
The VS score on the other hand evaluates how well the model captures the cross-dependence between $Y_1(\cdot)$ and $Y_2(\cdot)$; smaller values suggest a more realistic joint dependence.
For the proposed DCK model, we compute the samples from the estimated CDF given in \eqref{M-eq:biv_cdf} of the main text.
In contrast, DK draws samples independently from site-specific Gaussian distributions,
where the total variance combines model ensemble and residual uncertainties,
whereas CK derives samples from a joint Gaussian process characterized by a spatial covariance structure across all locations. We choose $M = 1000$ for all models to compute these matrices.

\begin{table}[ht]
\centering
\caption{ES and VS scores for DCK, CK, and DK models in the bivariate AQI and PM$_{2.5}$ analysis. Lower values indicate better predictive distribution performance.}
\label{tab:es_vs_results}
\begin{tabular}{||c c c||}
\hline
Method & ES & VS \\
\hline\hline
DCK & 5.2 & 0.5 \\
CK  & 8.5 & 1.3 \\
DK  & 9.2 & 0.9 \\
\hline
\end{tabular}
\end{table}

As shown in Table~\ref{tab:es_vs_results}, the proposed DCK model yields substantially lower ES and VS compared to both CK and DK, indicating that DCK generates predictive distributions that are substantially
closer to the true bivariate observations and more accurately captures the
cross-dependence between $Y_1(\cdot)$ and $Y_2(\cdot)$. Note that the ES from DK is suboptimal, partly because the DK model is trained only on the collocated observations in $U$, and neural networks typically do not perform well with such small sample sizes. In contrast, DCK is trained on the full fused dataset, which improves both marginal calibration and joint dependence modeling, resulting in a more realistic spatial predictive distribution and better overall performance.

\begin{singlespace}
\bibliographystyle{dcu}
\bibliography{ref}
\end{singlespace}